\documentclass[12pt]{article}
\def\llsymbol#1{\@llsymbol{\@nameuse{c@#1}}}
\def\@llsymbol#1{\ifcase#1\or {}\or {'}\or {''}\or {'''}\or
   {''''}\or {'''''}\or  \else\@ctrerr\fi\relaz}
\newcounter{contador}
\newcommand{\letra}{
   \stepcounter{equation}
   \setcounter{contador}{\value{equation}}
   \setcounter{equation}{0}
   \renewcommand{\theequation}{\thecontador\alph{equation}}}
\newcommand{\antiletra}{
   \renewcommand{\theequation}{\arabic{equation}}
   \setcounter{equation}{\value{contador}}}

\usepackage{color}
\usepackage{indentfirst}
\usepackage{eucal}
\usepackage{amsmath}
\usepackage{amsfonts}
\usepackage{amssymb}
\usepackage{mathrsfs}
\usepackage{bm}
\usepackage{color}
\usepackage{indentfirst}
\usepackage{eucal}
\usepackage{amsmath}
\usepackage{amsfonts}
\usepackage{amssymb}
\usepackage{mathrsfs}
\usepackage{bm}
\usepackage{fancyhdr}
\usepackage[nottoc,notlot,notlof]{tocbibind}
\begin{document}
\begin{center}
{\bf Solutions of  Heun's general equation and elliptic Darboux equation}\\
(An earlier version in:  Math. Meth. Appl. Sci., 44,  7165-7206 (2021))\\
%
\vskip 0,4cm {Bartolomeu D. B. Figueiredo \footnote{email:barto@cbpf.br}}\\
 %
{Centro Brasileiro de Pesquisas F\'{\i}sicas},\\
 {Rua Dr. Xavier Sigaud, 150, CEP 22290-180, Rio de Janeiro, RJ, Brasil}
\end{center}

%
\begin{abstract}
\noindent
New solutions for the elliptic Darboux equation are obtained as 
particular cases of 
solutions constructed for Heun's general equation. 
We consider two groups of power series 
expansions and two new groups of expansions in series of Gauss hypergeometric functions. 
The convergence of one group in power series is determined
by means of ratio tests for infinite series, while the other 
groups are designed to solve problems which admit finite-series solutions. 
Actually, we envisage periodic quasi-exactly 
solvable potentials for which the stationary one-dimensional  Schr\"odinger equation 
is reduced to the Darboux equation. In general, finite- and infinite-series
solutions are obtained from power series expansions for Heun's equation. However, we show   that 
the Schr\"odinger equation admits 
additional finite-series expansions in terms  of hypergeometric functions for a family of associated Lam\'e potentials used in band theory of solids. {For each finite-series solution
we find as well four infinite-series expansions which are  bounded and convergent
 for all values of the independent variable.}  {Finally we find that it is possible to use transformations of variables in order to generate new solutions for the Darboux equation out of other expansions in series of hypergeometric functions for the Heun equation.}

%
%
\end{abstract}
\tableofcontents
%
%
\section{Introduction }

We find new solutions for the 
elliptic Darboux equation 
from solutions of  Heun's general 
equation 
since this includes the former equation as a particular
case. By using known substitutions of variables for Heun's 
equation, 
we construct two groups of power series 
expansions and two groups of expansions in series of Gauss 
hypergeometric functions. 
Finally, we find that these groups are sufficient
to solve systematically the Darboux equations
 which emerge from the stationary one-dimensional 
Schr\"odinger equation for some periodic 
quasi-exactly solvable (QES) potentials.

The elliptic form of the Darboux equation can be written as  \cite{darboux,humbert}
\begin{equation}\label{darboux-0}
\begin{array}{l}
\frac{d^{2}U(u)}{du^{2}}+\left[h
- \mu(\mu+1)k^2{\rm sn}^2u
-\frac{\nu_1\left(\nu_1+1\right)}{{\rm sn^2}u}-
\frac{\nu_2\left(\nu_2+1\right){\rm dn^2}u}{{\rm cn^2}u}-
\frac{\lambda\left(\lambda+1\right)k^2{\rm cn}^2u}{{\rm dn^2}u}
\right]  U(u)=0,
\end{array}
%
\end{equation}
where $h$, $\mu$, $\nu_1$, $\nu_2$ and $\lambda $ are constants,  
while  $\operatorname{sn}{u}$,
$\operatorname{cn}{u}$ and $\operatorname{dn}{u}$ are the Jacobian elliptic functions \cite{nist} modulus $k$ ($0<k^2< 1$). 
Eq. (\ref{darboux-0}) reduces to the associated Lam\'e equation when $\nu_1(\nu_1+1)=\nu_2(\nu_2+1)=0$, and to 
the Lam\'e equation when $\nu_1(\nu_1+1)$$=\nu_2(\nu_2+1)=$$\lambda(\lambda+1)=0$.

The  Darboux equation is a particular case of 
the Heun  equation which, in an algebraic form \cite{heun,maier}, reads
\begin{equation}\label{heun}
\begin{array}{l}
\frac{d^{2}H(x)}{dx^{2}}+\left[\frac{\gamma}{x}+
\frac{\delta}{x-1}+\frac{\epsilon}{x-a}\right]\frac{dH(x)}{dx}+
 \frac{\alpha \beta x-q}{x(x-1)(x-a)} 
 H(x)=0,
 \quad  \epsilon=\alpha+\beta+1-\gamma-\delta,
 \end{array}
\end{equation}
where $a\in \mathbb{C}\setminus\{0,1\}$, 
and the points $x=0,1,a,\infty$ are regular singularities. 
If $a=0$ or $a=1$ the above equation 
can be reduced to the Gauss hypergeometric equation \cite{erdelyi1}
%
%
\begin{eqnarray}
\label{hypergeometric}
\begin{array}{l}
z(1-z)\frac{d^{2}Z(z)}{dz^{2}}+\big[\mathrm{c}-(\mathrm{a}+\mathrm{b}+1)z
\big]\frac{dZ(z)}{dz}-\mathrm{a}\mathrm{b}\;Z(z)=0,
\end{array}
\end{eqnarray}
which has only three singular points. In fact, there are several cases 
in which the Heun equation is reducible to the 
hypergeometric equation \cite{maier-2,vidunas}; some
of theses are written in Appendix B because they are relevant
for this article.
%

 %
We select the  form (\ref{heun}) because it is the one used for studying 
the substitutions of variables which keep invariant the form of the 
Heun equation \cite{maier}. 
These substitutions  
result from 8 homotopic transformations
(no change of the independent variable) 
and 24 linear fractional transformations of the independent variable. As in the case of the hypergeometric equation, the  substitutions can 
also be used  to transform an initial solution   
into new solutions which, for the Darboux equation,  can be rewritten
in terms of elliptic functions. 
Thus, from an initial solution, the homotopic transformations 
generate a group containing 8 expansions. New groups can be generated by firstly 
applying on the initial solution a fractional transformation
of the independent variable. 


As initial solutions  for Heun's equation we take  a familiar power series around the origin of 
the independent variable \cite{heun,kristensson} and an 
expansion in series of Gauss hypergeometric
functions. 
As a matter of fact, in Section 3 we consider two groups 
of power series whose initial solutions are
the expansions having the form 
\begin{eqnarray}\label{seires-0}
\mathring{H}^{(1)}(x)=\displaystyle \sum_{n=0}^{\infty}\mathring{b}_{n}^{(1)}
x^{n}\qquad \mbox{and}\qquad 
\mathring{\bm{H}}^{(1)}(x)= 
\displaystyle \sum_{n=0}^{\infty}\mathring{\bm{b}}_{n}^{(1)}
(1-x)^{n},
\end{eqnarray}
around the points $x=0$ and $x=1$. 
We will find that $\mathring{\bm{H}}^{(1)}$ results
from $\mathring{H}^{(1)}$ by means of the linear fractional transformation
$x\mapsto 1-x$. Furthermore, the homotopic transformations lead 
to 16 expansions, $\mathring{H}^{(i)}$ and $\mathring{\bm{H}}^{(i)}$ ($i=1,2,\cdots,8$).
The 
mathring symbol $\mathring{}$  over $\mathring{H}^{(i)}$ and $\mathring{\bm{H}}^{(i)}$
distinguishes these expansions from 
the expansions in series of hypergeometric functions, ${H}^{(i)}$ and ${\bm{H}}^{(i)}$.

On the other side, in Section 4 we 
consider two groups of expansions in series of hypergeometric 
functions, $F(\mbox{a}, \mbox{b};\mbox{c};z)$, generated from 
\begin{align} 
&
{{H}}^{(1)}(x)=
\begin{array}{l}
\left(1-\frac{x}{a}\right)^{-\alpha} 
\end{array}
\displaystyle \sum_{n=0}^{\infty}
\begin{array}{l}
\frac{b_{n}^{(1)}}{\Gamma(n+\gamma)}\left[\frac{(a-1)x}{a-x}\right]^{n}
F\left[n+\alpha,\gamma+\delta-\alpha-1;n+\gamma;\frac{(a-1)x}{a-x}\right],
\end{array}
\label{hiper-bar-zero-1-0}\\
&
{\bm{H}}^{(1)}(x)= \displaystyle \sum_{n=0}^{\infty}
\begin{array}{l}
\frac{\bm{b}_{n}^{(1)}}{\Gamma(n+\delta)}
(1-x)^{n}
F\left(n+\alpha,\gamma+\delta-\alpha-1;n+\delta;
1-x\right).
\end{array}
\label{h-bold-1-0}
\end{align}
The expansions (\ref{hiper-bar-zero-1-0}) and 
(\ref{h-bold-1-0}) will be 
obtained from an expansion having the form 
\begin{eqnarray}\label{inicial-0}
&&\begin{array}{l}
\bar{H}^{(1)}(x)= \displaystyle \sum_{n=0}^{\infty}
\bar{b}_{n}^{(1)}x^{n}
F\left(n+\alpha,\gamma+\delta-\alpha-1;n+\gamma;
x\right).\end{array}
\end{eqnarray}
In effect, we will find that 
the linear fractional substitutions
$x\mapsto{(a-1)x}/{(a-x)}$ and $x\mapsto 1-x$
transform $\bar{H}^{(1)}$ into ${{H}}^{(1)}$ and ${\bm{H}}^{(1)}$, respectively.
The homotopic transformations lead once again 
to 16 expansions, ${{H}}^{(i)}$ and 
${{H}}^{(i)}$. The arguments $z$ of the above 
$F(\mbox{a}, \mbox{b};\mbox{c};z)$ are appropriate for 
the Darboux equation because for this
case  $|z|\leq 1$, as we will see.  
For some constraints between the parameters, the hypergeometric-function expansions
reduce to power-series expansions; for example, $\bar{H}^{(1)}=
\mathring{H}^{(1)}$ if $\gamma+\delta=\alpha+1$. {
%
%
We will see as well that the initial solutions,  
$\mathring{H}(x)$ and $\bar{H}^{(1)}(x) $, lead to known solutions for the confluent Heun equation.}


The power-series and hypergeometric-function expansions 
are both given by one-sided 
infinite series ($n\geq 0$) whose coefficients obey three-term recurrence relations.
{These relations constitute a system of homogeneous linear equations which admits non
	trivial solutions only if the determinant of the corresponding matrix vanish ({characteristic equation}). 
	This fact requires a free parameter in the differential equation. The absence of free parameters  
	would demand two-sided infinite series ($-\infty<n<+\infty$)  having 
	a parameter which is not present in the differential equation, as in the case 
	of Mathieu equation (see chapter VI of Ref. \cite{nist}). }


Under certain 
conditions, the one-sided series truncate on the right at $n=N$, 
becoming finite series with $0\leq n\leq N$. 
For finite series the convergence is 
decided by examining each term of the series, 
provided  that the characteristic equation is satisfied. 
For infinite series the convergence 
requires some ratio test between successive terms
of the series.  In the present article, only the 
power expansions
$\mathring{H}^{(i)}$ are intended to be used as 
infinite-series solutions; by this reason, 
there is no concern about ratio tests for  
the other expansions.

Finite-series solutions occur in quasi-exactly solvable (QES) problems of 
quantum mechanics  \cite{turbiner1,ushveridze1}, 
for which one part of the energy spectrum and the corresponding 
eigenfunctions can be computed explicitly.  
In fact,  
a QES problem is sometimes defined as a problem which presents  solutions given 
by finite series with coefficients satisfying  three-term  or higher order
recurrence relations \cite{kalnins}. The ``quasi-exactly solvable"
part of the spectrum is obtained from the finite-series solutions,
but there are also infinite-series solutions which are convergent
and bounded, as we will see.

%


In Section 2 we 
explain how
to get solutions for the Darboux equation 
from solutions of the Heun equation and show that,
for some elliptic potentials, the  
Schr\"odinger equation 
reduces to the Darboux equation. In addition, we
collect mathematical informations used in subsequent sections:
properties of the hypergeometric functions, transformations of the 
Heun equation and some aspects of 
the three-term recurrence relations for the series coefficients.


In Section 3 we present the power series solutions 
$\mathring{H}^{(i)}$ and $\mathring{\bm{H}}^{(i)}$, 
while in Section 4 we construct the expansions  ${H}^{(i)}$ and $\bm{H}^{(i)}$ 
in series in hypergeometric functions. 
 The solutions of the 
 four groups are generated straightforward by applying 
 substitutions of variable on the expansion $\mathring{H}^{(1)}(x)$ written in 
(\ref{seires-0}), and on 
 $\bar{H}^{(1)}(x)$ written in (\ref{inicial-0}). Despite this, all the expansions are written
 explicitly for the purposes of applications.


The largest part of the present study is Section 5 which deals with a family of QES potentials 
which appear in band theory of solids. In this case,
the equation reduces to an associated Lam\'e equation and admits finite-series  solutions resulting from power series expansions and from expansions in series of   hypergeometric functions. In addition to
finite-series eigenfunctions (which characterize 
the quasi-exact solvability) we get as well infinite-series
eigenfunctions, convergent and bounded for any value
of the independent variable.

Final remarks and additional references are in Section 6, Appendix A gives the homotopic transformations of the Heun equation, while in Appendix B the Heun equation is reduced to hypergeometric equation when its parameters satisfy certain constraints. In Appendix C, we write degenerate solutions of the 
Schr\"odinger equation for some particular values
of the parameters of the associated Lam\'e potential. {
 Eventually, in Appendix D we discuss the possibility of developing new solutions for the Darboux equation by starting from expansions in series of hypergeometric functions (for the Heun equation) given by  Svartholm \cite{svartholm} and by  Erd\'elyi \cite{erdelyi2,erdelyi3}.}

%
%
\section{Mathematical Preliminaries} 
%

 In  this section  the Darboux equation is 
written as a particular case of the Heun equation, while 
the Schr\"odinger equation is written
 as a Darboux equation for four known elliptic potentials.  We write also some properties of the Gauss hypergeometric functions and select the transformations of variables of the Heun equation which will be used to generate 
solutions appropriate for the Darboux equation. 
Finally we {regard} three-term recurrence relations for the series coefficients and state a known theorem which gives conditions
sufficient for assuring real and distinct solutions for 
the characteristic equation of finite series.


\subsection{Elliptic Darboux equation and the Schr\"{o}dinger equation}

The basic Jacobian elliptic  functions   $\mathrm{sn}(u,k)$,
$\mathrm{cn}(u,k)$ and $\operatorname{dn}(u,k)$ are written as  $\operatorname{sn}{u}$,
$\operatorname{cn}{u}$ and $\operatorname{dn}{u}$, respectively \cite{nist,erdelyi-2}. Here the parameter $k$, called 
modulus of the functions, is supposed 
to be a real number such that $0<k^2<1$. These functions  
satisfy the relations 
\begin{equation}\label{imply}
{\rm sn}^2 u+{\rm cn}^{2}u=k^2{\rm sn}^2 u+{\rm dn}^2 u=1\;\Rightarrow\;
(k^2-1)  {\rm sn}^2 u=  {\rm cn}^2 u- {\rm dn}^2 u,\; 1-k^2={\rm dn}^2u-k^2{\rm cn}^2u.
\end{equation}
Beside these, we will use the functions $\operatorname {cd} u $
and $ \operatorname {sd} u$, defined by
\begin{eqnarray} \label{sd}
\operatorname {cd} u = {\operatorname{cn}u}/{\operatorname{dn}u},\qquad \operatorname {sd} u = {\operatorname{sn}u}/{\operatorname{dn}u}.
\end{eqnarray}
%
%

Now we obtain the Darboux equation from the Heun equation (\ref{heun}) 
by taking $a=k^{-2}$ and achieving the 
substitutions
\begin{eqnarray}\label{substituicoes}
\begin{array}{l}
 H\left[x(u)\right]=\left({
\rm sn}^2u\right)^{\frac{1-2\gamma}{4}}
 \left({\rm cn}^2u\right)^{\frac{	1-2\delta}{4}}
\left( {\rm dn}^2u\right)^{\frac{1-2\epsilon}{4}}U(u)\end{array},\quad
x(u)={\rm sn}^2u, \quad  \left[a=k^{-2}\right].
\end{eqnarray}
which lead to Eq. (\ref{darboux-0}) expressed in terms of the parameters of
the Heun equation, namely, 
\begin{eqnarray}\label{darboux}
&&\begin{array}{l}
\frac{d^{2}U(u)}{du^{2}}+\Big\{h
- \left[(\alpha-\beta)^{2}-\frac{1}{4}\right] k^2{\rm sn}^2u
-\left(\gamma-\frac{1}{2} \right) \left( \gamma-\frac{3}{2} \right)\frac{1}{{\rm sn^2}u}-\end{array}
\vspace{2mm}\nonumber\\
&&\begin{array}{l}
\left(\delta-\frac{1}{2} \right) \left( \delta-\frac{3}{2} \right) \frac{{\rm dn^2}u}{{\rm cn^2}u}-
\left(\epsilon-\frac{1}{2} \right) \left( \epsilon-\frac{3}{2} \right) 
\frac{k^2\;{\rm cn}^2u}{{\rm dn^2}u}
\Big\}  U(u)=0,\quad [\mbox{Darboux Eq,}]
\end{array}
\end{eqnarray}
%
%
%
where
$h=(\gamma+\delta)^2+1-2\gamma-2\delta-
\left[ 4q-(\gamma+\epsilon)^{2}-1+2\gamma+2\epsilon\right]k^2$. 
Therefore, solutions $U(u)$ for the Darboux equation (\ref{darboux}) 
follow from solutions
$H(x)$ for the Heun equation through (\ref{substituicoes}). 
The associated Lam\'e equation and the Lam\'e equation occur when, in Eq. (\ref{darboux}),
%
%
%
\begin{eqnarray}
\label{ass.lame-2}
\begin{array}{l}
(\gamma,\delta)=\left(\frac{1}{2},\frac{1}{2}\right),\;
\left(\frac{1}{2},\frac{3}{2}\right),\;
\left(\frac{3}{2},\frac{1}{2}\right),\;
\left(\frac{3}{2},\frac{3}{2}\right)\qquad [\text{associated Lam\'e equation}];
\end{array}
\end{eqnarray}
%
%
%
%
%
%
%
%
\begin{eqnarray}
(\gamma,\delta,\epsilon)&=&\begin{array}{l}\left(\frac{1}{2},\frac{1}
{2},\frac{1}{2}\right),\;
\left(\frac{1}{2},\frac{1}{2},\frac{3}{2}\right),\;
\left(\frac{1}{2},\frac{3}{2},\frac{1}{2}\right),\;
\left(\frac{1}{2},\frac{3}{2},\frac{3}{2}\right),\end{array}\nonumber
\\
&\mbox{}&\begin{array}{l}\left(\frac{3}{2},\frac{1}
{2},\frac{1}{2}\right),\;\left(\frac{3}{2},\frac{1}{2},\frac{3}{2}\right),\;
\left(\frac{3}{2},\frac{3}{2},\frac{1}{2}\right),\;
\left(\frac{3}{2},\frac{3}{2},\frac{3}{2}\right)\qquad
[\text{Lam\'e equation}].
\end{array}
\end{eqnarray}
%

Next we regard  
the one-dimensional 
stationary Schr\"{o}dinger 
equation,
%
\begin{eqnarray}
\label{schr}
\begin{array}{l}
\frac{d^2\psi(u)}{du^2}+\big[{\cal E}-{\cal V}(u)\big]\psi(u)=0, 
\quad u=\kappa \mbox{x}, \
\quad {\cal E}=\frac{2M }{\hbar^2 \kappa^2} E,
\quad {\cal V}(u)=\frac{2M }{\hbar^2 \kappa^2} {V}(\mbox{x}),
\end{array}
\end{eqnarray}
for a particle with mass $M$ and energy $E$ -- the constant $\kappa$ has inverse-length dimension, 
$\hbar$ is the Plank constant
divided by $2\pi$, $\mbox{x}$ is the spatial coordinate and 
${V}(\mbox{x})$ is the potential. 
Eq. (\ref{schr}) reduces to the Darboux equation (\ref{darboux}), with 
$\psi(u)=U(u)$,  
for the following  potentials 
${\cal V}_{i}(u)$ ($i=1,2,3,4$). 
The three Ganguly's potentials \cite{ganguly-1,ganguly-2}:
\begin{align}
&\begin{array}{l}
{\cal V}_1(u)=(1-k^2)\left[\frac{2}{\mathrm{cn}^2u}-\frac{(\bm{l}+2)(\bm{l}+3)}
{\mathrm{dn}^2u}\right]\end{array}\nonumber\vspace{2mm}\\
&\hspace{1cm}
\begin{array}{l}
 \stackrel{\text{Eq.(\ref{imply})}}{=\joinrel=\joinrel=}
-2k^2-(\bm{l}+2)(\bm{l}+3)
+2\;\frac{\mathrm{dn}^2u}{\mathrm{cn}^2u}+
(\bm{l}+2)(\bm{l}+3)
\frac{k^2\;\mathrm{cn}^2u}
{\mathrm{dn}^2u}.\end{array}\label{ganguly-1}
\vspace{3mm}\\
&\begin{array}{l}
{\cal V}_2(u)=\frac{2}{\mathrm{sn}^2u}-
\frac{(1-k^2)(\bm{l}+2)(\bm{l}+3)}{\mathrm{dn}^2u}\end{array}
%
\begin{array}{l}
\stackrel{\text{(\ref{imply})}}{=}
-(\bm{l}+2)(\bm{l}+3)+\frac{2}{\mathrm{sn}^2u}+(\bm{l}+2)(\bm{l}+3)
\frac{k^2\;\mathrm{cn}^2u}
{\mathrm{dn}^2u}.\end{array}\label{ganguly-2}\vspace{2mm}\\
&\begin{array}{l}
{\cal V}_3(u)=\bm{m}(\bm{m}+1)\;k^2 \mathrm{sn}^2u+\bm{l}(\bm{l}+1) \frac{k^2\;{\rm cn}^2u}{{\rm dn^2}u},\label{ganguly-3}\end{array}
\end{align}
where $\bm{l}$ and $\bm{m}$ are constants. 
For the potential ${\cal V}_3(u)$ 
the Schr\"odinger equation becomes an associated Lam\'e equation which 
was firstly considered by Khare and Sukhatme \cite{khare,khare2}. On the other side, there is the Ushveridze potential \cite{ushveridze1}
\begin{eqnarray*}
	&&\begin{array}{l}
		{\cal V}_{4}(u)=
		-4\left( \bm{a}-\frac{1}{4}\right) \left(  \bm{a}-\frac{3}{4}\right){\rm dn}^{2}u+
		4\left( \bm{b}-\frac{1}{4}\right)
		\left(  \bm{b}-\frac{3}{4}\right)\frac{{\rm dn^{2}u}}{{\rm sn^{2}u}}+4\left( \bm{c}-\frac{1}{4}\right)
		\left(  \bm{c}-\frac{3}{4}\right)\frac{{\rm dn^{2}u}}{{\rm cn^{2}u}}
	\end{array}
	\vspace{3mm}
	\nonumber\\
	&&\hspace{1.6cm}+\begin{array}{l}
		4\left(\bm{a}+\bm{b}+\bm{c}+\bm{l}-\frac{1}{4} \right)
		\left( \bm{a}+\bm{b}+\bm{c}+
		\bm{l}-\frac{3}{4} \right) k^2(k^2-1)\frac{{\rm sn^2u}}{{\rm dn^2u}},
	\end{array}
\end{eqnarray*}
where $\bm{a}$, $\bm{b}$, $\bm{c}$ and $\bm{l}$ are real constants. By rewriting ${\cal V}_{4}$ as
\begin{eqnarray}\label{ush-1}
&&\begin{array}{l}
{\cal V}_{4	}(u)\stackrel{\text{(\ref{imply})}}{=}
-4\left( \bm{a}-\frac{1}{4}\right) \left(  \bm{a}-\frac{3}{4}\right)-4\left(\bm{a}+\bm{c}+\bm{l} \right)
\left( \bm{a}+2\bm{b}+\bm{c}+
\bm{l}-1 \right)k^2
\end{array}
\nonumber\\
&&\hspace{.3cm}
\begin{array}{l}
-8\left(\bm{b}-\frac{1}{4}\right)\left(\bm{b}-\frac{3}{4}\right)k^2
+4\left( \bm{a}-\frac{1}{4}\right) \left(  \bm{a}-\frac{3}{4}\right)k^2{\rm sn}^{2}u+
4\left( \bm{b}-\frac{1}{4}\right)
\left(  \bm{b}-\frac{3}{4}\right)\frac{1}{{\rm sn^{2}u}}
\end{array}
\nonumber\\
&&\hspace{.3cm}
\begin{array}{l}
+4\left( \bm{c}-\frac{1}{4}\right)
\left(  \bm{c}-\frac{3}{4}\right)\frac{{\rm dn^{2}u}}{{\rm cn^{2}u}}
+4\left(\bm{a}+\bm{b}+\bm{c}+\bm{l}-\frac{1}{4} \right)
\left( \bm{a}+\bm{b}+\bm{c}+
\bm{l}-\frac{3}{4} \right) \frac{k^2\;{\rm cn}^2 u}{{\rm dn^2u}},
\end{array}\qquad
\end{eqnarray}
the Schr\"{o}dinger 
equation (\ref{schr}) once more takes the form (\ref{darboux}) for the 
Darboux equation. 

Thus, if we know solutions $H(x)$ for the Heun equation, the eigenfunctions $\psi(u)$ follow from 
\begin{eqnarray}\label{substituicoes-2}
\psi(u) =U(u)\stackrel{\text{(\ref{substituicoes})}}{=}\left({
\rm sn}{u}\right)^{\frac{2\gamma-1}{2}}
 \left({\rm cn}{u}\right)^{\frac{2\delta-1}{2}}
\left( {\rm dn}{u}\right)^{\frac{2\epsilon-1}{2}}H\left(x\right),
\qquad  \left[a=k^{-2},\;x={\rm sn}^2u \right],
\end{eqnarray}
provided that these $\psi(u)$ are bounded for
any admissible value of the variable $u$. The problem is
QES if there are solutions $H(x)$ given by finite series, 
a fact which occurs when the parameters of the Schr\"{o}dinger equation
satisfy certain conditions. For the associated Lam\'e potential
(\ref{ganguly-3}) the 
conditions stated in \cite{khare2} are: (i) either $\bm{m}-\bm{l}\neq 0$
or $\bm{l}+\bm{m}$ is an integer, (ii) either $\bm{l}$
or $\bm{m}$ is an integer and the other is half an odd integer.
In addition, the second case admits doubly degenerate eigenfunctions, that is, two independent eigenfunctions with the same value for the energy.

\subsection{The Gauss hypergeometric function}

The hypergeometric function
$F(\mathrm{a,b;c};z)$,  
solution of Eq. (\ref{hypergeometric}), is defined 
by the series  \cite{erdelyi1}
\begin{eqnarray*}
F\left(\mathrm{a,b;c};z\right)=F(\mathrm{b,a;c};z)=1+
\frac{\mathrm{ab}}{1!\ \mathrm{c}}z+
\frac{\mathrm{a(a+1)b(b+1)}}{2!\ \mathrm{c(c+1)}}z^2+\cdots
=
\sum_{n=0}^{\infty}
\frac{(\mathrm{a})_n(\mathrm{b})_n}{n!\;(\mathrm{c})_n}
z^n,
%
%
%
\end{eqnarray*}
where $(\mathrm{a})_n$, $(\mathrm{b})_n$ and $(\mathrm{c})_n$
are Pochhammer symbols: $(\mathrm{a})_0=1$, $\cdots$, $(\mathrm{a})_n=\mathrm{a}(\mathrm{a}+1)\cdots(\mathrm{a}+n-1)$ $=$
$\Gamma(\mathrm{a}+n)/\Gamma(\mathrm{a})$, etc. The 
above series is not defined when the parameter $\mathrm{c}$ is zero or negative integer. 
Besides this, if the parameter $\mathrm{a}$ or the parameter $\mathrm{b}$ is zero 
or negative integer, the series reduces to a polynomial. A infinite series 
$F\left(\mathrm{a,b;c};z\right)$ converges absolutely for $|z|<1$ and diverges for $|z|>1$;
by the Raabe test, the series converges on the circle $|z|=1$ if 
Re$(c-a-b)>0$. Thus,
\begin{eqnarray}\label{conver-hyper}
F\left(\mathrm{a,b;c};z\right)& & \mbox{converges absolutely for } |z|<1,\mbox{ diverges for } |z|>1,\nonumber\\
& &
\mbox{converges also on } |z|=1 \mbox{ if Re(c-a-b)}>0.
\end{eqnarray}
%
%
%
%
%
The following relations will be useful:
\begin{eqnarray}\label{euler-1}
&&\begin{array}{l}
F(\mathrm{a,b;c};z)=(1-z)^{\mathrm{c-a-b}}
F(\mathrm{c-a,c-b;c;z})\quad \mbox{for}\quad |z|<1,
\end{array}
\nonumber\vspace{2mm}\\
&&\begin{array}{l}
F\left(\mathrm{a,b;c};1\right)=
\frac{\Gamma(\mathrm{c})\Gamma(\mathrm{c-a-b})}{\Gamma(\mathrm{c-a})\Gamma(\mathrm{c-b})}
\quad \mbox{if}\quad \mbox{Re(c-a-b)}>0.
\end{array}
\end{eqnarray}
{
In Section 4 we also use the relations \cite{erdelyi1}
\begin{eqnarray}
\begin{array}{l}
\frac{d}{dz}\big[z^{\mathrm{c}-1}
(1-z)^{\mathrm{b}-\mathrm{c}+1}
F(\mathrm{a},\mathrm{b};\mathrm{c};z)\big]=
(\mathrm{c}-1)z^{\mathrm{c}-2}
(1-z)^{\mathrm{b}-\mathrm{c}}F(\mathrm{a}-1,\mathrm{b};\mathrm{c}-1;z),\vspace{3mm}\\
\frac{d}{dz}\big[(1-z)^{\mathrm{a}}
F(\mathrm{a},\mathrm{b};\mathrm{c};z)\big]=
-\frac{\mathrm{a}(\mathrm{c}-\mathrm{b})}
{\mathrm{c}}(1-z)^{\mathrm{a}-1}
F(\mathrm{a}+1,\mathrm{b};\mathrm{c}+1;z),
\end{array}
\end{eqnarray}
which imply
\begin{eqnarray}\label{relacao1}
\begin{array}{l}
(1-z)\frac{d}{dz}F(\mathrm{a,b;c};z)=
\left( \mathrm{b}-\frac{\mathrm{c}-1}{z}\right)
F(\mathrm{a,b;c};z)+\frac{\mathrm{c}-1}{z} F(\mathrm{a-1,b;c-1};z),\vspace{3mm}\\
%
(z-1)\frac{d}{dz}F(\mathrm{a,b;c};z) =
-\mathrm{a}F(\mathrm{a,b;c};z)+
\frac{\mathrm{a}(\mathrm{c}-\mathrm{b})}{\mathrm{c}}F(\mathrm{a+1,b;c+1};z).
\end{array}
\end{eqnarray}}
%
%

\subsection{Some transformations  of Heun's general equation}

In 2007 Maier
set down the 192 substitutions of variables which preserve the form of Heun's 
equation.  The table 2 of his paper  \cite{maier} 
displays the transformations of
each parameter and variable of the equation. 
These transformations follow from the composition of 8 homotopic 
substitutions of
the dependent variable $H(x)$ (which keep unaltered the 
independent variable
$x$) with   
24 linear fractional substitutions  
$x\mapsto y(x)=(Ax+B)/(Cx+D)$  (in general these demand a transformation of the independent variable as well).
We rewrite some of these substitutions in a form convenient
for applications.

We denote by $H(x)=H(q,a;\alpha,\beta,\gamma,\delta;x)$
a solution of the Heun equation (\ref{heun}); on the other hand,
the transformations of $H(x)$ are indicated by the symbol 
$M_{i}$ following the order in which they
appear in Maier´s table $ (i=1,2,\cdots,192)$. 
Hence, each $M_i$ transform $H(x)$
into $M_iH(x)$ according to
\begin{eqnarray}\label{prescription}
M_iH(x)=M_iH(a,q;\alpha,\beta,\gamma,\delta;x)=
f_i(x)H\left[\tilde{a},\tilde{q};\tilde{\alpha},\tilde{\beta},
\tilde{\gamma},\tilde{\delta};y(x)\right].
\end{eqnarray}
where $f_i(x)$ and $\left[\tilde{q},\tilde{a};\tilde{\alpha},\tilde{\beta},\tilde{\gamma},
\tilde{\delta};y(x)\right]$ depend on the transformation considered.
For example, the fractional substitutions  
\begin{eqnarray}\label{14}
\begin{array}{l}
 \frac{(1-a)x}{x-a},
\qquad 1-x, \qquad \frac{a(x-1)}{x-a}
\end{array}
\end{eqnarray}
may be represented by $M_{17}$,
$M_{49}$ and $M_{65}$, that is,
\begin{align} 
%
&\begin{array}{l}
M_{17}H(x)=\left(1-\frac{x}{a}\right)^{-\alpha}H \left[1-a,-q +\alpha\gamma; \alpha, -\beta+\gamma+\delta, \gamma, \delta; \frac{(1-a)x}{x-a}\right],
\label{17}\end{array}\vspace{2mm}\\
&\begin{array}{l}
M_{49}H(x)=H (1-a,-q +\alpha\beta; \alpha, \beta, \delta, \gamma; 1-x),\quad
\end{array}\label{49}\vspace{2mm}\\
&\begin{array}{l}
M_{65}H(x)=\left(1-\frac{x}{a}\right)^{-\alpha}H \left[a,q -\alpha(\beta-\delta); \alpha, -\beta+\gamma+\delta, \delta, \gamma; \frac{a(x-1)}{x-a}\right].
\label{65}\end{array}
\end{align}

As a matter of fact, for each substitution of the independent variable,  
there are 8 substitutions of the dependent variable on account of the homotopic transformations.  
These are given by the transformations $M_{1}-M_{4}$
and ${M}_{25}-M_{28}$ of Maier's table but here, as in \cite{arscott-ronveaux} and \cite{lea},  
they are denoted by 
$T_{i}$ ($i=1,\cdots,8$) according to
\begin{eqnarray}\label{arscott-maier}
&&T_1=M_1,\quad
T_2 =M_{25}, \quad
T_3=M_{2},\quad
T_4=M_{26},\nonumber\vspace{2mm}\\
&&T_5=M_3,\quad
T_6=M_{27},\quad
T_7=M_4,\quad
T_8 =M_{28}.
\end{eqnarray}
The 8 transformations are written in Appendix A.

Notice that for the Darboux equation 
(\ref{darboux})
\begin{eqnarray}\label{argumentos}
\begin{array}{l}
x=\mathrm{sn}^2u,\qquad 1-x=\mathrm{cn}^2u,\qquad
\frac{(a-1)x}{a-x}=(1-k^2)\;\mathrm{sd}^2u,\qquad\frac{a(x-1)}{x-a}=\mathrm{cd}^2u,
\end{array}
\end{eqnarray}
where $\operatorname{sd}u$ and $\operatorname{cd}u$ are the functions 
given in (\ref{sd}). From definitions \cite{nist,erdelyi-2}, we know 
that $0\leq \mathrm{sn}^2u\leq 1$, 
$0\leq \mathrm{cn}^2u\leq 1$ and $0< \mathrm{dn}^2u< 1$. Further, the relations \cite{nist}
\begin{eqnarray}\label{cd-sd}
&&\begin{array}{l}
\mathrm{cd}\;u=\mathrm{sn}(u+K)\quad\mbox{and}\quad 
(1-k^2)\;\mathrm{sd}^2u=1-\mathrm{cd}^2u,\end{array}
\vspace{3mm}\\ 
&&%
\mbox{where }K=K(k)=\int_0^{\frac{\pi}{2}}\left[1-k^2 \sin^2 
\theta\right]^{-\frac{1}{2}}d\theta,\nonumber
\end{eqnarray}
assure that 
$0\leq \mathrm{cd}^2u\leq 1$ and $0\leq (1-k^2)\;\mathrm{sd}^2u\leq 1$.
Thus, the variables (\ref{argumentos}) are
inside the interval $[0,1]$ which is suitable for arguments of the
hypergeometric functions, according to (\ref{conver-hyper}). Similarly, the variables $x/a$ and $1-(x/a)$ are also appropriate.
As we are interested in Darboux equation, 
in sections III and IV 
we shall use the homotopic transformations and the 
transformations $M_{17}$ and $M_{49}$ given in Eqs. (\ref{17})
and (\ref{49}).

Therefore, in most cases we will use the substitutions of variables 
as a procedure to generate new solutions out of a given solution
$H(x)$. However, the substitutions can be used to simply transform
 the Heun equation
(\ref{heun}) into another version of itself. We will use this viewpoint in Appendix B in order
to show that, for certain values of the parameters, 
the Heun equation (\ref{heun}) is transformed into 
an equation reducible to the hypergeometric equation (\ref{hypergeometric}).

 \subsection{Three-term recurrence relations and convergence}
%
The following are  known results which are also valid for the confluent cases of the Heun equation \cite{arscott-ronveaux}. 
The three-term recurrence relations for the coefficients $b_n$
of one-sided solutions 
have the form
\begin{eqnarray}\label{recurrence1}
\alpha_{0}b_{1}+\beta_{0}b_{0}=0,
\qquad
 \alpha_{n}b_{n+1}+\beta_{n}b_{n}+
\gamma_{n}b_{n-1}=0\qquad (n\geq1),
\end{eqnarray}
where $\alpha_{n}$, $\beta_{n}$ and $\gamma_{n}$ are 
functions of the parameters of the equation.
This system of homogeneous linear equations 
can be written in  matrix form.  
 %
   %
   %
   %
   %
   %
   %
   %
   %
   %
   %
   %
   %
   %
   %
   %
   %
 %
Then, we see that the system has non-trivial solutions 
only if the 
determinant of a tridiagonal coefficient matrix vanishes; such 
condition is called `characteristic 
equation' and determine the possible values
of some `free' parameter of the equation. 
The characteristic equation 
is also written as the continued fraction  \cite{erdelyi-3,gautschi}
  \begin{eqnarray}
  \label{characteristic}
  \beta_{0}=
  \frac{\alpha_{0}\gamma_{1}}{\beta_{1}-}
  \ \frac{\alpha_{1}\gamma_{2}}  
  {\beta_{2}-}\ \frac{\alpha_{2}\gamma_{3}}{\beta_{3}-}\cdots.
  \end{eqnarray}
%

A series whose coefficients satisfy three-term recurrence relations
like (\ref{recurrence1}) truncates  
on the right-hand side if $\gamma_n=0$ for some value of $n\geq$. 
Precisely, if $\gamma_{N+1}=0$, the series becomes a finite series 
with $0\leq n\leq N$ \cite{arscott}). So, we can write
\begin{eqnarray}\label{truncation}
\begin{array}{l}
\gamma_{N+1}=0\quad\Rightarrow\quad \text{finite series with } n\leq N.
\end{array}
\end{eqnarray}
For finite series, the characteristic equation
is equivalent to the vanishing of the determinant of a $(N+1)\times(N+1)$ matrix.  Precisely 
\begin{eqnarray}
\label{matriz}
\det\left[
\begin{array}{ccccccccc}
\beta_{0} & \alpha_{0} &0   &  &                   \\
\gamma_{1}&\beta_{1}   & \alpha_{1} &                \\
   &  \ddots       &     \ddots        &   \ddots             &          \\
                     & & \gamma_{N-1}& \beta_{N-1}&\alpha_{N-1}\\
                   &   &    0    &\gamma_{N} & \beta_{N}
\end{array}
\right]=0 
\end{eqnarray}
assures that there are  nontrivial solutions for the coefficients $b_n$ ($n=0,\;1,\cdots, N$).

%
%

The following theorem, 
due to Arscott \cite{arscott,arscott-2}, holds for finite series:

{\it Let $\beta_i=\mathcal{B}_i-\Lambda$ for $i=0,1,\cdots,N$, 
where $\Lambda$ is an arbitrary constant. If $\alpha_i$, $\mathcal{B}_i$
and $\gamma_i$ are all real and independent of $\Lambda$, and if each product 
$\alpha_{i-1}\;\gamma_i$ is positive, that is,} 
\begin{equation}\label{autovalores}
\mbox{\it if }\alpha_i, \;\mathcal{B}_i,\; \gamma_i \mbox{ \it are real and independent of }\Lambda, \;
\mbox{\it and if }\;\alpha_{i-1}\;\gamma_i>0\;\mbox{\it for}\; i=1,\cdots,N,\quad 
\end{equation} 
{\it then the vanishing of 
the determinant of the above matrix affords $N+1$ real and 
distinct values for $\Lambda$; if 
$\gamma_i\;\alpha_{i-1}\leq0$
nothing can be said about the nature of $\Lambda$.  }

The previous theorem will be useful to 
analyse the energy spectra of finite-series wavefunctions 
of quasi-exactly solvable problems, as the one discussed 
in Section 5. In these cases the characteristic equation  
becomes an algebraic equation of order $N+1$ in $\Lambda$, 
where $\Lambda $ is a linear function of the energy ${\cal E}$. For each 
value of ${\cal E}$, the series
coefficients can be computed by a forward application of the recurrence 
relations, from $b_0$ upward to $b_N$; this  
gives all the $b_n$ in terms of $b_0$. In this case the convergence is decided by
examining each term of the series.

On the other side, 
the convergence of infinite-series solutions comes
from ratio tests for successive terms of the series
\cite{watson,knopp}.  
In general the three-term relations (\ref{recurrence1}) present two linearly 
independent solutions \cite{kristensson,gautschi} and, consequently, two ratios for $b_{n+1}/b_n$ but, { in order to assure the convergence
	the continued fraction (\ref{characteristic}),
	we have to choose  the solution that Gautschi has called minimal (or minimal at infinity): 
	if  $f_n$ and $g_n$ denote such solutions and $\lim_{n\to\infty}(f_n/g_n)=0$, 
	then $f_n$ is the minimal solution. Alternatives to the term  {\it minimal}
	are give in a footnote on page 25 of Gautschi's paper \cite{gautschi}. 
	For practical purposes, if
	\begin{eqnarray}\label{poinkare-peron}
	\lim_{n\to\infty}\left({b_{n+1}}/{b_n}\right)=t_{1}, \qquad 
	\lim_{n\to\infty}\left({b_{n+1}}/{b_n}\right)=t_{2},\qquad t_1\neq t_2,
	\end{eqnarray}
	then} a Poincar\'e-Perron theorem \cite{gautschi} assures the existence of 
two linearly independent solutions presenting these limits, the minimal 
solution being the one with smaller modulus for the limit, $|t_1|$ or 
$|t_2|$. 

The use of a minimal solution demands  
a backward procedure to compute the ratios $b_{n+1}/b_{n}$, 
from a large value of $n$ downward to $n=0$. 
However, before this, 
it is necessary to solve the characteristic equation 
which affords relations between the parameters of the Heun equation. 
Some considerations concerning solutions of characteristic equation for  infinite series are given in \cite{liu,hodge,falloon}. {We just mention that solving Eq. (\ref{characteristic}) is an important part of the problem since that equation determines, for example, the energy spectra of hydrogen molecular ion
	(confluent Heun equation). }

\section{Power-series solutions}

This section {regards} the two groups 
of power series solutions,  
$\mathring{H}^{(i)}(x)$ around $x=0$ and 
$\mathring{\bm{H}}^{(i)}(x)$ around $x=1$ ($i=1,\cdots 8 $), each group being 
closed with respect to the homotopic transformations (\ref{t1}-\ref{t8}).
The series coefficients are denoted  by $\mathring{b}_{n}^{(i)}$
and $\mathring{\bm{b}}_{n}^{(i)}$, respectively. 

The starting-point solution $\mathring{H}^{(1)}(x)$    
generates the others by means of variable transformations as follows
\begin{eqnarray}
\begin{array}{l}
\mathring{H}^{(i)}(x)=T_i\mathring{H}^{(1)}(x),\qquad 
\mathring{ \bm{H}}^{(1)}(x)=M_{49}\mathring{H}^{(1)}(x),\qquad 
\mathring{ \bm{H}}^{(i)}(x)=T_i\mathring{\bm{H}}^{(1)}(x). 
 \end{array}
 \end{eqnarray}
Thus, the present section uses only the M\"obius substitution 
$M_{49}$ ($x\mapsto 1-x$) written in (\ref{49}). In order to determine the convergence of 
$\mathring{H}^{(i)}$ for infinite series it is necessary to suppose 
that $|a|\neq 1$.  
It will be seen as well that (for a fixed value of
$i$) both series $\mathring{H}^{(i)}$ and 
$\mathring{\bm{H}}^{(i)}$  reduce to finite series
under the same circumstances since $\mathring{\gamma}_n^{(i)}=\mathring{\bm{\gamma}}_n^{(i)}$ in condition (\ref{truncation}); 
however, one of the series 
may satisfy the Arscott condition (\ref{autovalores}) while 
the other may not satisfy.

All solutions $\mathring{H}^{(i)}$ and 
$\mathring{\bm{H}}^{(i)}$ will be used in 
Section 5. {The convergence} of infinite series  $\mathring{H}^{(i)}$ 
holds for minimal solutions of the recurrence relations for the  coefficients
$\mathring{b}_{n}^{(i)}$. At the end 
of Section 3.1, we {find} conditions for 
 linear dependence{ between the   expansions
$\mathring{H}^{(i)}$ and $\mathring{H}^{(i+4)}$  for $i=1,2,3,4$}.
{We show as well that the  expansion 
$\mathring{H}^{(1)}$, by means of 
a confluence procedure, leads to a known expansion
for solutions of the confluent Heun equation (CHE)}.

 \subsection{Series $\mathring{H}^{(i)}(x)$ around $x=0$}

By using the homotopic transformations  $T_i$, the solutions
around $x=0$ are written as 
\begin{eqnarray}\label{power}
\mathring{H}^{(1)}(x)=\displaystyle \sum_{n=0}^{\infty}
\mathring{b}_{n}^{(1)}
x^{n},\qquad \mathring{H}^{(i)}(x)=T_i\mathring{H}^{(1)}(x). 
\end{eqnarray}
The coefficients of the 
recurrence relations (\ref{recurrence1})
for $\mathring{b}_{n}^{(1)}$ are  \cite{arscott-ronveaux}
\begin{eqnarray}\label{rec1}
&&\begin{array}{l}
\mathring{\alpha}_n^{(1)}=a(n+1)(n+\gamma),\quad 
\mathring{\beta}_{n}^ {(1)}= -an(n+\gamma+\delta-1)-
n(n+\alpha+\beta-\delta)-q,\end{array}\nonumber
\vspace{2mm}\\
&&\begin{array}{l}
\mathring{\gamma}_{n}^{(1)}=(n+\alpha-1)(n+\beta-1).\end{array}
\end{eqnarray}
Dividing these relations 
by $n^ 2\mathring{b}_n^{(1)}$, we get
\begin{eqnarray*}
&&\begin{array}{l}
a\left[1+\frac{1+\gamma}{n}+O\left(\frac{1}{n^2}\right)\right]
\frac{\mathring{b}_{n+1}^{(1)}}{\mathring{b}_n^{(1)}}-
\left[a+1+\frac{a(\gamma+\delta-1)+\alpha+\beta-\delta}{n}+O\left(\frac{1}{n^2}\right)\right]
+\end{array}\vspace{2mm}\nonumber \\ 
&&\begin{array}{l}
\left[1+\frac{\alpha+\beta-2}{n}+O\left(\frac{1}{n^2}\right)\right]
\frac{\mathring{b}_{n-1}^{(1)}}{\mathring{b}_n^{(1)}}=0.
\end{array}
\end{eqnarray*}
Then,  the ratios  
$\mathring{b}_{n+1}^{(1)}/\mathring{b}_n^{(1)}$ for large $n$ are 
\begin{eqnarray*} 
\begin{array}{l}
  \frac{\mathring{b}_{n+1}^{(1)}}{\mathring{b}_n^{(1)}}\sim\frac{1}{a}\left[
1+\frac{\epsilon-2}{n}\right]\;
\Leftrightarrow \;
\frac{\mathring{b}_{n-1}^{(1)}}{\mathring{b}_n^{(1)}}\sim a\left[
1-\frac{\epsilon-2}{n}\right]\quad 
 \text{and}\quad
  \frac{\mathring{b}_{n+1}^{(1)}}{\mathring{b}_n^{(1)}}\sim 
1+\frac{\delta-2}{n}\;
\Leftrightarrow \;
\frac{\mathring{b}_{n-1}^{(1)}}{\mathring{b}_n^{(1)}}\sim
1-\frac{\delta-2}{n}.
\end{array}
\end{eqnarray*}
Thus, 
$\displaystyle\lim_{n\to\infty}\mathring{b}_{n+1}^{(1)}\big/\mathring{b}_n^{(1)}$ is $\frac{1}{a}$ or $1$. If $|a|\neq 1$, the two limits are distinct and by the Poincar\'e-Perron 
theorem, mentioned after Eq. (\ref{poinkare-peron}),
%
\begin{eqnarray} \label{minimal-positivo}
\begin{array}{l}
 \mbox{for minimal solution:}\;\;
  \frac{\mathring{b}_{n+1}^{(1)}}{\mathring{b}_n^{(1)}}\sim\frac{1}{a}\left[
1+\frac{\epsilon-2}{n}\right]\;
\text{ if } |a|>1,
\quad 
  \frac{\mathring{b}_{n+1}^{(1)}}{\mathring{b}_n^{(1)}}\sim 
1+\frac{\delta-2}{n}\; \text{ if } |a|<1.
%
%
\end{array}
\end{eqnarray}
Thus, for minimal solution, the regions of convergence follow from
\begin{eqnarray}\label{conve1}
\vline\frac{\mathring{b}_{n+1}^{(1)}x^{n+1}}{\mathring{b}_{n}^{(1)}x^{n}}\vline\sim
\begin{cases}
|x|\left[1+\frac{1}{n}
\text{Re}\left(\delta-2\right)\right], &\mbox{if}\ |a|<1,
\vspace{2mm}\\
\big| \frac{x}{a}\big|\left[1+\frac{1}{n}
\text{Re}\left(\epsilon-2\right)\right], &\mbox{if} \ |a|>1.
\end{cases}
\end{eqnarray}
Then, if $|a|<1$ the solution $ \mathring{H}^{\{1\}}(x)$ converges inside 
the circle $|x|=1$ which contains $x=0$ and $x=a$; by the Raabe's test 
\cite{watson,knopp}   
it converges also on $|x|=1$ if $\mbox{Re}\;\delta<1$.
On the other hand, if $|a|>1$ the solution    
converges inside 
the circle $|x|=|a|$ which contains $x=0$ and $x=1$; by the Raabe's test 
it converges also on $|x|=|a|$ if $\mbox{Re}\;\epsilon<1$.
In sum,  
\begin{eqnarray*}\label{convergence-S01}
 \begin{array}{c}\mbox{for minimal}\\
 \mbox{solution}
 \end{array}
 \begin{cases}
\mbox{if } |a|<1,\; \mathring{H}^{(1)}(x) \mbox{ converges
for } |x|<1 \mbox{ and on } |x|=1 \mbox{ if }
\mbox{Re}\;\delta<1; \vspace{2mm}\\
\mbox{if } |a|>1, \; \mathring{H}^{(1)}(x) \mbox{ converges
for } |x|<|a| \mbox{ and on } |x|=|a| \mbox{ if }
\mbox{Re}\;\epsilon<1.
\end{cases}
\end{eqnarray*}
{Observe that Eq. (\ref{poinkare-peron}) and the limits preceding Eqs. (\ref{minimal-positivo})
	imply that the recurrence relations also admit a non-minimal  solution.}

The remaining solutions $\mathring{H}^{(i)}(x)$ are generated
by the homotopic transformations $T_i$ as indicated in (\ref{power}).
This gives the expansions written below, where the coefficients 
$\mathring{\beta}_n^{(i)}$ are written as polynomials of degree 2 in $n$.
%
%
%
\begin{eqnarray}\label{S1}
 \begin{cases}
 \mathring{H}^{(1)}(x)=\displaystyle \sum_{n=0}^{\infty}
 \mathring{b}_{n}^{(1)}
x^{n},\\
 \mathring{\alpha}_n^{(1)}=a(n+\gamma)(n+1),\quad 
  \mathring{\beta}_n^{(1)}=
 -(a+1)n^2-[a(\gamma+\delta-1)+
\alpha+\beta-\delta)]n\\
-q,\qquad
 \mathring{\gamma}_n^{(1)}=(n+\alpha-1)(n+\beta-1).
\end{cases}
\end{eqnarray}
%
%
%
%
\begin{eqnarray}\label{S2}
 \begin{cases}
\mathring{H}^{(2)}(x)=x^{1-\gamma}\displaystyle \sum_{n=0}^{\infty}
\mathring{b}_{n}^{(2)}x^{n},\\
%
%
%
 \mathring{\alpha}_n^{(2)}=a(n+2-\gamma)(n+1),\quad 
 \mathring{\beta}_n^{(2)}=
- (a+1)n^2-[a(1-\gamma+\delta)+1+\epsilon-\gamma]n\\
-q-
(\delta a+\epsilon)(1-\gamma),\qquad
 \mathring{\gamma}_n^{(2)}=
(n+\alpha-\gamma)(n+\beta-\gamma).
\end{cases}
\end{eqnarray}
%
%
%
\begin{eqnarray}\label{S3}
\begin{cases}
\mathring{H}^{(3)}(x)=(1-x)^{1-\delta}\displaystyle \sum_{n=0}^{\infty}\mathring{b}_{n}^{(3)}
x^{n},\\
\mathring\alpha_n^{(3)}=a(n+\gamma)(n+1),\qquad 
\mathring\beta_n^{(3)}=- (a+1)n^2-[a(1+\gamma-\delta)+\alpha+\beta-\delta]n\\
-q-a\gamma (1-\delta),\qquad
\mathring\gamma_n^{(3)}=(n+\alpha-\delta)(n+\beta-\delta).
\end{cases}
\end{eqnarray}
%
%
%
%
\begin{eqnarray}\label{S4}
\begin{cases}
\mathring{H}^{(4)}(x)=x^{1-\gamma}(1-x)^{1-\delta}
\displaystyle \sum_{n=0}^{\infty}\mathring{b}_{n}^{(4)}
x^{n},\\
\mathring\alpha_n^{(4)}=a(n+2-\gamma)(n+1),\;\;
\mathring\beta_n^{(4)}=- (a+1)n^2-[a(3-\gamma-\delta)+1+\epsilon-\gamma]n\\
-q-a(2-\gamma-\delta)-\epsilon(1-\gamma),\;\;
\mathring\gamma_n^{(4)}=(n+1+\alpha-\gamma-\delta)(n+1+\beta-\gamma-\delta).
\end{cases}
\end{eqnarray}
%
%
%
\begin{eqnarray}\label{S5}
\begin{cases}
\mathring{H}^{(5)}(x)=\displaystyle\left[ 1-\frac{x}{a}\right]^{1-\epsilon} \sum_{n=0}^{\infty}\mathring{b}_{n}^{(5)}
x^{n},\\
\mathring\alpha_n^{(5)}=a(n+\gamma)(n+1),\;\;
\mathring\beta_n^{(5)}= -(a+1)n^2-[a(\gamma+\delta-1)+2\gamma+\delta-\alpha-\beta]n\\
-q-
\gamma(1-\epsilon), \qquad
\mathring\gamma_n^{(5)}=(n+\gamma+\delta-\alpha-1)(n+\gamma+\delta-\beta-1).
\end{cases}
\end{eqnarray}
%
%
%
%
\begin{eqnarray}\label{S6}
\begin{cases}
\mathring{H}^{(6)}(x)=\displaystyle x^{1-\gamma}\left[ 1-\frac{x}{a}\right]^{1-\epsilon}
 \sum_{n=0}^{\infty}\mathring{b}_{n}^{(6)}
x^{n},\\
\mathring\alpha_n^{(6)}=a(n+2-\gamma)(n+1),\quad
\mathring\beta_n^{(6)}= -(a+1)n^2-[a(1-\gamma+\delta)-\alpha-\beta+\\
\delta+2]n-q
-\delta a(1-\gamma)+\alpha+\beta-\delta-1,\quad
\mathring\gamma_n^{(6)}=
(n-\alpha+\delta)(n-\beta+\delta).
\end{cases}
\end{eqnarray}
%
%
%
\begin{eqnarray}\label{S7}
\begin{cases}
\mathring{H}^{(7)}(x)=\displaystyle (1-x)^{1-\delta}\left[ 1-\frac{x}{a}\right]^{1-\epsilon}
\sum_{n=0}^{\infty}\mathring{b}_{n}^{(7)}
x^{n},\\
\mathring\alpha_n^{(7)}=a(n+\gamma)(n+1),\qquad
\mathring\beta_n^{(7)}=- (a+1)n^2-[a(1+\gamma-\delta)-\alpha-\beta+\\
2\gamma+\delta]n
-q-\gamma a(1-\delta)-\gamma(1-\epsilon),\qquad
\mathring\gamma_n^{(7)}=(n-\alpha+\gamma)(n-\beta+\gamma).
\end{cases}
\end{eqnarray}
%
%
\begin{eqnarray}\label{S8}
\begin{cases}
\mathring{H}^{(8)}(x)=
\displaystyle  x^{1-\gamma}(1-x)^{1-\delta}\left[1-\frac{x}{a} \right]^{1-\epsilon}
\sum_{n=0}^{\infty}\mathring{b}_{n}^{(8)}
x^{n},\\
\mathring\alpha_n^{(8)}=a(n+2-\gamma)(n+1),\quad 
\mathring\beta_n^{(8)}=-(a+1)n^2-[a(3-\gamma-\delta)+2-\alpha
-\beta\\
+\delta]n
-q-a(2-\gamma-\delta) 
+\alpha+\beta-\delta-1,\qquad
\mathring\gamma_n^{(8)}=(n+1-\alpha)(n+1-\beta).
\end{cases}
\end{eqnarray}

Since $x$ and $a$ are kept fixed by homotopic transformations $T_i$,  all the solutions $\mathring{H}^{(i)}(x)$
are Heun functions which converge 
inside the circle $|x|=1$ if $|a|<1$
and inside the circle $|x|=|a|$ if $|a|>1$; 
in fact, by 
\begin{eqnarray}\label{convergence-primeiro}
\begin{array}{c}
\mbox{by D'Alembetert's}\\
 \mbox{test}\end{array}
\begin{cases}
\mbox{if } |a|<1\; \mathring{H}^{(i)} \mbox{ converges
	for } |x|<1\;(\Rightarrow x=0 \mbox{ and } x=a); \vspace{2mm}\\ 
\mbox{if } |a|>1 \; \mathring{H}^{(i)} \mbox{ converges
	for } |x|<|a|\;(\Rightarrow x=0 \mbox{ and } x=1). 
\end{cases}
\end{eqnarray}
In addition, by Raabe's test, the transformations 
(\ref{t1}-\ref{t8}) give the following   
conditions for convergence on $|x|=1$
and $|x|=|a|$:
\begin{eqnarray}
\mbox{if } |a|<1
& \begin{cases} 
\mathring{H}^{(1)}, \;\mathring{H}^{(2)} , 
\;\mathring{H}^{(5)} \text{ and } \mathring{H}^{(6)}) \text{ converge on }|x|=1 \mbox{ if }
\mbox{Re}\;\delta<1;\vspace{2mm}\\
\mathring{H}^{(3)}, \;\mathring{H}^{(4)} , 
\;\mathring{H}^{(7)} \text{ and } \mathring{H}^{(8)} \text{ converge on }|x|=1 \mbox{ if }
\mbox{Re}\;\delta>1;\end{cases}\label{S10}\vspace{2mm}\\
%
\mbox{if } |a|>1  
&\begin{cases} 
\mathring{H}^{(1)}, \;\mathring{H}^{(2)} , 
\;\mathring{H}^{(3)} \text{ and } \mathring{H}^{(4)} \text{ converge on }|x|=|a| \mbox{ if }
\mbox{Re}\;\epsilon<1;\vspace{2mm}\\
\mathring{H}^{(5)}, \;\mathring{H}^{(6)} , 
\;\mathring{H}^{(7)} \text{ and } \mathring{H}^{(8)} \text{ converge on }|x|=|a| \mbox{ if }
\mbox{Re}\;\epsilon>1.\end{cases}\label{S80}
\end{eqnarray}
The previous restrictions on $\delta$ and $\epsilon$   
hold only for infinite series since the convergence of finite series is determined by examining each term of the series. {Anyway, 
the behaviour of the solutions at the singular points depends
also on the multiplicative factors which preceed the series.} 

Since the recurrence relations admit two solutions,  Kristensson \cite{kristensson} used the Raabe test to 
determine the convergence of $\mathring{H}^{(1)}(x)$
by supposing that that $|a|>1$.
He found that 
\begin{eqnarray}
\mbox{if } |a|>1 \; \mathring{H}^{(1)} \mbox{ converges
	for} \left\{
\begin{array}{l}
|x|<1 \mbox{ and at } |x|=1 \mbox{ if }
\mbox{Re}\;\delta<1\; (\mbox{dominant solution});\vspace{2mm}\\
|x|<|a| \mbox{ and at } |x|=|a| \mbox{ if }
\mbox{Re}\;\epsilon<1\; (\mbox{minimal solution}).
\end{array}
\right.
\end{eqnarray}
Note the differences between the above treatments: we are considering
only minimal solutions for both $|a|>1$ and $|a|<1$; in addition, 
we are using the homotopic transformations.


In the previous expansions we see that
\begin{eqnarray}\label{S1-S8}
\begin{array}{lllll}
a=0,\quad&\Rightarrow & \mathring\alpha_n^{(i)}=0; \vspace{2mm}\\
a=-1,\quad & \epsilon=\delta,\qquad & q=0&\Rightarrow &
\mathring\beta_n^{(1)}=\mathring\beta_n^{(2)}=
\mathring\beta_n^{(7)}=\mathring\beta_n^{(8)}=0;\vspace{2mm}\\
a=-1,\quad & \alpha+\beta=1+\gamma,& 
q=\gamma(1-\delta) &\Rightarrow & \mathring\beta_n^{(3)}=\mathring\beta_n^{(4)}=
\mathring\beta_n^{(5)}=\mathring\beta_n^{(6)}=0.
\end{array}
\end{eqnarray}
For these sets of parameters, the recurrence relations (\ref{recurrence1}) 
become two-terms  
relations and the Heun equation (\ref{heun}) can be reduced to the 
hypergeometric equation 
(\ref{hypergeometric}) -- see Appendix B.

%
%
%

Finally, by considering the solutions $\mathring{H}^{(i+4)}(x)$ and 
$\mathring{H}^{(i)}(x)$
($i=1,2,3,4$) 
we find that
\begin{eqnarray}\label{ince}
\mbox{for infinite series,}\quad \mathring{H}^{(i+4)}(x)=
\mathring{H}^{(i)}(x)\quad\text{if}\quad |x|<|a|\quad \mbox{and}\quad 1-\epsilon\neq 0,1,2,\cdots.
\end{eqnarray}
%
Relations (\ref{ince}) results 
from \cite{Gradshteyn}
\begin{eqnarray}\label{binomial}
(1+y)^{\kappa}=\sum_{m=0}^\infty \frac{\Gamma(\kappa+m)}{m!\;\Gamma(\kappa)}\;y^m,\qquad
|y|<1,\qquad \kappa\neq 0,1,2,\cdots 
\end{eqnarray}
applied to the factor $[1-x/a]^{1-\epsilon}$, and from  
the Cauchy product for infinite series, namely, 
\begin{eqnarray}\label{cauchy}
\sum_{n=0}^\infty f_n\sum_{m=0}^\infty g_m = \sum_{j=0}^\infty h_j,
\quad\mathrm{where}\qquad h_j=\sum_{l=0}^j f_l\, g_{j-l}.
\end{eqnarray}
For instance, for $	\mathring{H}^{(5)}(x)$ we have
\begin{eqnarray*} 
	\begin{array}{l}
		\mathring{H}^{(5)}(x) \stackrel{\text{(\ref{S5})}}{=}
		\displaystyle \sum_{n=0}^{\infty}\mathring{b}_{n}^{(5)}
		x^{n}\left[ 1-\frac{x}{a}\right]^{1-\epsilon}
		%
		\stackrel{\text{(\ref{binomial})}}{=}
		\displaystyle \sum_{n=0}^{\infty}\mathring{b}_{n}^{(5)}x^n
		\sum_{m=0}^{\infty}
		\left(-\frac{1}{a}\right)^{m}\frac{\Gamma(1-\epsilon+m)}{\Gamma(1-\epsilon)m!}\;x^m
		\vspace{2mm}\\
		%
		\stackrel{\text{(\ref{cauchy})}}{=}\displaystyle \sum_{j=0}^{\infty}x^j
		\left[\frac{1}{\Gamma(1-\epsilon)}\sum_{l=0}^{j}
		\left(-\frac{1}{a}\right)^{j-l}\frac{\Gamma(1-\epsilon+j-l)}{(j-l)!}\;\mathring{b}_{j}^{(5)}\right],
		\;
		|x|<|a|, \;\; 1-\epsilon\neq 0,1,\cdots.
		\end{array}
\end{eqnarray*}
Thence, by taking $j=n$ and comparing the above equation with (\ref{S1}), we get
\begin{eqnarray*}\begin{array}{l}
\mathring{H}^{(5)}(x)=\mathring{H}^{(1)}(x)\quad \text{ with } \quad
\mathring{b}_n^{(1)}=\frac{1}{\Gamma(1-\epsilon)}\displaystyle \sum_{l=0}^{n}\left(-\frac{1}{a}\right)^{n-l}\frac{\Gamma(1-\epsilon+n-l)}{(n-l)!}
\mathring{b}_{l}^{(5)}.
\end{array}
\end{eqnarray*}
%
In fact, Eqs. (\ref{ince}) generalize an Ince result \cite{ince2}  
for solutions of the Lam\'e equation (in which case 
$\gamma=\delta=\epsilon=1/2$, $x=\mathrm{sn}^2{u}$ and $a=1/k^2$).


{
Finally we set the limits of $\mathring{H}^{(1)}$ for the confluent Heun equation (CHE) which, in the form used by Arscott 
\cite{arscott-ronveaux}, reads
\begin{eqnarray}\label{CHE}
x(x-1)\frac{d^2 S(x)}{dx^2}+[-\gamma+(\gamma+\delta)x+\rho x(x-1)]
\frac{dS(x)}{dx}+(\alpha\rho x-\sigma)S(x)=0,\vspace{2mm}\\
 \rho\neq 0,\quad (CHE),\nonumber
\end{eqnarray}
where $\rho$ and $\sigma$ are constants. The points $x=0$ and $x=1$ 
are regular singular  points while $z=\infty$ is an  irregular point. 
Eq. (\ref{CHE}) results from the Heun equation (\ref{heun}) 
by the limits 
\begin{eqnarray}\label{formal}
a, \; \beta,\; q\to\infty\;\; \mbox{such that}\quad \frac{\beta}{a} \to
\frac{\epsilon}{a}\to -\rho, \quad \frac{q}{a}\to -\sigma, \qquad H(x)\to S(x).
\end{eqnarray}}

{
Dividing the recurrence 
relations by $a$ and using the previous limits, the expansion (\ref{S1}) gives  ($ \mathring{b}_{-1}^{(1)}=0 $)
\begin{eqnarray}\label{baber}
 \begin{cases}
 \mathring{S}^{(1)}(x)=\displaystyle \sum_{n=0}^{\infty}
 \mathring{b}_{n}^{(1)}
x^{n},\\
 (n+\gamma)(n+1) \mathring{b}_{n+1}^{(1)}-
 \left[
 n^2+(\gamma+\delta-1
-\rho)n-\sigma\right] \mathring{b}_{n}^{(1)}-\rho
(n+\alpha-1) \mathring{b}_{n-1}^{(1)}=0,
\end{cases}
\end{eqnarray}
as solution to the CHE (\ref{CHE}).  From (\ref{baber}) we can generate
other expansions by means of transformations of the CHE \cite{lea,decarreau1}.  
Actually, expansion (\ref{baber}) includes  the one used in 1935 by Baber and Hass\'e  to obtain convergent solutions to the angular part of the two-center problem of quantum mechanics \cite{baber}.} 

\subsection{Series $\mathring{\bm{H}}^{(i)}(x)$ around $x=1$}

The following power series $\mathring{\bm{H}}^{(i)}(x)$ around 
$x=1$ come from $\mathring{\bm{H}}^{(1)}(x)=
M_{49}\mathring{H}^{(i)}(x)$, $\mathring{\bm{H}}^{(i)}(x)=
T_i\mathring{\bm{H}}^{(1)}(x)$. In the following the
coefficients $\mathring{\gamma}_n^{(i)}$ are the ones of section III.A.
%
%
%
\begin{eqnarray}\label{SS1}
 \begin{cases}
 \mathring{\bm{H}}^{(1)}(x)=\displaystyle 
 \sum_{n=0}^{\infty}\mathring{\bm{b}}_{n}^{(1)}
(1-x)^{n},\\
 \mathring{\bm{\alpha}}_n^{(1)}=(1-a)(n+\delta)(n+1),\quad 
  \mathring{\bm{\beta}}_n^{(1)}=
 (a-2)n^2+[(a-1)(\gamma+\delta-1)-\alpha-\beta
 \vspace{2mm}\\
%
+\gamma]n+q-\alpha\beta,\qquad
 \mathring{\bm{\gamma}}_n^{(1)}=
  \mathring{\gamma}_n^{(1)}.
\end{cases}
\end{eqnarray}
%
%
%
%
%
\begin{eqnarray}\label{SS2}
 \begin{cases}
\mathring{\bm{H}}^{(2)}(x)=x^{1-\gamma}\displaystyle 
\sum_{n=0}^{\infty}\mathring{\bm{b}}_{n}^{(2)}(1-x)^{n},\\
%
%
%
 \mathring{\bm{\alpha}}_n^{(2)}=(1-a)(n+\delta)(n+1),\qquad 
 \mathring{\bm{\beta}}_n^{(2)}=
(a-2)n^2+[(a-1)(1-\gamma+\delta)-
\alpha\\-\beta
+\gamma]n+q
-a(\gamma-1)\delta+(\gamma-1)\delta-\alpha\beta,\qquad \mathring{\bm{\gamma}}_n^{(2)}= \mathring{\gamma}_n^{(2)},
\end{cases}
\end{eqnarray}
%
%
%
\begin{eqnarray}\label{SS3}
\begin{cases}
\mathring{\bm{H}}^{(3)}(x)=(1-x)^{1-\delta}\displaystyle \sum_{n=0}^{\infty}\mathring{\bm{b}}_{n}^{(3)}
(1-x)^{n},\\
 \mathring{\bm{\alpha}}_n^{(3)}=(1-a)(n+2-\delta)(n+1),
 \quad
 \mathring{\bm{\beta}}_n^{(3)}=(a-2)n^2+\big[(a-1)(1+\gamma-\delta)\\
 -\alpha-\beta +\gamma+2\delta-2\big]n+q
-a\gamma(\delta-1)-\alpha\beta+(\delta-1)(\alpha+\beta-\delta+1),\\
 \mathring{\bm{\gamma}}_n^{(3)}=\mathring{\gamma}_n^{(3)}.
\end{cases}
\end{eqnarray}
%
%
%
%
\begin{eqnarray}\label{SS4}
\begin{cases}
\mathring{\bm{H}}^{(4)}(x)=x^{1-\gamma}(1-x)^{1-\delta}
\displaystyle \sum_{n=0}^{\infty}\mathring{\bm{b}}_{n}^{(4)}
(1-x)^{n},\\
 \mathring{\bm{\alpha}}_n^{(4)}=(1-a)(n+2-\delta)(n+1),\qquad
 \mathring{\bm{\beta}}_n^{(4)}= (a-2)n^2+\big[(a-1)(3-\gamma-\delta)\\
 -\alpha-\beta
 +\gamma+2\delta-2\big]n+q
-a(\gamma+\delta-2)-\alpha\beta+\gamma+\delta-2+(\delta-1)\epsilon,\\
 \mathring{\bm{\gamma}}_n^{(4)}=\mathring{\gamma}_n^{(4)}.
\end{cases}
\end{eqnarray}
%
%
%
\begin{eqnarray}\label{SS5}
\begin{cases}
\mathring{\bm{H}}^{(5)}(x)=\displaystyle\left[ 1-\frac{x}{a}\right]^{1-\epsilon} \sum_{n=0}^{\infty}\mathring{\bm{b}}_{n}^{(5)}
(1-x)^{n},\\
 \mathring{\bm{\alpha}}_n^{(5)}=(1-a)(n+\delta)(n+1),\qquad 
  \mathring{\bm{\beta}}_n^{(5)}= (a-2)n^2+
\big[(a-1)(\gamma+\delta-1)\\
+\alpha+\beta
-\gamma-2\delta\big]n+q
-\alpha\beta+(\alpha+\beta-\gamma-\delta)\delta,\qquad
 \mathring{\bm{\gamma}}_n^{(5)}=\mathring{\gamma}_n^{(5)}.
\end{cases}
\end{eqnarray}
%
%
%
%
\begin{eqnarray}\label{SS6}
\begin{cases}
\mathring{\bm{H}}^{(6)}(x)=\displaystyle x^{1-\gamma}\left[ 1-\frac{x}{a}\right]^{1-\epsilon}
 \sum_{n=0}^{\infty}\mathring{\bm{b}}_{n}^{(6)}
(1-x)^{n},\\
 \mathring{\bm{\alpha}}_n^{(6)}=(1-a)(n+\gamma)(n+1),\qquad
 \mathring{\bm{\beta}}_n^{(6)}= (a-2)n^2+\big[(a-1)(1-\gamma+\delta)\\
 +\alpha+\beta
 -\gamma-2\delta\big]n+q
-a(\gamma-1)\delta -\alpha\beta+(\alpha+\beta-\delta-1)\delta,
\qquad
\bm{\gamma}_n^{(6)}=\mathring{\gamma}_n^{(6)}.
\end{cases}
\end{eqnarray}
%
%
%
\begin{eqnarray}\label{SS7}
\begin{cases}
\mathring{\bm{H}}^{(7)}(x)=\displaystyle (1-x)^{1-\delta}\left[ 1-\frac{x}{a}\right]^{1-\epsilon}
\sum_{n=0}^{\infty}\mathring{\bm{b}}_{n}^{(7)}
(1-x)^{n},\\
 \mathring{\bm{\alpha}}_n^{(7)}=(1-a)(n+2-\delta)(n+1),\qquad
 \mathring{\bm{\beta}}_n^{(7)}= (a-2)n^2+\big[(a-1)(1+\gamma-\delta)\\
 +\alpha+\beta
 -\gamma-2\big]n+q -a\gamma(\delta-1)-\alpha\beta+\gamma(\delta-1)+
\alpha+\beta-\gamma-1,\\
 \mathring{\bm{\gamma}}_n^{(7)}=\mathring{\gamma}_n^{(7)}.
\end{cases}
\end{eqnarray}
%
%
\begin{eqnarray}\label{SS8}
\begin{cases}
\mathring{\bm{H}}^{(8)}(x)=
\displaystyle  x^{1-\gamma}(1-x)^{1-\delta}\left[1-\frac{x}{a} \right]^{1-\epsilon}
\sum_{n=0}^{\infty}\mathring{\bm{b}}_{n}^{(8)}
(1-x)^{n},\\
 \mathring{\bm{\alpha}}_n^{(8)}=(1-a)(n+2-\delta)(n+1),\qquad 
 \mathring{\bm{\beta}}_n^{(7)}= (a-2)n^2+\big[(a-1)(3-\gamma-\delta)\\
 +\alpha+\beta
 -\gamma-2\big]n+q
%
+a(2-\gamma-\delta)-\alpha\beta+\alpha+\beta+\delta-3,
\qquad %
\bm{\gamma}_n^{(8)}=\mathring{\gamma}_n^{(8)}.
\end{cases}
\end{eqnarray}
%
%

%
We find that
\begin{eqnarray}\label{SS1-SS8}
\begin{array}{l}
a=1 \quad \Rightarrow \quad  \mathring{\bm{\alpha}}_n^{(i)}=0;\vspace{2mm}\\
a=2,\; \alpha+\beta+1=2\gamma+\delta,\; q=\alpha\beta\quad\Rightarrow \quad
\mathring{\bm{\beta}}_n^{(1)}= \mathring{\bm{\beta}}_n^{(3)}=
\mathring{\bm{\beta}}_n^{(6)}= \mathring{\bm{\beta}}_n^{(8)}=0;\vspace{3mm}\\
a=2,\; \alpha+\beta=1+\delta,\; 
q=\alpha\beta+(\gamma-1)\delta \quad\Rightarrow
\quad \mathring{ \bm{\beta}}_n^{(2)}=\mathring{\bm{\beta}}_n^{(4)}=
\mathring{\bm{\beta}}_n^{(5)}=\mathring{\bm{\beta}}_n^{(7)}=0
\end{array}
\end{eqnarray}
%
%
%
%
%
in the expansions (\ref{SS1}-\ref{SS8}). Once more, 
for these cases, the Heun equation (\ref{heun}) can be reduced to the 
hypergeometric equation 
(\ref{hypergeometric}) -- see Appendix B.
\section{Solutions in series of hypergeometric functions}

This section presents the two groups of solutions in series of 
hypergeometric functions, $F(\mathrm{a,b;c};z)$,  which are
denoted by $H^{(i)}(x)$ and $\bm{H}^{(i)}(x)$
($i=1,2,\cdots,8$) and correspond, respectively, to the 
series coefficients denoted 
by ${b}_{n}^{(i)}$ and $\bm{b}_{n}^{(i)}$. The solutions are generated
from the solution $\bar{H}^{(1)}(x)$, written  
in Eqs. (\ref{inicial}) and (\ref{inicial-coef}), by means of the following transformations 
\begin{align}
&
{H}^{(1)}(x)=M_{17}\bar{H}^{(1)}(x), \qquad {H}^{(i)}(x)=
T_i{H}^{(1)}(x),\vspace{2mm}\\
&
\bm{H}^{(1)}(x)=M_{49}\bar{H}^{(1)}(x),\qquad \bm{H}^{(i)}(x)=
T_i\bm{H}^{(1)}(x),
\end{align}
where $M_{17}$ and  $M_{49}$ are the fractional transformations given 
in Eqs. (\ref{17}) and (\ref{49}), and $T_i$ are the homotopic 
transformations (\ref{t1}-\ref{t8}).

In section 4.1 we construct the initial solution $\bar{H}^{1}(x)$). In sections
4.2 and 4.3 we write the solutions $H^{(i)}(x)$ and $\bm{H}^{(i)}(x)$.
Formally the solutions are given by infinite series, but they are applied only
for the cases in which the series become finite. By this reason, there is
no discussion on convergence of the series.

To simplify the expressions of the recurrence relations, from Eq. 
(\ref{hiper-zero-1}) onwards, 
the solutions are 
written in terms of the functions $\tilde{F}(\mathrm{a,b;c};z)$ defined as
\begin{eqnarray}\label{Ftilde}
\tilde{F}(\mathrm{a,b;c};z)=\frac{1}{\Gamma(\mathrm{c})}{F}(\mathrm{a,b;c};z)
\end{eqnarray}
where $\Gamma$ stands for gamma function.

\subsection{The initial solution $\bar{H}^{(1)}(x)$}

The solution $\bar{H}^{(1)}$ in series of hypegeometric functions 
$F\left(n+\alpha,\gamma+\delta-\alpha-1;n+\gamma;
x\right)$ is 
\letra
\begin{eqnarray}\label{inicial}
&&\begin{array}{l}
\bar{H}^{(1)}(x)= \displaystyle \sum_{n=0}^{\infty}
\bar{b}_{n}^{(1)}x^{n}
F\left(n+\alpha,\gamma+\delta-\alpha-1;n+\gamma;
x\right).\end{array}
%
\end{eqnarray}
where the recurrence relations (\ref{recurrence1}) for $\bar{b}_{n}^{(1)}$ 
take the form $ \bar{\alpha}_{n}^{(1)}\bar{b}_{n+1}^{(1)}+
 \bar{\beta}_{n}^{(1)}\bar{b}_{n}^{(1)}+
\bar{\gamma}_{n}^{(1)}\bar{b}_{n-1}^{(1)}=0$ with $\bar{b}_{-1}^{(1)}=0$ and
\begin{eqnarray}\label{inicial-coef}
&&\begin{array}{l}
\bar{\alpha}_{n}^{(1)}=a(n+1)(n+\gamma),\quad 
\bar{\beta}_{n}^{(1)}=-(a+1)n^2-[a(2\alpha+1-\gamma-\delta)+\alpha+\beta-\delta]n \end{array}
\nonumber\vspace{2mm}\\
&&\begin{array}{l}
-q
 -a\alpha (\alpha+1-\gamma-\delta),
\qquad \bar{\gamma}_{n}^{(1)}=\frac{(n+\alpha-1)(n+\alpha-\delta)(n+\alpha+\beta-\gamma-\delta)}{n+\gamma-1}.
\end{array}
\label{F2}
\end{eqnarray}
%
%
%
%
%
%
The above hypergeometric functions converge for any $|x|< 1$ and on $|x|=1$ if $\mbox{Re }\delta<1$. For this reason, the solution 
$\bar{H}^{(1)}(x)$ will be valid only for $|x|\leq1$ in opposition to 
$\mathring{H}^{(1)}(x)$
which may be valid for $|x|>1$ if $|a|>1$ -- see conditions (\ref{convergence-primeiro}). In fact, $\bar{H}^{(1)}(x)=\mathring{H}^{(1)}(x)$ only if $\gamma+\delta-1=\alpha$.

The Heun equation (\ref{heun}) is equivalent to
\antiletra
\begin{equation}\label{heun-2}
(x-a)\left[x(x-1)\frac{d^{2}}{dx^{2}}+(\gamma+\delta)x
\frac{d}{dx}-\gamma\frac{d}{dx}\right]H+
\epsilon x(x-1)\frac{dH}{dx}+
\left[
 \alpha \beta x-q \right]H=0, 
\end{equation}
The solution (\ref{inicial}) is obtained by writing
\begin{eqnarray}\label{F3}
\begin{array}{l}
H(x)=\bar{H}^{(1)}(x)=\displaystyle 
\sum_{n=0}^{\infty}\bar{b}_n^{(1)}x^nF_{n}^{(1)}(x).
\quad \mbox{where} 
\vspace{2mm}\\
%
%
F_{n}^{(1)}(x)=F\left( \rho_{n},\mathrm{b};\sigma_{n};x\right),\qquad
 \rho_{n}=n+\alpha,\quad\mathrm{b}=\gamma+\delta-\alpha-1,\quad\sigma_{n}=n+\gamma.
\end{array}
\end{eqnarray}
Then,
\begin{eqnarray*}
\frac{dH}{dx}=\displaystyle \sum_{n=0}^{\infty}\bar{b}_n^{(1)}x^n
\left[\frac{d}{dx}+\frac{n}{x} \right] F_{n}^{(1)},\quad
 \frac{d^2H}{dx^2}=\displaystyle \sum_{n=0}^{\infty}\bar{b}_n^{(1)}x^n
\left[\frac{d^2}{dx^2}+\frac{2n}{x}\frac{d}{dx}+\frac{n(n-1)}{x^2} \right]F_{n}^{(1)}.
\end{eqnarray*}
By inserting the previous expressions into Eq. (\ref{heun-2})
and using the hypergeometric equation
\begin{eqnarray*}
x(1-x)\frac{d^2F_{n}^{(1)}(x)}{dx^2}+
[\sigma_{n}-(\rho_{n}+\mathrm{b}+1)x]\frac{dF_{n}^{(1)}(x)}{dx}-
\rho_{n}\mathrm{b} F_{n}^{(1)}(x)=0,
\end{eqnarray*}
for $F_{n}(x)$, we find
\begin{eqnarray}\label{F4}
&&\begin{array}{l}
\displaystyle \sum_{n=0}^{\infty}
\bar{b}_n^{(1)}x^n\left[ an
(1-x)\frac{dF_{n}^{(1)}}{dx}+(n+\epsilon)x
(x-1)\frac{dF_{n}^{(1)}}{dx}\right]\end{array}+
\nonumber\\
&&\begin{array}{l}
\displaystyle \sum_{n=0}^{\infty}\bar{b}_n^{(1)}
x^{n+1}\left[ n(n+\gamma+\delta-1+\epsilon)
+\alpha\beta-\mathrm{b}\rho_{n}\right] F_{n}^{(1)}+\end{array}
\nonumber\\
&&\begin{array}{l}
\displaystyle \sum_{n=0}^{\infty}
\bar{b}_n^{(1)}
x^{n}\left[ a\alpha(\gamma+\delta-\alpha-1)-q
-an(n+\alpha)-n(n+\gamma-1+\epsilon)\right] F_{n}^{(1)}+
\end{array}
\nonumber\\
&&\begin{array}{l}
\displaystyle \sum_{n=0}^{\infty}\bar{b}_n^{(1)}
an(n-1+\gamma)x^{n-1}F_{n}^{(1)}=0.\end{array}
\end{eqnarray}
On the other side, from the properties (\ref{relacao1}) of the hypergeometric  functions we get
\begin{eqnarray}\label{F6}
&&\begin{array}{l}
(1-x)\frac{dF_{n}^{(1)}}{dx}=
\left( \mathrm{b}-\frac{\sigma_{n}-1}{x}\right) F_{n}^{(1)}+\frac{\sigma_{n-1}}{x} F_{n-1}^{(1)},\qquad
x(x-1)\frac{dF_{n}^{(1)}}{dx} =\end{array}\nonumber\vspace{2mm}\\
&&\begin{array}{l}
-\rho_{n}xF_{n}^{(1)}+
\frac{\rho_{n}(\sigma_{n}-\mathrm{b})}{\sigma_{n}}xF_{n+1}^{(1)}
\end{array}.
\end{eqnarray}
By means of these equations, Eq. (\ref{F4}) becomes
\begin{eqnarray*}
\displaystyle \sum_{n=1}^{\infty}\bar{b}_n^{(1)}\sigma_{n-1}x^{n-1}
\ F_{n-1}^{(1)}
+\displaystyle \sum_{n=0}^{\infty}\bar{b}_n^{(1)}x^{n}
 \bar{\beta}_{n}F_{n}^{(1)}
+\displaystyle \sum_{n=0}^{\infty}
\bar{b}_n^{(1)}\frac{\rho_{n}(\sigma_{n}-\mathrm{b})(n+\epsilon)}{\sigma_{n}}x^{n+1}
F_{n+1}^{(1)}=0.
\end{eqnarray*}
Now, by setting $n=m+1$ and $n=m-1$ in the 
first and last
terms of the above equation, one finds
\begin{eqnarray*}
&&\left[a\sigma_{0}\;\bar{b}_{1}^{(1)}+\bar{\beta}_{0}\;\bar{b}_{0}^{(1)}\right]F_{0}^{(1)}(x)+
\\
&&
\displaystyle \sum_{n=1}^{\infty}
\begin{array}{l}
\left[ a(n+1)\sigma_{n}\;\bar{b}_{n+1}^{(1)}
+\bar{\beta}_{n}\;\bar{b}_n^{(1)}+
\frac{\rho_{n-1}(\sigma_{n-1}-\mathrm{b})(n+\epsilon-1)}{\sigma_{n-1}}\;\bar{b}_{n-1}^{(1)}\right]x^{n} F_{n}^{(1)}(x)
=0.\end{array}
\end{eqnarray*}
Thence, the recurrence relations for $\bar{b}_n^{(1)}$  
follow by requiring that the coefficients of the functions $x^n F_{n}^{(1)}(x)$ 
vanish. 
%
%

%
%

Finally, by taking $\tilde{b}_{n}^{(1)}=\Gamma(n+\gamma)\bar{b}_{n} $, 
the expansion $\bar{H}^{(1)}$ is re-expressed in terms of the functions $\tilde{F}$
defined in (\ref{Ftilde}), namely,
\begin{eqnarray}%
\label{hiper-zero-1}\begin{cases}
\bar{H}^{(1)}(x)= \displaystyle \sum_{n=0}^{\infty}\tilde{b}_{n}^{(1)}x^{n}
\tilde{F}\left(n+\alpha,\gamma+\delta-\alpha-1;n+\gamma;
x\right)\quad \mbox{where, in (\ref{recurrence1} ) for }\tilde{b}_{n}^{(1)}, \vspace{3mm}\\
%
¨
\tilde{\alpha}_{n}^{(1)}=a(n+1),\qquad
\tilde{\beta}_{n}^{(1)}=-(a+1)n^2-[a(2\alpha+1-\gamma-\delta)+\alpha+\beta-\delta]n
-q \vspace{2mm}\\
-a\alpha (\alpha+1-\gamma-\delta),\qquad
%
 \tilde{\gamma}_{n}^{(1)}=(n+\alpha-1)(n+\alpha-\delta)(n+\alpha+\beta-\gamma-\delta)
 . \end{cases}
\end{eqnarray}
Other solutions are given $\bar{H}^{(i)}(x)=T_i\bar{H}^{(1)}(x)$.
Some of these expansions yield finite-series solutions for the problem discussed
in Section 5; however, such solutions do not fulfil the Arscott conditions  
(\ref{autovalores}). By this reason, we use $\bar{H}^{(1)}(x)$
to obtain the expansions ${H}^{(i)}(x)$ and ${\bm{H}}^{(i)}(x)$ 
which are
used in Sections 5.3 and 5.4, respectively.


{As a by-product we find that the expansion (\ref{hiper-zero-1}) admits the limits (\ref{formal}) for the CHE (\ref{CHE}), leading to
\begin{eqnarray}
\label{hiper-che-1}
\begin{cases}
\bar{S}^{(1)}(x)= \displaystyle \sum_{n=0}^{\infty}\tilde{b}_{n}^{(1)}x^{n}
\tilde{F}\left(n+\alpha,\gamma+\delta-\alpha-1;n+\gamma;
x\right)\qquad \mbox{with}\vspace{3mm}\\
(n+1)\;\tilde{b}_{n+1}^{(1)}-\left[n^2+(2\alpha+1-\gamma-\delta-\rho)n
-\sigma+\alpha (\alpha+1-\gamma-\delta)\right]\;\tilde{b}_{n}^{(1)}-
\vspace{3mm}\\
\rho(n+\alpha-1)(n+\alpha-\delta)\;
 \tilde{b}_{n-1}^{(1)}=0, \qquad \left[  \tilde{b}_{-1}^{(1)}=0 \right]
 . \end{cases}
\end{eqnarray}
This $\bar{S}^{(1)}$ is equivalent to a solution for the CHE proposed 1937 by E. Fisher \cite{fisher}}.

\subsection{The solutions ${H}^{(i)}(x)$ around ${(a-1)x}/{(a-x)}=0$}

In this section ${{H}}^{(1)}(x)= M_{17}\bar{H}^{(1)}(x)$  and 
${{H}}^{(i)}(x)=T_i{{H}}^{(1)}(x)$, where $\bar{H}^{(1)}$ is given
in (\ref{hiper-zero-1}). The 
recurrence relations for the coefficients ${b}_{n}^{(i)}$ once more  have the form (\ref{recurrence1}).
%
%
\begin{eqnarray}\label{adendos-1}\begin{cases}
{{H}}^{(1)}(x)=\left(1-\frac{x}{a}\right)^{-\alpha} 
\displaystyle \sum_{n=0}^{\infty}
\begin{array}{l}
b_{n}^{(1)}\left[\frac{(a-1)x}{a-x}\right]^{n}
\tilde{F}\left[n+\alpha,\gamma+\delta-\alpha-1;n+\gamma;\frac{(a-1)x}{a-x}\right],\vspace{2mm}
\end{array}\\
%
¨
{{\alpha}}_{n}^{(1)}=(1-a)(n+1),\qquad
{\beta}_{n}^{(1)}=(a-2)n^2+[(a-1)(2\alpha+1-\gamma-\delta)-\alpha+\vspace{3mm}\\
\beta-\gamma]n
+q
+(a-1)\alpha (\alpha+1-\gamma-\delta)-\alpha\gamma,
\qquad
 {\gamma}_{n}^{(1)}
 =(n+\alpha-1)\times\\
 \hspace{8mm}
 	(n+\alpha-\delta)(n+\alpha-\beta). \end{cases}
\end{eqnarray}
%
%
%
\begin{eqnarray}\label{adendos-2}\begin{cases}
{H}^{(2)}(x)= x^{1-\gamma}\left(1-\frac{x}{a}\right)^{\gamma-\beta-1} \times\vspace{2mm}\\
\hspace{1.8cm}
\displaystyle \sum_{n=0}^{\infty}
\begin{array}{l}
{b}_{n}^{(2)}\left[\frac{(a-1)x}{a-x}\right]^{n}
\tilde{F}\left[n+\beta+1-\gamma,\delta-\beta;n+2-\gamma;\frac{(a-1)x}{a-x}\right]
\end{array},\vspace{3mm}\\
{\alpha}_{n}^{(2)}=(1-a)(n+1),\qquad 
{\beta}_{n}^{(2)}=(a-2)n^2+[(a-1)(2\beta+1-\gamma-\delta)-2+
\vspace{3mm}\\
\alpha-\beta+\gamma]n+q+a\beta (\beta+1-\gamma-\delta)-\beta(\beta-\delta+1)+(\gamma-1)(1-\alpha+\beta)
,
\vspace{3mm}\\
 {\gamma}_{n}^{(2)}={(n+\beta-\gamma)(n+\beta+1-\gamma-\delta)(n-\alpha+\beta)}\end{cases}
\end{eqnarray}
%
%
The solutions ${H}^{(3)}$ and ${H}^{(4)}$ 
take the forms
\begin{eqnarray}\label{adendos-3}
&&{H}^{(3)}(x)=
 (1-x)^{1-\delta}\left(1-\frac{x}{a}\right)^{\delta-\beta-1} \times\nonumber\vspace{2mm}\\
&&
\displaystyle \sum_{n=0}^{\infty}
\begin{array}{l}
{b}_{n}^{(3)}\left[\frac{(a-1)x}{a-x}\right]^{n}
\tilde{F}\left[n+\beta+1-\delta,\gamma-\beta;n+\gamma;\frac{(a-1)x}{a-x}\right]
\end{array}
%
\stackrel{\text{(\ref{euler-1}})}{=} {H}^{(1)}(x)\Big|_{\alpha\leftrightarrow\beta},\qquad
\end{eqnarray}
\begin{eqnarray}\label{adendos-4}
&&{H}^{(4)}(x)=x^{1-\gamma}
 (1-x)^{1-\delta}\left(1-\frac{x}{a}\right)^{\gamma+\delta-\alpha-2}
 \displaystyle \sum_{n=0}^{\infty}{b}_{n}^{(4)} \times\nonumber\vspace{2mm}\\
&&
\begin{array}{l}
\left[\frac{(a-2)x}{a-x}\right]^{n}
\tilde{F}\left[n+2+\alpha-\gamma-\delta,1-\alpha;n+2-\gamma;\frac{(a-1)x}{a-x}\right]
\end{array}
%
{=} {H}^{(2)}(x)\Big|_{\alpha\leftrightarrow\beta},
\end{eqnarray}
where $\alpha$ and $\beta$ must be interchanged also in the recurrence
relations.
%
%
\begin{eqnarray}\label{adendos-5}
\begin{cases}
{H}^{(5)}(x)= \left(1-\frac{x}{a}\right)^{-\beta} 
\displaystyle \sum_{n=0}^{\infty}
\begin{array}{l}
{b}_{n}^{(5)}\left[\frac{(a-1)x}{a-x}\right]^{n}
\tilde{F}\left[n-\alpha+\gamma+\delta,\alpha-1;n+\gamma;\frac{(a-1)x}{a-x}\right],
\end{array}\\
{\alpha}_{n}^{(5)}=(1-a)(n+1),\qquad 
{\beta}_{n}^{(5)}=(a-2)n^2+[(a-1)(1-2\alpha+\gamma+\delta)+
\vspace{3mm}\\
%
\alpha-\beta-\gamma]n +q+(a-1)(\alpha-1)(\alpha-\gamma-\delta)-\beta\gamma,
\vspace{3mm}\\
{\gamma}_{n}^{(5)}={(n-\alpha+\gamma+\delta-1)(n-\alpha+\gamma)(n-\alpha+\beta)}. \end{cases}
\end{eqnarray}
%
%
%
\begin{eqnarray}\label{adendos-6}
\begin{cases}
{H}^{(6)}(x)= x^{1-\gamma} \left(1-\frac{x}{a}\right)^{\gamma-\alpha-1}
\displaystyle 
\sum_{n=0}^{\infty}\begin{array}{l}
{b}_{n}^{(6)}\left[\frac{(a-1)x}{a-x}\right]^{n}
\end{array}
\times\vspace{2mm}\\
\hspace{5cm}\begin{array}{l}
\tilde{F}\left[n-\beta+\delta+1,\beta-\gamma;n+2-\gamma;\frac{(a-1)x}{a-x}
\right], \end{array}\vspace{3mm}\\
 {\alpha}_{n}^{(6)}=(1-a)(n+1),\quad
{\beta}_{n}^{(6)}=(a-2)n^2+[(a-1)(1-2\beta+\gamma+\delta)-\alpha+
\vspace{3mm}\\ 
\beta+\gamma-2]n+q-\alpha
+a(\beta-1)(\beta-\gamma-\delta)-(\beta-1)(\beta-\delta-1),\qquad\vspace{3mm}\\ 
{\gamma}_{n}^{(6)}={(n-\beta+\delta)(n-\beta+1)(n+\alpha-\beta)}. 
\end{cases}
\end{eqnarray}
%
%
%
%
The solutions ${H}^{(7)}(x)$ and ${H}^{(8)}(x)$ 
are expressed as
\begin{eqnarray}\label{adendos-7}
&&{H}^{(7)}(x)
= \begin{array}{l}(1-x)^{1-\delta}\left(1-\frac{x}{a}\right)^{\delta-\alpha-1}
\end{array}
\times \nonumber\\
&&
\displaystyle \sum_{n=0}^{\infty}\begin{array}{l}
{b}_{n}^{(7)}
\left[\frac{(a-1)x}{a-x}\right]^{n}
\tilde{F}\left[n+1-\beta+\gamma,\beta-\delta;n+\gamma;
\frac{(a-1)x}{a-x}\right]
\stackrel{\text{(\ref{euler-1}})}{=} H^{(5)}(x)\Big|_{\alpha\leftrightarrow\beta},
\end{array}
\end{eqnarray}
\begin{eqnarray}\label{adendos-8}
&&{H}^{(8)}(x)
= \begin{array}{l}x^{1-\gamma}(1-x)^{1-\delta}
\left(1-\frac{x}{a}\right)^{\gamma+\delta-\beta-2}
\displaystyle \sum_{n=0}^{\infty}
{b}_{n}^{(8)}\times\end{array}
\nonumber\\
&&
\begin{array}{l}
\left[\frac{(a-1)x}{a-x}\right]^{n}
\tilde{F}\left[n+
2-\alpha,\alpha+1-\gamma-\delta;n+2-\gamma;
\frac{(a-1)x}{a-x}\right]\end{array}
%
%
\stackrel{\text{(\ref{euler-1}})}{=}H^{(6)}(x)\Big|_{\alpha\leftrightarrow\beta}.
\end{eqnarray}

In this group the hypergeometric functions converge in the region 
$|{(a-1)x}/{(a-x)}|<1$ which includes only one singular point ($x=0$);
in accordance with (\ref{conver-hyper}),  
a restriction on $\delta$ assure convergence also on  
$|{(a-1)x}/{(a-x)}|=1$ (includes $x=1$). Precisely,
\begin{eqnarray}\label{adendos-1-conv}
\begin{array}{l}
\mbox{hypergeometric functions in }
{H}^{(i)}  \text{ converge on }
\left|\frac{(a-1)x}{(a-x)}\right|=1 \mbox{ if } \mbox{Re}\;\delta<1.\end{array} 
\end{eqnarray}
%
%
{In fact,  in view of (\ref{euler-1}) there are two forms for each expansion in  series of hypergeometric functions, but only one of them is suitable for 
$|z|=1$. For example,
%
%
%
%
\begin{eqnarray*}
&&{{H}}^{(1)}(x)=\left(1-\frac{x}{a}\right)^{-\alpha} 
\displaystyle \sum_{n=0}^{\infty}
\begin{array}{l}
b_{n}^{(1)}\left[\frac{(a-1)x}{a-x}\right]^{n}
\tilde{F}\left[n+\alpha,\gamma+\delta-\alpha-1;n+\gamma;\frac{(a-1)x}{a-x}\right]
\end{array}\vspace{2mm}\\
&&\stackrel{\text{(\ref{euler-1}})}{=}(1-x)^{1-\delta}\left(1-\frac{x}{a}\right)^{\delta-1-\alpha} 
\displaystyle \sum_{n=0}^{\infty}
\begin{array}{l}
	b_{n}^{(1)}\left[\frac{(a-1)x}{a-x}\right]^{n}	\tilde{F}\left[n+1+\alpha-\delta,\gamma-\alpha;
	n+\gamma;\frac{(a-1)x}{a-x}\right].
	\end{array}
\end{eqnarray*}
In the first expression, the hypergeometric function converges 
on $x=1$ if $\mbox{Re}\;\delta<1$ while the factor $(1-x/a)^{-\alpha}$ is finite because $a\neq 0,1$.
In the second expression, the hypergeometric function converges 
on $x=1$ only if $\mbox{Re}\;\delta>1$, but in this case the factor
$(1-x)^{1-\delta}$ in general becomes infinite at $x=1$. Thus, the second form is inappropriate to get the behaviour at $x=1$.}

We find that, for the expansions ${H}^{(i)}(x)$,
\begin{eqnarray}\label{HH1-HH8}
\begin{array}{l}
a=1\; \quad\Rightarrow \quad {\alpha}_n^{(i)}=0;\vspace{2mm}\\
a=2,\;  \alpha+\beta+1=2\gamma+\delta,\;  q=\alpha\beta \quad \Rightarrow \quad
{\beta}_n^{(1)}={\beta}_n^{(3)}=
{\beta}_n^{(6)}={\beta}_n^{(8)}=0;\vspace{3mm}\\
a=2,\;  \alpha+\beta=1+\delta,\; 
q=\alpha\beta+(\gamma-1)\delta \quad\Rightarrow \quad {\beta}_n^{(2)}={\beta}_n^{(4)}=
{\beta}_n^{(5)}={\beta}_n^{(7)}=0, 
\end{array}
\end{eqnarray}
which reproduce the conditions given in (\ref{SS1-SS8})
for reducing the Heun equation to a hypergeometric equation.
%

%
%
%


%
\subsection{The  solutions ${\bm{H}}^{(i)}(x)$  around $x=1$}
%

These arise from the solution $\bar{H}^{(1)}(x)$, given
in (\ref{hiper-zero-1}), by means of the 
following transformations: ${\bm{H}}^{(1)}(x)=M_{49}\bar{H}^{(1)}(x)$
and ${\bm{H}}^{(i)}(x)=T_i{\bm{H}}^{(1)}(x)$. 
%
%
%
%
\begin{eqnarray}\label{h-bold-1}
\begin{cases}
{\bm{H}}^{(1)}(x)= \displaystyle \sum_{n=0}^{\infty}{\bm{b}}_{n}^{(1)}
(1-x)^{n}
\tilde{F}\left(n+\alpha,\gamma+\delta-\alpha-1;n+\delta;
1-x\right),
\vspace{2mm}\\
{\bm{\alpha}}_{n}^{(1)}=(1-a)(n+1),\qquad 
{\bm{\beta}}_{n}^{(1)}=
(a-2)n^2+[(a-1)(2\alpha+1-\gamma-\delta)-
\vspace{3mm}\\
\qquad\alpha-\beta+\gamma]n+q+a\alpha (\alpha+1-\gamma-\delta) -\alpha \epsilon,
\\
{\bm{\gamma}}_{n}^{(1)}={(n+\alpha-1)(n+\alpha-\gamma)
	(n+\alpha+\beta-\gamma-\delta)}. 
	\end{cases} 
\end{eqnarray}
%
%
\begin{eqnarray}\label{h-bold-2}
{\bm{H}}^{(2)}(x)&= &x^{1-\gamma}\displaystyle 
\sum_{n=0}^{\infty}{\bm{b}}_{n}^{(2)}(1-x)^{n}
\tilde{F}\left(n+\beta+1-\gamma,\delta-\beta;n+\delta;
1-x\right)\nonumber\\
&\stackrel{\text{(\ref{euler-1}})}{=}&{\bm{H}}^{(1)}(x)\Big|_{\alpha\leftrightarrow\beta}.
\end{eqnarray}
%
%
%
\begin{eqnarray}\label{h-bold-3}\begin{cases}
{\bm{H}}^{(3)}(x)
=(1-x)^{1-\delta} \displaystyle \sum_{n=0}^{\infty}{\bm{b}}_{n}^{(3)}
(1-x)^{n}\times\\
\hspace{5.5cm}
\tilde{F}\left(n+1+\beta-\delta,\gamma-\beta;n+2-\delta;
1-x\right),\vspace{2mm}\\
{\bm{\alpha}}_{n}^{(3)}=(1-a)(n+1),\quad 
{\bm{\beta}}_{n}^{(3)}=(a-2)n^2+[(a-1)(1+2\beta-\gamma-\delta)-\alpha-
\vspace{2mm}\\
\beta+\gamma+2\delta-2]n+q+
a\beta(\beta-\gamma-\delta+1)-(\beta-\delta+1)(\alpha+\beta+1-\gamma-\delta),\quad
\vspace{2mm}\\
{\bm{\gamma}}_{n}^{(3)}=
{(n+\beta-\delta)(n+1+\beta-\gamma-\delta)(n+\alpha+\beta-\gamma-\delta)},
\end{cases}
\end{eqnarray}
%
%
\begin{eqnarray}\label{h-bold-4}
&&{\bm{H}}^{(4)}(x)=
x^{1-\gamma}(1-x)^{1-\delta} \displaystyle \sum_{n=0}^{\infty}
{\bm{b}}_{n}^{(4)}
(1-x)^{n}\times \nonumber\\
&&\hspace{16mm}\tilde{F}\left(n+2+\alpha-\gamma-\delta,1-\alpha;n+2-\delta;
1-x\right)
\stackrel{\text{(\ref{euler-1}})}{=}{\bm{H}}^{(3)}(x)\Big|_{\alpha\leftrightarrow\beta}.
\end{eqnarray}
%
%
\begin{eqnarray}\label{h-bold-5}
\begin{cases}
{\bm{H}}^{(5)}(x)= \left(1-\frac{x}{a}\right)^{1-\epsilon}
\displaystyle \sum_{n=0}^{\infty}{\bm{b}}_{n}^{(5)}(1-x)^{n}
%
\tilde{F}\left(n-\alpha+\gamma+\delta,\alpha-1;n+\delta;
1-x\right),\vspace{2mm}\\
\bm{\alpha}_n^{(5)}=(1-a)(n+1),\quad
{\bm{\beta}}_{n}^{(5)}=
(a-2)n^2+[(a-1)(1-2\alpha+\gamma+\delta)+\alpha+
\vspace{2mm}\\
\beta-\gamma-2\delta]n+q+
a(\alpha-\gamma-\delta)(\alpha-1)-(\alpha-\delta)(\alpha+\beta-\gamma-\delta-1)-\gamma,
 \quad
 \vspace{2mm}\\
{\bm{\gamma}}_{n}^{(5)}={(n-1-\alpha+\gamma+\delta)
(n-\alpha+\delta)(n-\alpha-\beta+\gamma+\delta)}.
\end{cases}
\end{eqnarray}
%
%
\begin{eqnarray}\label{h-bold-6}
{\bm{H}}^{(6)}(x)&=&x^{1-\gamma}\left(1-\frac{x}{a}\right)^{1-\epsilon}
\displaystyle \sum_{n=0}^{\infty}{\bm{b}}_{n}^{(6)}(1-x)^{n}
\tilde{F}\left(n+1-\beta+\delta,\beta-\gamma;n+\delta;
1-x\right)
\nonumber\\
%
&\stackrel{\text{(\ref{euler-1}})}{=}&
{\bm{H}}^{(5)}(x)\Big|_{\alpha\leftrightarrow\beta}.
\end{eqnarray}
%
%
%
%
\begin{eqnarray}\label{h-bold-7}
\begin{cases}
{\bm{H}}^{(7)}(x)= (1-x)^{1-\delta} \left[1-\frac{x}{a}\right]^{1-\epsilon}
%
\displaystyle 
\sum_{n=0}^{\infty}
{\bm{b}}_{n}^{(7)}(1-x)^{n}\times\\
\hspace{17mm}\tilde{F} \left(n-\beta+\gamma+1,\beta-\delta;n+2-\delta;
1-x\right),\vspace{2mm}\\
{\bm{\alpha}}_{n}^{(7)}=(1-a)(n+1),\quad 
{\bm{\beta}}_{n}^{(7)}=
(a-2)n^2+[(a-1)(1-2\beta+\gamma+\delta)+\alpha+
\vspace{2mm}\\
\beta-\gamma-2]n+q+
a(\beta-\gamma-\delta)(\beta-1)-(\beta-1)(\alpha+\beta-\gamma-\delta-1)-\gamma,\quad
\vspace{2mm}\\
{\bm{\gamma}}_{n}^{(7)}={(n-\beta+\gamma)(n+1-\beta)(n-\alpha-\beta+
\gamma+\delta)}.
\end{cases}
\end{eqnarray}
%
%
\begin{eqnarray}\label{h-bold-8}
&&\begin{array}{l}
{\bm{H}}^{(8)}(x)=x^{1-\gamma}(1-x)^{1-\delta} \left[1-\frac{x}{a}\right]^{1-\epsilon}
\displaystyle\sum_{n=0}^{\infty}
{\bm{b}}_{n}^{(8)}(1-x)^{n}
\end{array}
\times\nonumber\\
&&\hspace{15mm}
\begin{array}{l}
\tilde{F}\left(n+2-\alpha,\alpha+1-\gamma-\delta;n+2-\delta;
1-x\right)\end{array}
\stackrel{\text{(\ref{euler-1}})}{=} {\bm{H}}^{(7)}(x)\Big|_{\alpha\leftrightarrow\beta}.
\end{eqnarray}
The hypergeometric functions in ${\bm{H}}^{(i)}(x)$ converge in the region 
$|x-1|<1$ but, by (\ref{conver-hyper}), 
\begin{eqnarray}\label{adendos-2-conv}
\begin{array}{l}\mbox{hypergeometric } 
\mbox{functions in }
{\bm{H}}^{(i)} \text{ converge on }|x-1|=1  \mbox{ if }
\mbox{Re}\;\gamma<1.\end{array}
\end{eqnarray}
For the above expansions, $\bm{H}^{(i)}(x)$,
\begin{eqnarray}\label{bbH}
\begin{array}{lllll}
a=1\quad &\qquad\Rightarrow &\bm{\alpha}_n^{(i)}= 0;\vspace{2mm}\\
a=2,\quad & \beta=\alpha+1-\delta,\qquad & q=\alpha\gamma&\Rightarrow & 
\bm{\beta}_n^{(1)}=\bm{\beta}_n^{(3)}=
\bm{\beta}_n^{(6)}=\bm{\beta}_n^{(8)}=0;\vspace{3mm}\\
a=2,\quad & \alpha=\beta+1-\delta,& 
q=\beta\gamma&\Rightarrow & \bm{\beta}_n^{(2)}=\bm{\beta}_n^{(4)}=
\bm{\beta}_n^{(5)}=\bm{\beta}_n^{(7)}=0.
\end{array}
\end{eqnarray}
Thence, Eqs.(\ref{recurrence1}) become two-term recurrence relations for the series coefficients and the Heun equation (\ref{heun}) is 
reducible to the hypergeometric equation 
(\ref{hypergeometric}) (Appendix B).

\section{Schr\"odinger equation for associated Lam\'e  potentials}

There are some periodic potentials 
for which the Schr\"{o}dinger equation leads to equations of 
Heun family \cite{ganguly-1,ganguly-2,khare,khare2,slater,matveev}. {Solutions for these equations in general demand numerical computations}. In this section we{ consider only some analytical aspects
concerning the} solutions of the Schr\"{o}dinger equation 
for the associated Lam\'e  potential 
 (\ref{ganguly-3})
which depends on the parameters $\bm{m}$ and $\bm{l}$.

%
%
 We will find that
the solutions of the general Heun equation allow to
solve this problem as follows:
 \begin{itemize}
 \itemsep-1pt
\item  
if $\bm{m}-\bm{l}$
or $\bm{l}+\bm{m}$ is an integer, we get finite series from
the power series expansions given in Section 3;  further 
restrictions on $\bm{m}$ and $\bm{l}$ assure that these solutions satisfy the Arscott criterion (\ref{autovalores});
 \item   
 if either $\bm{l}$
or $\bm{m}$ is an integer and the other is half an odd integer, we get finite series from
the expansions in series of hypergeometric functions given in Section 4; such solutions are degenerate and satisfy the {Arscott criterion only for special values of $\bm{l}$
and $\bm{m}$};
\item
bounded infinite-series solutions are obtained from the expansions
given in section 3.1.
 \end{itemize}
The finite-series solutions are valid under general restrictions on 
the parameters, that is, without regarding particular values for 
$\bm{m}$ and $\bm{l}$ (some specific values 
are discussed in Appendix C).  {Initially,  we explain how we obtain the several  types of solutions and set up some notations for the solutions.}

The Schr\"{o}dinger
equation (\ref{schr}) for the associated Lam\'e potencial (\ref{ganguly-3}) is 
\begin{eqnarray}\label{associado-gan}
		\frac{d^{2}\psi(u)}{du^{2}}+	
		\left[
		{\cal E}-\bm{m}(\bm{m}+1)k^ 2\;\mathrm{sn}^2u-\bm{l}(\bm{l}+1)k^2\;
		\frac{\mathrm{cn}^2u}
		{\mathrm{dn}^2u}\right]\psi(u)=0,
\end{eqnarray}
which is the Darboux Eq. (\ref{darboux}) with parameters ($a=1/k^2$)
\begin{eqnarray}\label{parametros-ganguly-3}
\begin{array}{l}
\alpha=\frac{\bm{l}-\bm{m}+1}{2}, \quad
\beta=\frac{\bm{l}+\bm{m}+2}{2}, \quad \gamma=
\delta=\frac{1}{2},\quad 
q=\frac{(\bm{l}+1)^2}{4}-\frac{{\cal E}}{4k^2},
\quad 
\left(\epsilon=\bm{l}+\frac{3}{2}\right).
\end{array}
\end{eqnarray}
Thence, the eigenfunctions $\psi(u)$
follow from solutions $H(x)$ of the Heun equation through
\begin{eqnarray}\label{second-ganguly-3}
\psi(u)\stackrel{\text{Eq.(\ref{substituicoes-2})}}{=\joinrel=}
\mathrm{dn}^{\bm{l}+1}u \;H[x(u)],\qquad 0\leq  x(u)={\rm sn}^2 u\leq 1,
\end{eqnarray}
provided that $H(x)$ is a convergent series and  $\psi(u)$ is bounded for 
any value of $u$.  In addition, we impose the Arscott 
condition $\alpha_{n-1}\gamma_{n}> 0$ 
for finite series with two or more terms ($n\geq 1$). 
%
%

%
%

The associated Lam\'e  equation (\ref{associado-gan}) remains invariant under the
simultaneous substitutions
\begin{eqnarray}\label{subs}
\bm{m}\mapsto -\bm{m}-1,\qquad \bm{l}\mapsto -\bm{l}-1
\qquad (\mbox{leave Eq.  (\ref{associado-gan})   invariant})
\end{eqnarray}
which also leave invariant  the solutions constructed according
to (\ref{second-ganguly-3}), as we will see. 
Alternatively, (\ref{subs}) can be used 
to set up a new solution out of a given solution.
{Further, according to (\ref{argumentos}),  the  interchange 
	of $\bm{l}$ 
	and $\bm{m}$ in conjunction with the shift of $u$ to $u+K$ leaves 
	the associated Lam\`e equation invariant, that is, 
	\begin{eqnarray}\label{associado-gan-2}
	u\mapsto u+K, \qquad \bm{m} \leftrightarrow \bm{l}
	\qquad (\mbox{leave Eq.  (\ref{associado-gan})   invariant}).
	\end{eqnarray}
	This substitutions may be used to get new solutions to the associated Lam\'e  equation by
	means of the relations \cite{nist,khare}
	\begin{eqnarray}\label{sd-cd-5}
	\begin{array}{l}
	\mathrm{sn}(u+K)=\mathrm{cd}\;u, \quad
	\mathrm{cn}(u+K)=-\sqrt{1-k^2}\;\mathrm{sd}\;u, 
	\quad
	\mathrm{dn}(u+K)=\frac{\sqrt{1-k^2}}{\mathrm{dn}\;u}.
	\end{array}
	\end{eqnarray}
	Thence,  power series of $\operatorname{sn}{u}$ and $\operatorname{cn}{u}$ are transformed, respectively,
	in power series of  $\operatorname{cd}{u}$ and $\operatorname{sd}{u}$.
	To obtain these series directly from solutions of the Heun equation 
	it would be necessary to construct two additional groups by the fractional  
	transformations 
	$M_{17}$ and $M_{65}$ as we see from (\ref{argumentos}).} Notice that the particular cases
 %
 %
 \begin{align}
 &\begin{array}{l}
 \label{first-case}
 \frac{d^{2}\psi(u)}{du^{2}}+	
 \left[
 {\cal E}-\bm{l}(\bm{l}+1)k^2\;
 \frac{\mathrm{cn}^2u}
 {\mathrm{dn}^2u}\right]\psi(u)=0,\qquad [\bm{m}=0 \mbox{ or } \bm{m}=-1], \quad\mbox{and}
 \end{array}\vspace{3mm}\\
 &\begin{array}{l}\label{second-case}
 \frac{d^{2}\psi(u)}{du^{2}}+	
 \left[
 {\cal E}-\bm{m}(\bm{m}+1)k^ 2\;\mathrm{sn}^2u\right]\psi(u)=0, \quad
 [\bm{l}=0 \mbox{ or } \bm{l}=-1: \mbox{Lam\'e equation}],\end{array}
 %
 \end{align}
 are connected by (\ref{associado-gan-2}). We will see that finite-series solutions which satisfy the 
 Arscott condition for the first case do not satisfy for the second case, and vice-versa.

For comparison purposes, we write the Khare-Sukhatme notation \cite{khare,khare2}
for the parameters $\bm{l}$, $\bm{m}$ and $k^2$, namely,
\begin{eqnarray}
\bm{l}= b,\qquad \bm{m}=a, \qquad k^2=m:\qquad 
\mbox{Khare-Sukhatme \cite{khare,khare2}}.
\end{eqnarray}
%

%
%

According to  (\ref{truncation}), finite series result when 
$\gamma_n=0$ for some value of $n$. For each power-series expansion the  condition $\gamma_n=0$ gives two possibilities for truncating the 
series on the right-hand side. 
For solutions obtained from $\mathring{{H}}^{(1)}$ and  $\mathring{\bm{H}}^{(1)}$, the two possibilities result from the fact that ($\mathring{{\gamma}}_n^{(1)}=\mathring{\bm{\gamma}}_n^{(1)}$)
\begin{eqnarray}\label{truncation-2}
 \begin{array}{l}
\mathring{{\gamma}}_n^{(1)}=\left(n-\frac{\bm{m}-\bm{l}+1}{2}\right)\left(n+\frac{\bm{l}+\bm{m}}{2}\right)=0 \mbox{ if } \begin{cases}\bm{m}-\bm{l}= 1,3,5,\cdots \; \Rightarrow\; 0\leq n\leq \frac{\bm{m-l}-1}{2} \;\mbox{or}\vspace{2mm}\\
\bm{m}+\bm{l}= -2,-4,\cdots \; \Rightarrow\; 0\leq n
\leq -\frac{\bm{m+l}+2}{2}.
\end{cases}
\end{array}
\end{eqnarray}
{For $\bm{m}=-1/2$, the two cases become identical. For $\bm{m}\neq-1/2$, if  the above conditions on $\bm{m-l}$ and $\bm{m+l}$ are satisfied for the same values of $\bm{m}$ and 
$\bm{l}$, thence we have to consider only the truncation which gives the small interval for $n$. The conditions for this  will be  written  in relations (\ref{p-1}), (\ref{p-5}),
$\cdots$,  
(\ref{p-8}).  Such conditions imply that both $\bm{l}$ and $\bm{m}$ are half an integer number and, consequently, they do not stand for solutions of Eqs. (\ref{first-case}) and (\ref{second-case}).}

Solutions resulting from $\mathring{{H}}^{(i)}$ and  
$\mathring{\bm{H}}^{(i)}$ are expansions in series
of $\mathrm{sn}^2u$ and $\mathrm{cn}^2u$, respectively. 
We will find that finite series in terms of $\mathrm{sn}^2u$ and $\mathrm{cn}^2u$ 
cannot satisfy simultaneously the Arscott condition 
because this condition requires that $\mathring{{\gamma}}_n^{(i)}>0$ and 
$\mathring{{\gamma}}_n^{(i)}<0$, respectively. For the present case, the Arscott condition is important for assuring  that the characteristic equation (\ref{matriz}) yields real and distinct values for the energy ${\cal E}$.

%
%
{
 We  use the following notations for the solutions of Eq. (\ref{associado-gan}):
 %
%
\begin{itemize}
 \itemsep-3pt
\item [(1)] 
the pairs $\left(\mathring{{\psi}}^{(i)},  \tilde{{\psi}}^{(i)}\right)$
 denote finite series of $\mathrm{sn}^2u,$ obtained from truncation of the powers
 series  $\mathring{{H}}^{(i)} $ ; 
 \item [(2)] 
the pairs $\left(\mathring{\bm{\Psi}}^{(i)},\tilde{\bm{\Psi}}^{(i)}\right)$
  denote finite series of  $\mathrm{cn}^2u,$ obtained from truncation of the powers
 series  $\mathring{\bm{H}}^{(i)} $ ;
\item[(3)]  $\psi^{(i)}$  ($i=1,2,\cdots,8$)
 denote finite series of hypergeometric functions obtained from  expansions
 $H^{(i)}$ 
 of Sec. 4.2;
\item[(4)] $  \bm{\Psi}^{(i)}$  
  denote finite series of hypergeometric functions resulting from expansions
 $  \bm{H}^{(i)}$ of Sec. 4.3;
\item[(5)] $\mathring{\Phi}^{(i)}  $
 denote infinite-series solutions in terms of 
 $ \mathrm{sn}^2u $ whose linear dependence requires further investigation.
 \end{itemize}
 %
%


%
Each of the above items are discussed in the next subsections. 
We will use the following parity and periodicity properties for elliptic functions \cite{nist}:
\begin{align}
&\begin{array}{l}
\mathrm{sn}(-u)=-\mathrm{sn}{\;u},\qquad\qquad
\mathrm{cn}(-u)=\mathrm{cn}{\;u},\qquad\qquad\;
\mathrm{dn}(-u)=\mathrm{dn}{\;u}.
\end{array}
\vspace{2mm}\\
&\begin{array}{l}
\mathrm{sn}(u+2K)=-\mathrm{sn}{\;u},\qquad
\mathrm{cn}(u+2K)=-\mathrm{cn}{\;u},\qquad
\mathrm{dn}(u+2K)=\mathrm{dn}{\;u}.
\end{array}
\end{align}
{As in the case of Lam\'e equation \cite{ince2}, these properties
give  
four types of solutions for the associated Lam\'e equation:
even and odd solutions with period $2K$, even and odd solutions with period $4K$. In Sections 5.1 and 5.2,
solutions of each type are put together.}
We also take into account the special values 
\begin{eqnarray}
\begin{array}{l}
\mathrm{sn}\;0=0,\quad \mathrm{sn}\;K=1,\quad
\mathrm{cn}\;0=1,\quad \mathrm{cn}{\;K}=0,\quad
\mathrm{dn}\;0=1,\quad \mathrm{dn}{\;K}=1-k^2.
\end{array}
\end{eqnarray}
%
 %
%
%
%
\subsection{Finite-series eigenfunctions $\left(\mathring{\psi}^{(i)},
\tilde{\psi}^{(i)}\right)$ 
in terms of 
$\mathrm{sn}^2u$}
{
For the associated Lam\'e equation (\ref{associado-gan}), 
the power series solutions  
$\mathring{{H}}^{(i)}(x)$ lead  
to eight pairs of
finite-series expansions
$\left(\mathring{\psi}^{(i)},
\tilde{\psi}^{(i)}\right)$ 
by means of the truncation procedure indicated in (\ref{truncation-2}). 
To write down the four types of expansions
we observe that the pairs   $\left(\mathring{\psi}^{(j+4)} ,
\tilde{\psi}^{(j+4)}\right)$  ($j=1,2,3,4$) can be generated from   
$\left(\mathring{\psi}^{(j)},
\tilde{\psi}^{(j)}\right)$ by the substitutions  
$\bm{l}\mapsto -\bm{l}-1$ and $\bm{m}\mapsto -\bm{m}-1$ which 
preserve the period and parity of the expansions.}

{
The series coefficients for  $\left(\mathring{\psi}^{(i)},
\tilde{\psi}^{(i)}\right)$ 
are denoted 
by $\left( \mathring{b}_{n}^{(i)},\tilde{b}_{n}^{(i)}\right)$.
However, in the recurrence relations 
$\mathring{\alpha}_{n}\mathring{b}_{n+1}+\mathring{\beta}_{n}\mathring{b}_{n}+
\mathring{\gamma}_{n}\mathring{b}_{n-1}=0$ and $\tilde{\alpha}_{n}\tilde{b}_{n+1}+\tilde{\beta}_{n}\tilde{b}_{n}+
\tilde{\gamma}_{n}\tilde{b}_{n-1}=0$ , the  coefficients are formally identical, that is,
\begin{eqnarray}\label{coef-tilde}
\tilde{{\alpha}}_{n}^{(i)}=\mathring{{\alpha}}_n^{(i)},\qquad
\tilde{{\beta}}_{n}^{(i)}=\mathring{{\beta}}_n^{(i)},\qquad
\tilde{{\gamma}}_{n}^{(i)}=\mathring{{\gamma}}_n^{(i)}.
\end{eqnarray}
On the  right-hand side of each expansion, inside square brackets, we write the 
restrictions on the parameters $\bm{l}$ and $\bm{m}$ that assure the Arscott condition (\ref{autovalores}). }

{
First we write the four types of expansions, then we discuss  the Arscott condition (\ref{autovalores}) and, finally, we exhibit 
some problems which can be solved by these expansions.}
%
%
%
%
%
\subsubsection{First type: Even solutions, period  $2K$}
%
\letra
\begin{eqnarray}\label{ps1-A}  
\begin{cases}             		
\mathring{{\psi}}^{(1)}(u)=
\mathrm{dn}^{\bm{l}+1} u \;\displaystyle
\sum_{n=0}^{\frac{\bm{m-l}-1}{2}}\mathring{{b}}_{n}^{(1)}\;
\mathrm{sn}^{2n}u
\quad\mbox{ if}\quad
 \bm{m}-\bm{l}= 1,3,5,\cdots ,\;\\
%
\hspace{8cm}\left[\;\bm{m}\leq-\frac{1}{2} ,\;
\bm{m}+\bm{l}<-2\;\right],
\vspace{3mm}\\
\tilde{{\psi}}^{(1)}(u)=
\mathrm{dn}^{\bm{l}+1} u \;\displaystyle
\sum_{n=0}^{-\frac{\bm{l+m}+2}{2}}\tilde{{b}}_{n}^{(1)}\quad
\mathrm{sn}^{2n}u
\;\mbox{ if } \quad
\bm{m}+\bm{l}= -2,-4,-6,\cdots, \;\\
\hspace{8cm}\left[\bm{m}\geq-\frac{1}{2},\;
\bm{m}-\bm{l}>1\; \right] ,
\end{cases}
\end{eqnarray}
%
%
\begin{eqnarray}\label{ps1-A-2}
&&\begin{array}{l}
\mathring{{\alpha}}_n^{(1)}=
\frac{1}{k^2}\left(n+\frac{1}{2}\right)(n+1),\qquad
 \mathring{{\beta}}_n^{(1)}=
\left(\frac{1}{k^2}+1\right)n^2-
(\bm{l}+1)n+
\end{array}			\vspace{2mm}\nonumber\vspace{3mm}\\
%
&&
\begin{array}{l}\frac{1}{4k^2}{\cal E}-\frac{1}{4}(\bm{l}+1)^2,	\quad
\mathring{{\gamma}}_n^{(1)}=
\left(n-\frac{\bm{m}-\bm{l}+1}{2}\right)\left(n+\frac{\bm{l}+\bm{m}}{2}\right);
\end{array}
\end{eqnarray}
{
Only one of the above expansions is valid when (from  now on, $M_i$ and $N_i$ denote positive integers)
\begin{eqnarray}\label{p-1}
&&\begin{array}{l}
\bm{m-l}-1=2M_1\; \mbox{ and }\;
\bm{m+l}+2=-2N_1,\quad
\left(\Leftrightarrow\bm{m}=M_1-N_1-\frac{1}{2},\;  \right.\end{array}
\nonumber\vspace{2mm}\\
&&
\begin{array}{l}
\left.  \bm{l}=-M_1-N_1-\frac{3}{2}\right):\mathring{\psi}^{(1)}\mbox{ if } \; M_1<N_1,\quad
\tilde{\psi}^{(1)}\mbox{ if }  N_1<M_1.\end{array}
\end{eqnarray}
For series with two or more terms ($M_1\geq 1$ or $N_1\geq 1$),
we see that these expansions satisfy the Arscott criterion. The same holds for the other seven cases given below: (\ref{p-5}), (\ref{p-2}),$\cdots$, 
(\ref{p-8}).}

%
%
%
\antiletra\letra
\begin{eqnarray}\label{ps5-B}  
\begin{cases}             		
\mathring{{\psi}}^{(5)}(u)=
\mathrm{dn}^{-\bm{l}} u \;\displaystyle
\sum_{n=0}^{\frac{\bm{l-m}-1}{2}}\mathring{{b}}_{n}^{(5)}\;
\mathrm{sn}^{2n}u
\quad\mbox{ if }\quad
\begin{array}{l}
 \bm{l}-\bm{m}= 1,3,5,\cdots ,\hspace{8mm}\end{array}\\
\hspace{8cm}\begin{array}{l}
\left[\;\bm{m}\geq-\frac{1}{2} ,\;
\bm{m}+\bm{l}>0\;\right], \end{array}
\vspace{3mm}\\
\tilde{{\psi}}^{(5)}(u)=
\mathrm{dn}^{-\bm{l}} u \;\displaystyle
\sum_{n=0}^{\frac{\bm{l+m}}{2}} \;\tilde{{b}}_{n}^{(5)}\;
\mathrm{sn}^{2n}u
\qquad\mbox{ if } \quad 
\begin{array}{l}
\bm{m}+\bm{l}= 0,2,4,\cdots, \end{array}\qquad\\
\begin{array}{l}
\hspace{8cm}\left[\bm{m}\leq-\frac{1}{2},\;
\bm{l}-\bm{m}>1\; \right] ,\end{array}
\end{cases}
\end{eqnarray}
%
%
\begin{eqnarray}\label{ps5-A-2}
&&
\begin{array}{l}
\mathring{{\alpha}}_n^{(5)}=\frac{1}{k^2}\left(n+\frac{1}{2}\right)(n+1),\quad\end{array}
\begin{array}{l}
 \mathring{{\beta}}_n^{(5)}=-
\left(\frac{1}{k^2}+1\right)n^2+
\bm{l}n+\frac{1}{4k^2}{\cal E}-\frac{1}{4}\bm{l}^2,	
\end{array}\quad      \vspace{2mm}\nonumber\vspace{3mm}\\
&&
\begin{array}{l}
\mathring{{\gamma}}_n^{(5)}=\left(n-\frac{\bm{l}-\bm{m}+1}{2}\right)\left(n-\frac{\bm{l}+\bm{m}+2}{2}\right).
\end{array}
\end{eqnarray}
Only one expansion when 
\begin{eqnarray}\label{p-5}
&&\begin{array}{l}
\bm{l-m}-1=2M_2\; \mbox{ and }\;
\bm{m+l}=2N_2,\quad
\left(\Leftrightarrow\bm{m}=N_2-M_2-\frac{1}{2},\;  \right.\end{array}
 \nonumber\vspace{2mm}\\
&&\begin{array}{l}
\left.
\bm{l}=-M_2+N_2+\frac{1}{2}\right):
\mathring{\psi}^{(5)}\mbox{ if } \; M_2<N_2,\quad 
\tilde{\psi}^{(5)}\mbox{ if } \; N_2<M_2.\end{array}
\end{eqnarray}
%
%
%
\subsubsection{Second type: Odd solutions, period  $4K$}
%
%
%
\antiletra\letra
\begin{eqnarray}\label{ps2-A}
\begin{cases}
\mathring{{\psi}}^{(2)}(u)=
\mathrm{sn}u\;\mathrm{dn}^{\bm{l}+1}u \displaystyle
\sum_{n=0}^{\frac{\bm{m-l}-2}{2}}\mathring{{b}}_{n}^{(2)}\;
\mathrm{sn}^{2n}u
\quad\mbox{ if }\quad
\begin{array}{l}
\bm{m}-\bm{l}= 2,4,6,\cdots,\quad
\hspace{1.2cm}\end{array}\\
\begin{array}{l}\hspace{8cm}
\left[\;\bm{m}\leq-\frac{1}{2} ,\;\;
\bm{m}+\bm{l}<-3\;\right]\end{array}
, \vspace{3mm}\\
\tilde{{\psi}}^{(2)}(u)=
\mathrm{sn}u\;\mathrm{dn}^{\bm{l}+1}u \displaystyle
\sum_{n=0}^{-\frac{\bm{m+l}+3}{2}}\tilde{{b}}_{n}^{(2)}\;
\mathrm{sn}^{2n}u
\quad\mbox{ if }\quad
\begin{array}{l}
\bm{m}+\bm{l}= -3,-5,-7,\cdots,\quad\end{array}\\
\begin{array}{l}\hspace{8cm}
\;\;\left[\bm{m}\geq-\frac{1}{2},\;\;
\bm{m}-\bm{l}>2\; \right],\end{array}
\end{cases}
\end{eqnarray}
\begin{eqnarray}\label{ps2-A-2}
&&\begin{array}{l}
\mathring{{\alpha}}_n^{(2)}=\frac{1}{k^2}\left(n+\frac{3}{2}\right)(n+1),\quad 
\mathring{{\beta}}_n^{(2)}=-
\left(\frac{1}{k^2}+1\right)
n^2-\left(\frac{1}{k^2}+\bm{l}+2\right)n+\end{array}\nonumber\vspace{3mm}\\
%
&&\begin{array}{l}
\frac{1}{4k^2}({\cal E}-1)
-\frac{1}{4}(\bm{l}+2)^2,\quad
\mathring{{\gamma}}_n^{(2)}=\left(n-\frac{\bm{m}-\bm{l}}{2}\right)\left(n+\frac{\bm{l}+\bm{m}+1}{2}\right);
\end{array},
\end{eqnarray}
Only one expansion when 
\begin{eqnarray}\label{p-2}
&&\begin{array}{l}
\bm{m-l}-2=2M_3\; \mbox{ and }\;
\bm{m+l}+3=-2N_3,\quad\left(\Leftrightarrow\bm{m}=M_3-N_3-\frac{1}{2}, \right.  \end{array}
 \nonumber\vspace{2mm}\\
&&\begin{array}{l}
\left.
\bm{l}=-M_3-N_3-\frac{5}{2}\right):\mathring{\psi}^{(2)}\mbox{ if } \; M_3<N_3,\;
\tilde{\psi}^{(2)}\mbox{ if } \; N_3<M_3.\end{array}
\end{eqnarray}
%
%
%
%
%
%
\antiletra\letra
\begin{eqnarray}\label{ps6-B}
\begin{cases}
\mathring{{\psi}}^{(6)}(u)=
\mathrm{sn}u\;\mathrm{dn}^{-\bm{l}}u \displaystyle
\sum_{n=0}^{\frac{\bm{l-m}-2}{2}}\mathring{{b}}_{n}^{(6)}\;
\mathrm{sn}^{2n}u
\quad\mbox{ if }\quad
\begin{array}{l}
\bm{l}-\bm{m}= 2,4,6,\cdots,\quad
\hspace{8mm}\end{array}\\
\begin{array}{l}\hspace{8cm}
\left[\;\bm{m}\geq-\frac{1}{2} ,\;\;
\bm{m}+\bm{l}>1\;\right]\end{array}
, \vspace{3mm}\\
\tilde{{\psi}}^{(6)}(u)=
\mathrm{sn}u\;\mathrm{dn}^{-\bm{l}}u \displaystyle
\sum_{n=0}^{\frac{\bm{m+l}-1}{2}}\tilde{{b}}_{n}^{(6)}\;
\mathrm{sn}^{2n}u
\quad\mbox{ if }\quad
\begin{array}{l}
\bm{m}+\bm{l}= 1,3,5,\cdots,\qquad\end{array}\\
\begin{array}{l}\hspace{8cm}
\;\;\left[\bm{m}\leq-\frac{1}{2},\;\;
\bm{l}-\bm{m}>2\; \right],\end{array}
\end{cases}
\end{eqnarray}
\begin{eqnarray}\label{ps6-A-2}
&&\begin{array}{l}
\mathring{{\alpha}}_n^{(6)}=\frac{1}{k^2}\left(n+\frac{3}{2}\right)(n+1),\quad 
\mathring{{\beta}}_n^{(6)}=-
\left(\frac{1}{k^2}+1\right)
n^2-\left(\frac{1}{k^2}-\bm{l}+1\right)n+\end{array}\nonumber\vspace{3mm}\\
%
&&\begin{array}{l}
\frac{1}{4k^2}({\cal E}-1)
-\frac{1}{4}(\bm{l}-1)^2,\qquad
\mathring{{\gamma}}_n^{(6)}=\left(n-\frac{\bm{l}-\bm{m}}{2}\right)\left(n-\frac{\bm{l}+\bm{m}+1}{2}\right)
\end{array},
\end{eqnarray}
Only one expansion when 
\begin{eqnarray}\label{p-6}
&&
\begin{array}{l}
\bm{l-m}-2=2M_4\; \mbox{ and }\;
\bm{m+l}-1=2N_4\quad
\left(\Leftrightarrow\bm{m}=N_4-M_4-\frac{1}{2},\right. \end{array}\nonumber\vspace{2mm}\\
&&\begin{array}{l}
\left.\bm{l}=M_4+N_4+\frac{3}{2}\right):
\mathring{\psi}^{(6)}\mbox{ if } \; M_4<N_4,\quad 
\tilde{\psi}^{(6)}\mbox{ if } \; N_4<M_4.\end{array}
\end{eqnarray}
%
%
%
\subsubsection{Third type:  Even solutions, period  $4K$}
\antiletra\letra
\begin{eqnarray}\label{ps3-A}
\begin{cases}
\mathring{{\psi}}^{(3)}(u)=
\mathrm{cn}u\;\mathrm{dn}^{\bm{l}+1}u \;\displaystyle
\sum_{n=0}^{\frac{\bm{m-l}-2}{2}}\mathring{{b}}_{n}^{(3)}
\;\mathrm{sn}^{2n}u
\quad\mbox{ if }\quad
\begin{array}{l}
\bm{m}-\bm{l}=2,4,6,\cdots,
\hspace{1.3cm}\end{array}\\
\begin{array}{l}\hspace{8cm}
\left[\;\bm{m}\leq-\frac{1}{2} ,\;\;
\bm{m}+\bm{l}<-3\;\right] 
\end{array}\vspace{3mm}\\
\tilde{{\psi}}^{(3)}(u)=
\mathrm{cn}u\;\mathrm{dn}^{\bm{l}+1}u \;\displaystyle
\sum_{n=0}^{-\frac{\bm{m+l}+3}{2}}\tilde{{b}}_{n}^{(3)}
\;\mathrm{sn}^{2n}u
\quad\mbox{ if }\quad
\begin{array}{l}
\bm{m}+\bm{l}= -3,-5,-7,\cdots,\quad;\end{array}\\
\begin{array}{l}\hspace{8cm}
\left[\bm{m}\geq -\frac{1}{2},\;\
\bm{m}-\bm{l}>2\; \right],\end{array}
\end{cases}\end{eqnarray}
\begin{eqnarray}\label{ps3-A-2}
&&\begin{array}{l}
\mathring{{\alpha}}_n^{(3)}=\frac{1}{k^2}\left(n+\frac{3}{2}\right)(n+1),\qquad 
\mathring{{\beta}}_n^{(3)}=
-\left(\frac{1}{k^2}+1\right) n^2-\left(\frac{1}{k^2}+\bm{l}+1\right)n+
\end{array}
\nonumber\vspace{2mm}\\
&&\begin{array}{l}
\frac{1}{4k^2}({\cal E}-1)-\frac{1}{4}(\bm{l}+1)^2,\qquad
\mathring{\gamma}_n^{(3)}=\mathring{\gamma}_n^{(2)}.
\end{array}
\end{eqnarray}
Only one expansion when 
\begin{eqnarray}\label{p-3}
&&\begin{array}{l}
\bm{m-l}-2=2M_3\; \mbox{ and }\;
\bm{m+l}+3=-2N_3\quad
\left(\Leftrightarrow\bm{m}=M_3-N_3-\frac{1}{2},\right. 
\end{array}\nonumber\vspace{2mm}\\
&&\begin{array}{l}
 \left. 
\bm{l}=-M_3-N_3-\frac{5}{2}\right):
\mathring{\psi}^{(3)}\mbox{ if } \; M_3<N_3,\quad 
\tilde{\psi}^{(3)}\mbox{ if } \; N_3<M_3.\end{array}
\end{eqnarray}
%
%
%
%
\antiletra\letra
\begin{eqnarray}\label{ps7-B}
\begin{cases}
\mathring{{\psi}}^{(7)}(u)=
\mathrm{cn}u\;\mathrm{dn}^{-\bm{l}}u \;\displaystyle
\sum_{n=0}^{\frac{\bm{l-m}-2}{2}}\mathring{{b}}_{n}^{(7)}
\;\mathrm{sn}^{2n}u
\quad\mbox{ if }\quad
\begin{array}{l}
\bm{l}-\bm{m}=2,4,6,\cdots,
\hspace{7mm};\end{array}\\
\begin{array}{l}\hspace{8cm}
\left[\;\bm{m}\geq-\frac{1}{2} ,\;\;
\bm{m}+\bm{l}>1\;\right] 
\end{array},\vspace{3mm}\\
\tilde{{\psi}}^{(7)}(u)=
\mathrm{cn}u\;\mathrm{dn}^{-\bm{l}}u \;\displaystyle
\sum_{n=0}^{\frac{\bm{m+l}-1}{2}}\tilde{{b}}_{n}^{(7)}
\;\mathrm{sn}^{2n}u
\quad\mbox{ if }\quad
\begin{array}{l}
\bm{m}+\bm{l}= 1,3,5,\cdots,\qquad;\end{array}\\
\begin{array}{l}\hspace{8cm}
\left[\bm{m}\leq -\frac{1}{2},\;\;
\bm{l}-\bm{m}>2\; \right],\end{array}
\end{cases}\end{eqnarray}
\begin{eqnarray}\label{ps7-A-2}
&&\begin{array}{l}
\mathring{{\alpha}}_n^{(7)}=\frac{1}{k^2}\left(n+\frac{3}{2}\right)(n+1),\qquad 
\mathring{{\beta}}_n^{(7)}=
-\left(\frac{1}{k^2}+1\right) n^2-\left(\frac{1}{k^2}-\bm{l}\right)n+\end{array}\nonumber\vspace{3mm}\\
&&\begin{array}{l}
\frac{1}{4k^2}({\cal E}-1)-\frac{1}{4}\bm{l}^2,\qquad
\mathring{\gamma}_n^{(7)}=\mathring{\gamma}_n^{(6)}.
\end{array}
\end{eqnarray}
Only one expansion when 
\begin{eqnarray}\label{p-7}
&&\begin{array}{l}
\bm{l-m}-2=2M_4\; \mbox{ and }\;
\bm{m+l}-1=2N_4\quad
\left(\Leftrightarrow\bm{m}=N_4-M_4-\frac{1}{2}, \right.\end{array}\nonumber\vspace{2mm}\\
&&
\begin{array}{l}
\left.\bm{l}=M_2+N_4+\frac{3}{2}\right): 
\mathring{\psi}^{(7)}\mbox{ if } \; M_4<N_4,\quad 
\tilde{\psi}^{(7)}\mbox{ if } \; N_4<M_4.\end{array}
\end{eqnarray}
%
%
%
%

 \subsubsection{Fourth type: Odd solutions, period  $2K$}%
\antiletra\letra
\begin{eqnarray}\label{ps4-A}
\begin{cases}
\mathring{{\psi}}^{(4)}(u)=
\mathrm{sn}u\;\mathrm{cn}u\;
\mathrm{dn}^{\bm{l}+1}u \displaystyle
\sum_{n=0}^{\frac{\bm{m-l}-3}{2}}
\mathring{{b}}_{n}^{(4)}\;
\mathrm{sn}^{2n}u
\quad\mbox{ if }\quad 
\begin{array}{l}
 \bm{m}-\bm{l}=3,5,7,\cdots,\hspace{1.3cm};\end{array}\\
\begin{array}{l}\hspace{8cm}
\left[\;\bm{m}\leq-\frac{1}{2} ,\;\;
\bm{m}+\bm{l}<-4\;\right] \end{array}
\vspace{3mm}\\
\tilde{{\psi}}^{(4)}(u)=
\mathrm{sn}u\;\mathrm{cn}u\;
\mathrm{dn}^{\bm{l}+1}u \displaystyle
\sum_{n=0}^{-\frac{\bm{m+l}+4}{2}}
\tilde{{b}}_{n}^{(4)}\;
\mathrm{sn}^{2n}u
\quad\mbox{ if }\quad
\begin{array}{l}
 \bm{m}+\bm{l}=-4,-6,-8,\cdots, 
\;;\end{array}\\
\begin{array}{l}\hspace{8cm}\left[\bm{m}\geq-\frac{1}{2},\;\;
\bm{m}-\bm{l}>3\; \right]
 \end{array}
\end{cases}
\end{eqnarray}
\begin{eqnarray}\label{ps4-A-2}
&&\begin{array}{l}
\mathring{{\alpha}}_n^{(4)}=\frac{1}{k^2}\left(n+\frac{3}{2}\right)(n+1),\qquad
\mathring{{\beta}}_n^{(4)}= -\left(\frac{1}{k^2}+1\right)n^2-
\left(\frac{2}{k^2}+\bm{l}+2\right)n+\end{array}\nonumber\vspace{2mm}\\
&&\begin{array}{l}
\frac{1}{4k^2}({\cal E}-4)-
\frac{1}{4}(\bm{l}+2)^2,\qquad
\mathring{\bm{\gamma}}_n^{(4)}=\left(n-\frac{\bm{m}-\bm{l}-1}{2}\right)\left(n+\frac{\bm{l}+\bm{m}+2}{2}\right);
\end{array}
\end{eqnarray}
Only one expansion when 
\begin{eqnarray}\label{p-4}
&&
\begin{array}{l}
\bm{m-l}-3=2M_5\; \mbox{ and }\;
\bm{m+l}+4=-2N_5,\quad
\left(\Leftrightarrow\bm{m}=M_5-N_5-\frac{1}{2},\right. \end{array}\nonumber\vspace{2mm}\\
&&
\begin{array}{l}
\left. \bm{l}=-M_5-N_5-\frac{3}{2}\right):
\mathring{\psi}^{(4)}\mbox{ if } \; M_5<N_5,\quad 
\tilde{\psi}^{(4)}\mbox{ if } \; N_5<M_5.\end{array}
\end{eqnarray}
%
%
%
%
%
%
%
\antiletra\letra
\begin{eqnarray}\label{ps8-B}
\begin{cases}
\mathring{{\psi}}^{(8)}(u)=
\mathrm{sn}u\;\mathrm{cn}u\;
\mathrm{dn}^{-\bm{l}}u \displaystyle
\sum_{n=0}^{\frac{\bm{l-m}-3}{2}}
\mathring{{b}}_{n}^{(8)}\;
\mathrm{sn}^{2n}u
\quad\mbox{ if }\quad 
\begin{array}{l}
 \bm{l}-\bm{m}=3,5,7,\cdots,\hspace{7mm};\end{array}\\
\begin{array}{l}\hspace{8cm}
\left[\;\bm{m}\geq-\frac{1}{2} ,\;\;
\bm{m}+\bm{l}>2\;\right], \end{array}
\vspace{3mm}\\
\tilde{{\psi}}^{(8)}(u)=
\mathrm{sn}u\;\mathrm{cn}u\;
\mathrm{dn}^{-\bm{l}}u \displaystyle
\sum_{n=0}^{\frac{\bm{m+l}-2}{2}}
\tilde{{b}}_{n}^{(8)}\;
\mathrm{sn}^{2n}u
\quad\mbox{ if }\quad
\begin{array}{l}
 \bm{m}+\bm{l}=2,4,6,\cdots, ;\end{array}\\
\begin{array}{l}\hspace{8cm}
\qquad\left[\bm{m}\leq-\frac{1}{2},\;\;
\bm{l}-\bm{m}>3\; \right],
 \end{array}
\end{cases}
\end{eqnarray}
\begin{eqnarray}\label{ps8-A-2}
&&\begin{array}{l}
\mathring{{\alpha}}_n^{(8)}=\frac{1}{k^2}\left(n+\frac{3}{2}\right)(n+1),\qquad
\mathring{{\beta}}_n^{(5)}= -\left(\frac{1}{k^2}+1\right)n^2-
\left(\frac{2}{k^2}-\bm{l}+1\right)n+\end{array}\nonumber\vspace{2mm}\\
&&\begin{array}{l}
\frac{1}{4k^2}({\cal E}-4)-
\frac{1}{4}(\bm{l}-1)^2,\qquad\mathring{\bm{\gamma}}_n^{(8)}=\left(n-\frac{\bm{l}-\bm{m}-1}{2}\right)\left(n-\frac{\bm{l}+\bm{m}}{2}\right).
\end{array}
\end{eqnarray}
Only one expansion when 
\begin{eqnarray}\label{p-8}
&&
\begin{array}{l}
\bm{l-m}-3=2M_6\; \mbox{ and }\;
\bm{m+l}-2=-2N_6\quad
\left(\Leftrightarrow\bm{m}=N_6-M_6-\frac{1}{2},\right. \end{array}\nonumber\vspace{2mm}\\
&&
\begin{array}{l}\left.
\bm{l}=M_6+N_6+\frac{5}{2}\right): 
\mathring{\psi}^{(8)}\mbox{ if } \; M_6<N_6,\quad 
\tilde{\psi}^{(8)}\mbox{ if } \; N_6<M_6.\end{array}
\end{eqnarray}
\antiletra
%
%
%
%

{Due to the symmetry under 
the substitutions (\ref{subs}), we discuss the Arscott condition (\ref{autovalores}) only for  $\left(\mathring{\psi}^{(i)},
\tilde{\psi}^{(i)}\right)$ with $i=1,2,3,4$ .} As the $\mathring{{\alpha}}_{n}^{(i)}$ 
are positive,  the  condition  
$\mathring{\alpha}_{n-1}\mathring\gamma_n>0$
is satisfied if
$\mathring\gamma_n>0$ for any admissible $n$. So,
for $ \mathring{{\psi}}^{(i)}$ we requires that 
\begin{align*}
&\begin{array}{r}
\mathring{{\gamma}}
_n^{(1)}=\left(n-\frac{\bm{m}-\bm{l}+1}{2}\right)
\left(n+\frac{\bm{l}+\bm{m}}{2}\right)>0
\;\mbox{for}\; 1\leq n\leq \frac{\bm{m}-\bm{l}-1}{2},
\;
 \left[\bm{m}-\bm{l}=3,5,7,\cdots\right],
\end{array}\\
&\begin{array}{r}
\mathring{{\gamma}}
_n^{(2)}=  \mathring{\bm{\gamma}}
_n^{(3)}=   \left(n-\frac{\bm{m}-\bm{l}}{2}\right)
\left(n+\frac{\bm{l}+\bm{m}+1}{2}\right)>0
\; \mbox{ for }\;1\leq n\leq \frac{\bm{m}-\bm{l}-2}{2},
\;
 \left[{m}-\bm{l}= 4,6,8,\cdots\right],
\end{array}\\
&\begin{array}{r}
\mathring{\bm{\gamma}}
_n^{(4)}=\left(n-\frac{\bm{m}-\bm{l}-1}{2}\right)
\left(n+\frac{\bm{l}+\bm{m}+2}{2}\right)>0
\;\mbox{for}\;1\leq n\leq \frac{\bm{m}-\bm{l}-3}{2},
\;
 \left[\bm{m}-\bm{l}= 5,7,9,\cdots\right].
\end{array}
\end{align*}
Thus, since the first term is negative,  $\mathring\gamma_n>0$ if  the last term on the right
of the above equations is also negative for each value of $n$.
By considering  the maximum value of $n$ we find that $\bm{m}<1/2$
for the four expansions; {to include all
values of $n$,  we take  
$\bm{m}\leq-1/2$ due to 
(\ref{nova}) and similar relations for each $\mathring\gamma_n^ {(i)}$ . On the other side, for series with several terms,
$\mathring\gamma_n^ {(i)}>0$ implies $\mathring\gamma_1^ {(i)}>0$: 
this imposes the following restrictions on $\bm{m}+\bm{l}$
\begin{eqnarray}\label{A-1}
&&\begin{array}{l}
\bm{m}\leq-\frac{1}{2} \mbox{ for } 
\mathring{{\psi}}^{(i)},\quad
\bm{m}+\bm{l}+2<0\; \mbox{ for } \;\mathring{{\psi}}^{(1)},\;\; \bm{m}+\bm{l}+3<0\; \mbox{ for }\; \mathring{{\psi}}^{(2)}
\mbox{ and }\mathring{{\psi}}^{(3)},\end{array}
\nonumber
\vspace{3mm}\\
&&\begin{array}{l}
\bm{l}+\bm{m}+4<0\; \mbox{ for }\; \mathring{{\psi}}^{(4)}
\end{array},
\end{eqnarray}
($i=1,2,3,4$) in addition to the restrictions on $\bm{m}-\bm{l}$. Actually, firstly we find  conditions  to get $\mathring{\gamma}_n^{(i)}\leq 0$ by considering $n$ as a continuous variable. For example,
\begin{eqnarray}\label{nova}
\begin{array}{l}
\mathring{\bm{\gamma}}
_n^{(1)}\leq 0 \; \mbox{ if }
\begin{cases}
        \frac{\bm{m}-\bm{l}+1}{2}  \leq n\leq-\frac{\bm{m}+\bm{l}}{2}  \qquad \mbox{ when }\qquad \bm{m}\leq -\frac{1}{2},\vspace{2mm}\\
  -\frac{\bm{m}+\bm{l}}{2} \leq n \leq \frac{\bm{ m}-\bm{l}+1}{2}
 \qquad \mbox{ when }\qquad \bm{m}\geq -\frac{1}{2}.
\end{cases}\end{array}
\end{eqnarray}
Thence, $\mathring{\gamma}_n^{(i)}>0$ when $n$ lies outside the above intervals. In particular,
\begin{eqnarray*}\begin{array}{l}
\mathring{\bm{\gamma}}
_n^{(1)}>0 \; \mbox{ for }
n<\frac{\bm{m}-\bm{l}+1}{2}, \mbox{ with }\; \bm{m}\leq -\frac{1}{2},
\end{array}
\end{eqnarray*}
an  interval which includes the subinterval $n\leq (\bm{m}-\bm{l}-1)/2$. }


%
On the other side, for $ \tilde{{\psi}}^{(i)}$, we demand that 
($\mathring{{\gamma}}
_n^{(2)}=  \mathring{{\gamma}}
_n^{(3)}$)
\begin{align*}
&\begin{array}{r}
\mathring{{\gamma}}
_n^{(1)}=\left(n+\frac{\bm{l}+\bm{m}}{2}\right)
\left(n+\frac{\bm{l}-\bm{m}-1}{2}\right)>0
\;\mbox{for}\; 1\leq n\leq -\frac{\bm{m}+\bm{l}+2}{2},
\;
 \left[\bm{m}+\bm{l}=-4,-6,-8,\cdots\right],
\end{array}\\
&\begin{array}{r}
\mathring{{\gamma}}
_n^{(2)}=  
\left(n+\frac{\bm{l}+\bm{m}+1}{2}\right)\left(n+\frac{\bm{l}-\bm{m}}{2}\right)>0
\; \mbox{for}\;1\leq n\leq -\frac{\bm{m}+\bm{l}+3}{2},
\;
 \left[{m}+\bm{l}= -5,-7,-9,\cdots\right],
\end{array}\\
&\begin{array}{r}
\mathring{{\gamma}}
_n^{(4)}=
\left(n+\frac{\bm{l}+\bm{m}+2}{2}\right)\left(n+\frac{\bm{l}-\bm{m}+1}{2}\right)>0
\; \mbox{for}\;1\leq n\leq -\frac{\bm{m}+\bm{l}+4}{2},
\;
 \left[\bm{m}-\bm{l}= -6,-8,\cdots\right].
\end{array}
\end{align*}
As the first term is negative, {then $\mathring\gamma_n^{(i)}>0$ 
if the second term on the right
of these equations is also negative for each value of $n$. From (\ref{nova}) and similar relations for each $\mathring\gamma_n^ {(i)}$, we find that $\bm{m}\geq -{1}/{2}$. 
Besides this, $\mathring\gamma_1^ {(i)}>0$ restricts  the values of  $\bm{m}-\bm{l}$, namely,
\begin{eqnarray}\label{B-1}
&&\begin{array}{l}
\bm{m}\geq-\frac{1}{2} \mbox{ for } 
\tilde{{\psi}}^{(i)},\quad
\bm{l}-\bm{m}+1<0\; \mbox{ for } \;\tilde{{\psi}}^{(1)},\;\; \bm{l}-\bm{m}+2<0\; \mbox{ for }\; \tilde{{\psi}}^{(2)}
\mbox{ and }\tilde{{\psi}}^{(3)},\end{array}\nonumber\vspace{3mm}\\
&&\begin{array}{l}
\bm{l}-\bm{m}+3<0\;\mbox{ for }\;
\tilde{{\psi}}^{(4)},
\end{array}
\end{eqnarray}
($i=1,2,3,4$) in addition to the restrictions on $\bm{m}+\bm{l}$. 
Using  (\ref{nova})  we can check, for example, that
\begin{eqnarray*}\begin{array}{l}
\mathring{\bm{\gamma}}
_n^{(1)}>0 \; \mbox{ for }
n<-\frac{\bm{l}+\bm{m}}{2}, \mbox{ when }\; \bm{m}\geq -\frac{1}{2},
\end{array}
\end{eqnarray*}
where the above interval includes all the possible 
values for $n$, that is, $n\leq -(\bm{l}+\bm{m}+2)/2$.}


%
%
%
%
%
%

{
To proceed  we mention some examples,  Obviously, the cases (\ref{p-1}), (\ref{p-2}), $\cdots$,  (\ref{p-8}) are solved by finite series of $\mathrm{sn}^{2}u $. Furthermore, we find that each of these cases admits a second expansion having the same period and opposite parity. 
For instance, if $0<M_1<N_1$ in  (\ref{p-1}), we have 
%

%
%
\begin{eqnarray}\label{ex-01}
\begin{array}{l}
\mbox{for }\bm{m}=M_1-N_1-\frac{1}{2}\\
\mbox{and }\bm{l}=-M_1-N_1-\frac{3}{2}\end{array}:
\begin{cases}\mathring{\psi}^{(1)} \mbox { with } 0\leq n\leq M_1<N_1
\mbox{, even, period }2K ;\vspace{2mm} \\
 \mathring{\psi}^{(4)} \mbox { with } 0\leq n\leq (M_1-1), 
\mbox{  odd, period }2K;
\end{cases}
\end{eqnarray}
%
%
%
%
and so far on. In each case there are only two distinct expansions satisfying the Arscott condition.}


{
As a second example we take $\bm{m}= -\frac{1}{2}$ 
in which case $\mathring{{\psi}}^{(i)}(u)=\tilde{{\psi}}^{(i)}(u) $.
The eight expansions satisfy the Arscott condition with 
the following values for $\bm{l}$:
\begin{align}
%
&\begin{array}{l}
\mbox{even solutions, period } 2K \; (\bm{m}=-1/2):
\left\{
\begin{array}{l}
\mathring{\psi}^{(1)} \;\mbox{ if }\; \bm{l}=-\frac{3}{2},-\frac{7}{2}, -\frac{11}{2},\cdots; \vspace{3mm}
\\
\mathring{{\psi}}^{(5)}\; \mbox{ if }\;\bm{l}= \frac{1}{2},\frac{5}{2}, \frac{9}{2},\cdots;
\end{array}
\right.
\end{array}
\label{m1/2-1}\vspace{3mm}\\
%
%
&\begin{array}{l}
\mbox{odd solutions, period } 2K\; (\bm{m}=-1/2):
\left\{
\begin{array}{l}
\mathring{\psi}^{(4)}\;\mbox{ if } \;
\bm{l}=-\frac{7}{2},-\frac{11}{2}, -\frac{15}{2}\cdots;
\vspace{2mm}\\
\mathring{{\psi}}^{(8)}\; \mbox{ if }\; \bm{l}=
\frac{5}{2},\frac{9}{2}, \frac{13}{2}\cdots;
 \end{array}
 \right.
 \vspace{2mm}\end{array}
 \label{m1/2-4}\\
%
%
&\begin{array}{l}
\mbox{odd solutions, period } 4K \; (\bm{m}=-1/2):
\left\{
\begin{array}{l}
\mathring{\psi}^{(2)}\;\mbox{ if }\;\bm{l}=-\frac{5}{2},-\frac{9}{2}, -\frac{13}{2}\cdots;
\\
\mathring{{\psi}}^{(6)}\;\mbox{ if }\; \bm{l}=\frac{3}{2},\frac{7}{2}, \frac{11}{2}\cdots;
 \end{array}
 \right.
 \vspace{2mm}\end{array}\label{m1/2-2}\\
 %
 %
&\begin{array}{l}
\mbox{even solutions, period } 4K\; (\bm{m}=-1/2):
\left\{
 \begin{array}{l}
\mathring{\psi}^{(3)}\;\mbox{ if } \; \bm{l}=-\frac{5}{2},-\frac{9}{2}, -\frac{13}{2}\cdots;\vspace{2mm}\\
\mathring{{\psi}}^{(7)}\;\mbox{ if }\;\bm{l}=\frac{3}{2},\frac{7}{2}, \frac{11}{2},\cdots,
 \end{array}
 \right.
 \end{array}
 \label{m1/2-3}
\end{align}
In the above expressions, there are at the most two distinct expansions for a given
value of $\bm{l}$ (only one solution when
$\bm{l}=-3/2$ or $\bm{l}=1/2$). Notice that the Arscott condition
is applicable for series with two or more terms; further, if the condition is not
fulfilled, we cannot in advance state anything on the eigenvalues of the 
characteristic equation.  


{
As a last illustration, we consider solutions for Eq. (\ref{first-case}), that is,  for  
$\bm{m}=0$ or $\bm{m}=-1$. There are finite-series solutions in series of
 $\mathrm{sn}^{2n}u $ for any integer value of $\bm{l}$ (excepting 
  $\bm{l}=-1$), but  only the following eight expansions
satisfy the Asrcott condition 
\begin{align}
%
&\begin{array}{l}
\mbox{even solutions, period } 2K\; (\bm{m}=0):
\left\{
\begin{array}{l}
\tilde{\psi}^{(1)} \quad \mbox{ if }\quad \bm{l}=-2,-4,-6,\cdots;\vspace{2mm}\\
\mathring{{\psi}}^{(5)}\quad \mbox{ if }\quad\bm{l}=1,3,5,\cdots;\end{array}
\right.
\end{array}
\label{m0-1}\\
%
%
&\begin{array}{l}
\mbox{odd solutions, period } 2K\; (\bm{m}=0):
\left\{
\begin{array}{l}
\tilde{\psi}^{(4)}\quad\mbox{ if } \quad
\bm{l}=-4,-6-8,\cdots ;\vspace{2mm}\\
\mathring{{\psi}}^{(8)}\quad \mbox{ if }\quad \bm{l}=3,5,7,\cdots;
\end{array}
\right.
\end{array}
\label{m0-4}\\
%
&
\begin{array}{l}
\mbox{odd solutions, period } 4K\; (\bm{m}=0):
\left\{
\begin{array}{l}
\tilde{\psi}^{(2)}\quad\mbox{ if }\quad\bm{l}=-3,-5,-7,\cdots
;\vspace{2mm}\\
\mathring{{\psi}}^{(6)}\quad\mbox{ if }\quad \bm{l}=-2,-4,-6,\cdots;
\end{array}
\right.
\end{array}\label{m0-2}\\
 %
 %
&\begin{array}{l}
\mbox{even solutions, period } 4K\; (\bm{m}=0):
\left\{
\begin{array}{l}
\tilde{\psi}^{(3)}\quad\mbox{ if } \quad \bm{l}=-3,-5,-7,\cdots
;\vspace{2mm}\\
\mathring{{\psi}}^{(7)}\quad\mbox{ if }\quad\bm{l}=-2,-4,-6,\cdots.
\end{array}
\right.
\end {array}\label{m0-3}
\end{align}
On the other hand, for $\bm{m}=-1$ the remaining expansions also lead to solutions satisfying the Arscott condition; nevertheless, such solutions are identical to the above ones.  For example, 
we find that
\begin{eqnarray*}\begin{array}{l}
\mathring{\psi}^{(1)} \big|_{\bm{m}=-1}=\tilde{\psi}^{(1)} \big|_{\bm{m}=0}, \qquad    
\tilde{\psi}^{(5)} \big|_{\bm{m}=-1}=\mathring{\psi}^{(5)} \big|_{\bm{m}=0},\qquad\mbox{and so on.}
\end{array}
\end{eqnarray*}
From relations  (\ref{m0-1}-\ref{m0-3}), we
find that there are four types of finite-series solutions if  $\bm{l}=-4,-6,-8,\cdots$, unlike the expansions of the previous examples.}

%
%

%
%

%
\subsection{Finite-series eigenfunction $ \left( \mathring{\bm{\Psi}}^{(i)},\tilde{\bm{\Psi}}^{(i)}\right)$ in  terms of 
$\mathrm{cn}^2u$}
%
%
{This is similar to Section 5.1. 
The 
finite-series expansions
$\left(\mathring{\Psi}^{(i)},
\tilde{\Psi}^{(i)}\right)$ now result from the power-series solutions 
$\mathring{\bm{H}}^{(i)}(x)$ for the Heun equation, 
truncated as in (\ref{truncation-2}). 
The pairs   $\left(\mathring{\Psi}^{(j+4)} ,
\tilde{\psi}^{(j+4)}\right)$  ($j=1,2,3,4$) can be generated from   
$\left(\mathring{\psi}^{(j)},
\tilde{\Psi}^{(j)}\right)$ by the substitutions  
$\bm{l}\mapsto -\bm{l}-1$ and $\bm{m}\mapsto -\bm{m}-1$. The Arscott condition is satisfied if 
$\mathring{\bm{\gamma}}_n^{(i)}=\mathring{{\gamma}}_n^{(i)}<0$, in opposition 
to solutions of Section 5.1.}

{
In the recurrence relations   $\mathring{\bm{\alpha}}_{n}\mathring{\bm{b}}_{n+1}+\mathring{\bm{\beta}}_{n}
\mathring{\bm{b}}_{n}+
\mathring{\bm{\gamma}}_{n}\mathring{\bm{b}}_{n-1}=0$ and 
$\tilde{\bm{\alpha}}_{n}\tilde{\bm{b}}_{n+1}+\tilde{\bm{\beta}}_{n}
\tilde{\bm{b}}_{n}+
\tilde{\bm{\gamma}}_{n}\tilde{\bm{b}}_{n-1}=0$,  formally we have 
\begin{eqnarray}
\tilde{\bm{\alpha}}_{n}^{(i)}=\mathring{\bm{\alpha}}_n^{(i)},\qquad
\tilde{\bm{\beta}}_{n}^{(i)}=\mathring{\bm{\beta}}_n^{(i)},\qquad
\tilde{\bm{\gamma}}_{n}^{(i)}=\mathring{\bm{\gamma}}_n^{(i)}.
\end{eqnarray}
After  writing the four types of expansions,  we discuss the Arscott condition (\ref{autovalores}) and comment on the 
Lam\'e equation.}
%
%
%
%

 \subsubsection{First  type: even solutions, period  $2K$}
\letra
\begin{eqnarray}\label{ps1-C}
\begin{cases}
\mathring{\bm{\Psi}}^{(1)}(u)=
\mathrm{dn}^{\bm{l}+1} u \;\displaystyle
\sum_{n=0}^{\frac{\bm{m-l}-1}{2}}\mathring{\bm{b}}_{n}^{(1)}\;
\mathrm{cn}^{2n}u  \quad\mbox{ if }\quad 
\begin{array}{l}
\bm{m}-\bm{l}= 1,3,5,\cdots,\end{array}\\
\begin{array}{l}\hspace{7cm}
\left[\; \bm{m}>-\frac{1}{2},\; \bm{m+l}>-2\; \right],\end{array}\vspace{2mm}\\
\tilde{\bm{\Psi}}^{(1)}(u)=
\mathrm{dn}^{\bm{l}+1} u \;\displaystyle
\sum_{n=0}^{-\frac{\bm{m+l}+2}{2}}\tilde{\bm{b}}_{n}^{(1)}\;
\mathrm{cn}^{2n}u  \quad\mbox{ if }\quad 
\begin{array}{l}
\bm{m}+\bm{l}= -2,-4,-6,\cdots,\end{array}\\
\begin{array}{l}\hspace{7cm}
\left[\; \bm{m}< -\frac{1}{2},\; \bm{m-l}<1\; \right],\end{array}
\end{cases}
\end{eqnarray}
\begin{eqnarray}\label{ps1-C-1}
&&\begin{array}{l}
\mathring{\bm{\alpha}}_n^{(1)}=\left(1-\frac{1}{k^2}\right)\left(n+\frac{1}{2}\right)(n+1),\qquad \mathring{\bm{\beta}}_n^{(1)}=
\left(\frac{1}{k^2}-2\right)n^2-
(\bm{l}+1)n-\frac{{\cal E}}{4k^2}+,\end{array}
\nonumber\vspace{2mm}\\
&&\begin{array}{l}\frac{\bm{m}(\bm{m}+1)-\bm{l}-1}{4},\qquad
\mathring{\bm{\gamma}}_n^{(1)}=\left(n-\frac{\bm{m}-\bm{l}+1}{2}\right)\left(n+\frac{\bm{l}+\bm{m}}{2}\right);
\end{array}
\end{eqnarray}
{
As in the expansions in series of $ \mathrm{sn}^{2}u $, there is only one expansion when ($M_1$ and $N_1$ denote positive integers)
\begin{eqnarray*}\label{P-1}
&\bm{m-l}-1=2M_1\; \mbox{ and }\;
\bm{m+l}+2=-2N_1\quad
\left(\Leftrightarrow\bm{m}=M_1-N_1-\frac{1}{2},\right.: \nonumber\vspace{2mm}\\
&\left.\bm{l}=-M_1-N_1-\frac{3}{2}\right): \mathring{\bm\psi}^{(1)}\mbox{ if } \; M_1<N_1,\qquad
\tilde{\bm\psi}^{(1)}\mbox{ if } \; N_1<M_1.
\end{eqnarray*}
These expansions do not satisfy the Arscott condition. Actually, the 
Arscott condition is not  satisfied by series of $ \mathrm{cn}^{2}u $ for any 
of the cases given in (\ref{p-1}), (\ref{p-5}), $\cdots$, (\ref{p-8}).
Thus,  these cases will not be considered from now on.}
%
%
\antiletra\letra
\begin{eqnarray}\label{ps5-D}
\begin{cases}
\mathring{\bm{\Psi}}^{(5)}(u)=
\displaystyle \frac{1}{\mathrm{dn}^{\bm{l}}u}\; 
\sum_{n=0}^{\frac{\bm{l}-\bm{m}-1}{2}}\mathring{\bm{b}}_{n}^{(5)}\;
\mathrm{cn}^{2n}u \quad\mbox{ if }\quad
\begin{array}{l}
\bm{l}-\bm{m}= 1,3,5,\cdots\quad
\hspace{5mm};\end{array}\\
\begin{array}{l}\hspace{8cm}
\left[\; \bm{m}<-\frac{1}{2},\; \bm{m+l}<0\; \right],
\end{array}
\vspace{2mm}\\
\tilde{\bm{\Psi}}^{(5)}(u)=
\displaystyle \frac{1}{\mathrm{dn}^{\bm{l}}u}\; 
\sum_{n=0}^{\frac{\bm{l}+\bm{m}}{2}}\;\tilde{\bm{b}}_{n}^{(5)}\;
\mathrm{cn}^{2n}u \qquad\mbox{if }\quad
\begin{array}{l}
\bm{l}+\bm{m}= 0,2,4,\cdots,
\qquad;\end{array}\\
\begin{array}{l}\hspace{8cm}
\left[\; \bm{m}> -\frac{1}{2},\; \bm{l-m}<1\; \right],\end{array}
\end{cases}
\end{eqnarray}
\begin{eqnarray}\label{ps5-C-1}
&&\begin{array}{l}
\mathring{\bm{\alpha}}_n^{(5)}=\left(1-\frac{1}{k^2}\right)\left(n+\frac{1}{2}\right)(n+1),\qquad \mathring{\bm{\beta}}_n^{(5)}= \left(\frac{1}{k^2}-2\right)n^2+
\bm{l}n-\frac{{\cal E}}{4k^2}+ \end{array}\nonumber
\vspace{2mm}\\
&&\begin{array}{l}\frac{\bm{l}+\bm{m}(\bm{m}+1)}{4}
,\qquad
\mathring{\bm{\gamma}}_n^{(5)}=\left(n´-\frac{\bm{l}-\bm{m}+1}{2}\right)
\left(n-\frac{\bm{l}+\bm{m}+2}{2}\right).
\end{array}
\end{eqnarray}
%
%
%
%
%
%
%
\subsubsection{Second  type: odd solutions, period  $4K$}
\antiletra\letra
\begin{eqnarray}\label{ps2-C}
\begin{cases}
\mathring{\bm{\Psi}}^{(2)}(u)=
\mathrm{sn}u\;\mathrm{dn}^{\bm{l}+1}u \displaystyle
\sum_{n=0}^{\frac{\bm{m-l}-2}{2}}\mathring{\bm{b}}_{n}^{(2)}\;
\mathrm{cn}^{2n}u 
\qquad\mbox{ if }\quad
\begin{array}{l}
\bm{m}-\bm{l}= 2,4,6,\cdots,\end{array}\\
\begin{array}{l}\hspace{8cm}
\left[\; \bm{m}> -\frac{1}{2},\; \bm{m+l}>-3\; \right],\end{array} \vspace{2mm}\\
{\tilde{\bm\Psi}}^{(2)}(u)=
\mathrm{sn}u\;\mathrm{dn}^{\bm{l}+1}u \displaystyle
\sum_{n=0}^{-\frac{\bm{m+l}+3}{2}}\tilde{\bm{b}}_{n}^{(2)}\;
\mathrm{cn}^{2n}u \quad\mbox{ if }\quad
\begin{array}{l}
\bm{m}+\bm{l}= -3,-5,-7,\cdots,\end{array}\\
\begin{array}{l}\hspace{8cm}
\left[\; \bm{m}< -\frac{1}{2},\; \bm{m-l}<2\; \right],
\end{array}
\end{cases}
\end{eqnarray}

\begin{eqnarray}\label{ps2-C-1}
&&\begin{array}{l}
\mathring{\bm{\alpha}}_n^{(2)}=\left(1-\frac{1}{k^2}\right)\left(n+\frac{1}{2}\right)(n+1),\qquad 
\mathring{\bm{\beta}}_n^{(2)}=
\left(\frac{1}{k^2}-2\right)
n^2+\left(\frac{1}{k^2}-\bm{l}-2\right)n+
\end{array}\nonumber
\vspace{2mm}\\
&&\begin{array}{l}
\frac{1-{\cal E}}{4k^2}+\frac{\bm{m}(\bm{m}+1)-\bm{l}-2}{4},\qquad
\mathring{\bm{\gamma}}_n^{(2)}=\left(n-\frac{\bm{m}-\bm{l}}{2}\right)\left(n+\frac{\bm{l}+\bm{m}+1}{2}\right);
\end{array}
\end{eqnarray}
%
%
%
%
%
\antiletra\letra
\begin{eqnarray}\label{ps6-D}
\begin{cases}
\mathring{\bm{\Psi}}^{(6)}(u)=
\displaystyle
\frac{\mathrm{sn}{u}}{{\rm dn}^{\bm{l}}u} \;
\sum_{n=0}^{\frac{\bm{l}-\bm{m}-2}{2}}\mathring{\bm{b}}_{n}^{(6)}
\;{\rm cn}^{2n}u 
\qquad  \mbox{ if }\qquad
\bm{l}-\bm{m}=2,4,6,\cdots,\\
\begin{array}{l}\hspace{8cm}
 \left[\; \bm{m}<-\frac{1}{2},\; \bm{m+l}<1\; \right],
 \end{array}
 \vspace{2mm}\\
\tilde{\bm{\Psi}}^{(6)}(u)=
\displaystyle
\frac{\mathrm{sn}{u}}{{\rm dn}^{\bm{l}}u} \;
\sum_{n=0}^{\frac{\bm{l}+\bm{m}-1}{2}}\tilde{\bm{b}}_{n}^{(6)}
\;{\rm cn}^{2n}u 
\qquad  \mbox{ if }\qquad
\bm{l}+\bm{m}=1,3,5,\cdots,\\
\begin{array}{l}\hspace{8cm}
%
\left[\; \bm{m}>-\frac{1}{2},\; \bm{l-m}<2\; \right],
\end{array}
\end{cases}
\end{eqnarray}
\begin{eqnarray}\label{ps6-C-1}
&&\begin{array}{l}
\mathring{\bm{\alpha}}_n^{(6)}=\left(1-\frac{1}{k^2}\right)\left(n+\frac{1}{2}\right)(n+1),
\qquad
\mathring{\bm{\beta}}_n^{(6)}= \left(\frac{1}{k^2}-2\right)n^2+\left(\frac{1}{k^2}
+\bm{l}-1\right)n+
\end{array}\nonumber
\vspace{3mm}\\
&&\begin{array}{l}
\frac{1-{\cal E}}{4k^2}+\frac{\bm{m}(\bm{m}+1)+\bm{l}-1}{4},\qquad
\mathring{\bm{\gamma}}_n^{(6)}=\left(n-\frac{\bm{l}-\bm{m}}{2}\right)\left(n-\frac{\bm{l}+\bm{m}+1}{2}\right).
\end{array}
\end{eqnarray}
%
%
%
%
\subsubsection{Third  type: even solutions, period  $4K$}
\antiletra\letra
\begin{eqnarray}\label{ps3-C}
\begin{cases}
\mathring{\bm{\Psi}}^{(3)}(u)=
\mathrm{cn}u\;\mathrm{dn}^{\bm{l}+1}u \;\displaystyle
\sum_{n=0}^{\frac{\bm{m-l}-2}{2}}\mathring{\bm{b}}_{n}^{(3)}
\;\mathrm{cn}^{2n}u  \quad\mbox{ if }\quad
\begin{array}{l}
\bm{m}-\bm{l}=2,4,6,\cdots,\quad\end{array}
%
\\
\begin{array}{l}\hspace{8cm}
\left[\; \bm{m}> -\frac{1}{2},\; \bm{m+l}>-3\; \right],
\end{array}
\vspace{2mm}\\
{\tilde{\bm\Psi}}^{(3)}(u)=
\mathrm{cn}u\;\mathrm{dn}^{\bm{l}+1}u \;\displaystyle
\sum_{n=0}^{-\frac{\bm{m+l}+3}{2}}\tilde{\bm{b}}_{n}^{(3)}
\;\mathrm{cn}^{2n}u \quad\mbox{ if }\quad
\begin{array}{l}
\bm{m}+\bm{l}=-3,-5,-7,\cdots
,\qquad\end{array}
%
\\
\begin{array}{l}\hspace{8cm}
\left[\; \bm{m}<-\frac{1}{2},\; \bm{m-l}<2\; \right],
\end{array}
\end{cases}
\end{eqnarray}
\begin{eqnarray}\label{ps3-C-1}
&&\begin{array}{l}
\mathring{\bm{\alpha}}_n^{(3)}=\left(1-\frac{1}{k^2}\right)\left(n+\frac{3}{2}\right)(n+1),\quad 
\mathring{\bm{\beta}}_n^{(3)}=\left(\frac{1}{k^2}-2\right) n^2+\left(\frac{1}{k^2}-\bm{l}-3\right)n+
\end{array}\nonumber\vspace{2mm}\\
&&\begin{array}{l}
\frac{1-{\cal E}}{4k^2}
-\frac{1}{4}(3\bm{l}+5)+\frac{1}{4}\bm{m}(\bm{m}+1),\qquad
\mathring{\bm{\gamma}}_n^{(3)}=\left(n-\frac{\bm{m}-\bm{l}}{2}\right)\left(n+\frac{\bm{l}+\bm{m}+1}{2}\right);
\end{array}
\end{eqnarray}
%
%
%
%
\antiletra\letra
\begin{eqnarray}\label{ps7-D}
\begin{cases}
\mathring{\bm{\Psi}}^{(7)}(u)=
\displaystyle\; 
\frac{\mathrm{cn}{u}}{ {\rm dn}^{\bm{l}}u}\; 
\sum_{n=0}^{\frac{\bm{l}-\bm{m}-2}{2}}
\mathring{\bm{b}}_{n}^{(7)}\;
{\rm cn}{u}^{2n}u 
\quad\mbox{ if }\quad
\bm{l}-\bm{m}=2,4,6,\cdots,
%
\\
\begin{array}{l}\hspace{7cm}
%
%
\left[\; \bm{m}<-\frac{1}{2},\; \bm{m+l}<1\; \right],
\end{array}
\vspace{2mm}\\
\tilde{\bm{\Psi}}^{(7)}(u)=
\displaystyle\; 
\frac{\mathrm{cn}{u}}{ {\rm dn}^{\bm{l}}u}\; 
\sum_{n=0}^{\frac{\bm{l}+\bm{m}-1}{2}}
\tilde{\bm{b}}_{n}^{(7)}\;
{\rm cn}{u}^{2n}u
\quad\mbox{ if }\quad
\bm{l}+\bm{m}=1,3,5,\cdots,\quad\\
\begin{array}{l}\hspace{7cm}
%
%
\left[\; \bm{m}>-\frac{1}{2},\; \bm{l-m}<2\; \right],
\end{array}
\end{cases}
\end{eqnarray}
\begin{eqnarray}\label{ps7-C-1}
&&
\begin{array}{l}
\mathring{\bm{\alpha}}_n^{(7)}=\left(1-\frac{1}{k^2}\right)\left(n+\frac{3}{2}\right)(n+1),
\qquad
\mathring{\bm{\beta}}_n^{(7)}= \left(\frac{1}{k^2}-2\right)n^2+\left(\frac{1}{k^2}+
\bm{l}-2\right)n+\end{array}
\nonumber\vspace{2mm}\\
&&\begin{array}{l}
\frac{1-{\cal E}}{4k^2}
+\frac{1}{4}(3\bm{l}-2)+\frac{1}{4}\bm{m}(\bm{m}+1),\qquad
\mathring{\bm{\gamma}}_n^{(7)}=
\left(n-\frac{\bm{l}-\bm{m}}{2}\right)\left(n-\frac{\bm{l}+\bm{m}+1}{2}\right).
\end{array}
\end{eqnarray}
%
%
%
%
%
%
\subsubsection{Fourth type:  odd solutions, period  $2K$}
\antiletra\letra
\begin{eqnarray}\label{ps4-C}
\begin{cases}
\mathring{\bm{\Psi}}^{(4)}(u)=
\mathrm{sn}u\;\mathrm{cn}u\;
\mathrm{dn}^{\bm{l}+1}u \displaystyle
\sum_{n=0}^{\frac{\bm{m-l}-3}{2}}
\mathring{\bm{b}}_{n}^{(4)}\;
\mathrm{cn}^{2n}u \quad\mbox{if}\quad
\begin{array}{l}
 \bm{m}-\bm{l}=3,5,7,\cdots,\quad\end{array}
\\
\begin{array}{l}\hspace{7cm}
\left[ \bm{m}> -\frac{1}{2},\; \bm{m+l}>-4\; \right],\end{array} \vspace{2mm}\\
\tilde{\bm{\Psi}}^{(4)}(u)=
\mathrm{sn}u\;\mathrm{cn}u\;
\mathrm{dn}^{\bm{l}+1}u \displaystyle
\sum_{n=0}^{-\frac{\bm{m+l}+4}{2}}
\tilde{\bm{b}}_{n}^{(4)}\;
\mathrm{cn}^{2n}u \;\mbox{ if }\;
\begin{array}{l}
 \bm{m}+\bm{l}=-4,-6,-8,\cdots,\end{array}\\
\begin{array}{l}\hspace{7cm}
\left[\; \bm{m}< -\frac{1}{2},\; \bm{m-l}<3\; \;\right],
\end{array}
\end{cases}
\end{eqnarray}
\begin{eqnarray}\label{ps4-C-1}
&&\begin{array}{l}
\mathring{\bm{\alpha}}_n^{(4)}=\left(1-\frac{1}{k^2}\right)\left(n+\frac{3}{2}\right)(n+1),\quad
\mathring{\bm{\beta}}_n^{(4)}= \left(\frac{1}{k^2}-2\right)n^2+
\left(\frac{1}{k^2}-\bm{l}-4\right)n+
\end{array}
\nonumber\vspace{3mm}\\
&&\begin{array}{l}
\frac{4-{\cal E}}{4k^2}
-\frac{1}{4}(3\bm{l}+8)+\frac{1}{4}\bm{m}(\bm{m}+1),\quad
\mathring{\bm{\gamma}}_n^{(4)}=\left(n-\frac{\bm{m}-\bm{l}-1}{2}\right)\left(n+\frac{\bm{l}+\bm{m}+2}{2}\right);
\end{array}
\end{eqnarray}
%
%
%
\antiletra\letra
\begin{eqnarray}\label{ps8-D}
\begin{cases}
\mathring{\bm{\Psi}}^{(8)}(u)
\displaystyle\; 
=
\frac{\mathrm{sn}{u}\; \mathrm{cn}{u}}{{\rm dn}^{\bm{l}}u}
\sum_{n=0}^{\frac{\bm{l}-\bm{m}-3}{2}}\mathring{\bm{b}}_{n}^{(8)}
\;{\rm cn}^{2n}u
\quad\mbox{ if }\quad
\bm{l}-\bm{m}=3,5,7,\cdots,\quad \\
\begin{array}{l}\hspace{7cm}
%
\left[\; \bm{m}< -\frac{1}{2},\; \bm{m+l}<2\; \right],
\end{array}
\vspace{2mm} \\
\tilde{\bm{\Psi}}^{(8)}(u)=
\displaystyle\; 
\frac{\mathrm{sn}{u}\; \mathrm{cn}{u}}{{\rm dn}^{\bm{l}}u}
\sum_{n=0}^{\frac{\bm{l}+\bm{m}-2}{2}}\tilde{\bm{b}}_{n}^{(8)}
\;{\rm cn}^{2n}u 
\quad\mbox{ if }\quad
\bm{l}+\bm{m}=2,4,6,\cdots,\quad \\
\begin{array}{l}\hspace{7cm}
%
%
\left[\; \bm{m}> -\frac{1}{2},\; \bm{l-m}<3\; \;\right],
\end{array}
\end{cases}
\end{eqnarray}
\begin{eqnarray}\label{ps8-C-2}
&&\begin{array}{l}
\mathring{\bm{\alpha}}_n^{(8)}=\left(1-\frac{1}{k^2}\right)\left(n+\frac{3}{2}\right)(n+1),\qquad 
\mathring{\bm{\beta}}_n^{(8)}= \left(\frac{1}{k^2}-2\right)n^2+\left(
\frac{2}{k^2}+\bm{l}-3\right)n+
\end{array}
\vspace{2mm}
\nonumber\\
&&\begin{array}{l}\frac{4-{\cal E}}{4k^2}
+\frac{3\bm{l}-5}{4}+\frac{\bm{m}}{4}(\bm{m}+1),\qquad
\mathring{\bm{\gamma}}_n^{(8)}=
\left(n-\frac{\bm{l}-\bm{m}-1}{2}\right)\left(n-\frac{\bm{l}+\bm{m}}{2}\right).
\end{array}
\end{eqnarray}
\antiletra
%

%
{
Concerning the 
Arscott condition (\ref{autovalores}), we notice that $\left(1-{1}/{k^2}\right)<1$
implies that the coefficients 
$\mathring{\bm{\alpha}}_{n}^{(i)}$ 
are negative}. Then,  the  condition  
$\mathring{\alpha}_{n-1}\mathring\gamma_n>0$
is satisfied if
$\mathring\gamma_n<0$ for any admissible value of $n$, that is,
\begin{align*}
&\begin{array}{r}
\mathring{\bm{\gamma}}
_n^{(1)}=\left(n-\frac{\bm{m}-\bm{l}+1}{2}\right)
\left(n+\frac{\bm{l}+\bm{m}}{2}\right)<0
\; \mbox{for}\;1\leq n\leq \frac{\bm{m}-\bm{l}-1}{2},
\;
 \left[\bm{m}-\bm{l}=3,5,7,\cdots\right],
\end{array}\\
&\begin{array}{r}
\mathring{\bm{\gamma}}
_n^{(2)}=  \mathring{\bm{\gamma}}
_n^{(3)}=   \left(n-\frac{\bm{m}-\bm{l}}{2}\right)
\left(n+\frac{\bm{l}+\bm{m}+1}{2}\right)<0
\;\mbox{for}\; 1\leq n\leq \frac{\bm{m}-\bm{l}-2}{2},
\;
 \left[\bm{m}-\bm{l}= 4,6,8,\cdots\right],
\end{array}\\
&\begin{array}{r}
\mathring{\bm{\gamma}}
_n^{(4)}=\left(n-\frac{\bm{m}-\bm{l}-1}{2}\right)
\left(n+\frac{\bm{l}+\bm{m}+2}{2}\right)<0
\; \mbox{for}\qquad 1\leq n\leq \frac{\bm{m}-\bm{l}-3}{2},
\;
 \left[\bm{m}-\bm{l}= 5,7,9,\cdots\right],
\end{array}
\end{align*}
for ${\mathring{\Psi}}^{(i)}$. 
As the first term of these $\mathring\gamma_n^{(i)}$ is negative, 
 $\mathring\gamma_n^{(i)}<0$ if the the last term is positive for each value of $n$. {From (\ref{nova}) and similar relations for each $\mathring\gamma_n^ {(i)}$, it follows  that $\bm{m}>-{1}/{2}$.  In addition, from 
  $\mathring\gamma_{n=1}^{(i)}<0$ we get restritions on 
  $\bm{m}+\bm{l}$, namely,
\begin{eqnarray}\label{C-1}
&&\begin{array}{l}
\bm{m}>	-\frac{1}{2} \mbox{ for } 
\mathring{\bm{\Psi}}^{(i)},\quad
\bm{m}+\bm{l}+2> 0\; \mbox{ for } \;\mathring{\bm{\Psi}}^{(1)},\;\; \bm{m}+\bm{l}+3> 0\; \mbox{ for }\; \mathring{\bm{\Psi}}^{(2)}
\mbox{ and }\mathring{\bm{\Psi}}^{(3)},\end{array}\nonumber\vspace{3mm}\\
&&\begin{array}{l}
 \bm{m}+\bm{l}+4> 0\;\mbox{ for }\;
\mathring{\bm{\Psi}}^{(4)}. 
\end{array}
\end{eqnarray}
We see, for example, that the interval 
\begin{eqnarray*}\begin{array}{l}
   -\frac{\bm{m}+\bm{l}}{2}< n<\frac{\bm{m}-\bm{l}-1}{2}+1,\quad \mbox{for } \;  \bm{m}>	-\frac{1}{2}\;  
   \; \mbox{ and }
   -\bm{m}-\bm{l}<2
\end{array}
\end{eqnarray*}
assures that  $\mathring{\bm{\gamma}}_n^{(1)}<0 $
for all the possible values for $n$. }

%
{
On the other hand, for ${\tilde{\Psi}}^{(i)}$ we must  require  that
($\mathring{\gamma}
_n^{(2)}=  \mathring{\gamma}
_n^{(3)}$)
\begin{align*}
&\begin{array}{r}
{\mathring{\gamma}}
_n^{(1)}=
\left(n+\frac{\bm{l}+\bm{m}}{2}\right)
\left(n+\frac{\bm{l}-\bm{m}-1}{2}\right)<0
\; \mbox{for}\; 1\leq n\leq -\frac{\bm{m}+\bm{l}+2}{2},
\;
 \left[\bm{m}+\bm{l}=-4,-6,-8,\cdots\right],
\end{array}\\
&\begin{array}{r}
{\mathring{\gamma}}
_n^{(2)}=  
\left(n+\frac{\bm{l}+\bm{m}+1}{2}\right)
 \left(n+\frac{\bm{l}-\bm{m}}{2}\right)<0
\; \mbox{for}\; 1\leq n\leq -\frac{\bm{m}+\bm{l}+3}{2},
\;
 \left[\bm{m}+\bm{l}= -5,-7,-9,\cdots\right],
\end{array}\\
&\begin{array}{r}
{\mathring{\gamma}}
_n^{(4)}=\left(n+\frac{\bm{l}+\bm{m}+2}{2}\right)
\left(n+\frac{\bm{l}-\bm{m	}+1}{2}\right)<0
\; \mbox{for}\; 1\leq n\leq -\frac{\bm{m}+\bm{l}+4}{2},
\;
 \left[\bm{m}+\bm{l}= -6.-8,\cdots\right].
\end{array}
\end{align*}
Since the first term is negative, 
$\tilde\gamma_n^{(i)}<0$ if the last term on the right
of these equations is positive. Thence, 
we find that the Arscott condition is assured by 
the following restrictions on the parameters 
\begin{eqnarray}\label{D-1}
&&\begin{array}{l}
\bm{m}<-\frac{1}{2} \mbox{ for } 
{\tilde{\Psi}}^{(i)},\;\;
\bm{l}-\bm{m}+1>0\; \mbox{ for } \;{\tilde{\Psi}}^{(1)},\;\; 
\bm{l}-\bm{m}+2> 0\; \mbox{ for }\; {\tilde{\Psi}}^{(2)}
\mbox{ and }{\tilde{\Psi}}^{(3)},\nonumber\end{array}\vspace{3mm}\\
&&\begin{array}{l}
\bm{l}-\bm{m}+3> 0\;\mbox{ for }\;
{\tilde{\Psi}}^{(4)};\; 
\end{array}
\end{eqnarray}
These conditions imply that $\mathring\gamma_n^{(i)}<0$ 
for any admissible $n$. For example, by (\ref{nova}) 
and  (\ref{D-1}), we see that 
$\mathring\gamma_n^{(1)}<0$ in the interval 
\begin{eqnarray*}\begin{array}{l}
   \frac{\bm{m}-\bm{l}+1}{2}< n<-\frac{\bm{m}+\bm{l}}{2}\quad\mbox{for}\quad \bm{l}-\bm{m}+1>0 \quad
   \mbox{and}\quad \bm{m}<-\frac{1}{2}
\end{array}
\end{eqnarray*}
which includes all the possible values of $n$
for the solutions ${\tilde{\Psi}}^{(1)}$. 
}

%

{
The eigenfunctions $ \left( \mathring{\bm{\Psi}}^{(i)},\tilde{\bm{\Psi}}^{(i)}\right)$  do not satisfy the Arscott condition  
for the examples discussed in Section 5.1. However, we can check that the sixteen  
eigenfunctions give different types of solutions for Lam\'e equation (\ref{second-case}) when $\bm{m}=0,\pm1,\pm2,\cdots$; each solution satisfies the Arscott condition. Except if $\bm{m}=-2$, $\bm{m}=-1$, $\bm{m}=0$ and $\bm{m}=1$,
for each value of $\bm{m}$ there are four types  of solutions. 
\begin{align}
%
%
&\begin{array}{l}
\mbox{even solutions, period } 2K \;(\bm{l}=0):\left\{ \begin{array}{l}
\mathring{\bm{\Psi}}^{(1)} \quad\mbox{if} \quad\bm{m}=1,3,5,\cdots;\vspace{2mm}\\
%
\tilde{\bm{\Psi}}^{(1)} \quad\mbox{if} \quad\bm{m}=-2,-4,-6,\cdots;\vspace{2mm}\\
\mathring{\bm{\Psi}}^{(5)} \quad\mbox{if} \quad\bm{m}=-1,-3,-5,\cdots
;
\vspace{2mm}\\
\tilde{\bm{\Psi}}^{(5)} \quad\mbox{if} \quad\bm{m}=0,2,4,\cdots;
 \end{array}
\right.\vspace{2mm}\end{array}
\label{even2K}\\
%
%
&\begin{array}{l}
\mbox{odd solutions, period } 2K \;(\bm{l}=0):\left\{ \begin{array}{l}
\mathring{\bm{\Psi}}^{(4)} \quad\mbox{if} \quad\bm{m}=3,5,7,\cdots;
\vspace{2mm}\\
%
\tilde{\bm{\Psi}}^{(4)} \quad\mbox{if} \quad\bm{m}=-4,-6,-8,\cdots;
\vspace{2mm}\\
\mathring{\bm{\Psi}}^{(8)} \quad\mbox{if} \quad\bm{m}=-3,-5,-7,\cdots;\vspace{2mm}\\
%
\tilde{\bm{\Psi}}^{(8)} \quad\mbox{if} \quad\bm{m}=2,4,6,\cdots;
\end{array}\right.
\vspace{2mm}
\end{array}
\label{odd2K}\\
%
%
&\begin{array}{l}
\mbox{odd solutions, period } 4K \;(\bm{l}=0):
\left\{ 
\begin{array}{l}
\mathring{\bm{\Psi}}^{(2)} \;\mbox{ if } \;\bm{m}=2,4,6,\cdots;\vspace{2mm}\\
%
\tilde{\bm{\Psi}}^{(2)} 
\;\mbox{ if } \;\bm{m}=-3,-5,-7,\cdots;
\qquad \vspace{2mm}\\
\mathring{\bm{\Psi}}^{(6)} \;\mbox{ if } \;\bm{m}=-2,-4,-6,\cdots;
\vspace{2mm}\\
\tilde{\bm{\Psi}}^{(6)} \;\mbox{ if } \;\bm{m}=1,3,5,\cdots;
\end{array}
\right.
\end{array}
\label{odd4K}\vspace{2mm}\\
%
%
&\begin{array}{l}
\mbox{even solutions, period } 4K \;(\bm{l}=0):
\left\{
 \begin{array}{l}
\tilde{\bm{\Psi}}^{(3)} \quad\mbox{if} \quad\bm{m}=2,4,6,\cdots;\vspace{2mm}\\
%
\mathring{\bm{\Psi}}^{(3)} \quad\mbox{if} \quad\bm{m}=-3,-5,-7,\cdots;
\vspace{2mm}\\
\mathring{\bm{\Psi}}^{(7)} \quad\mbox{if} \quad\bm{m}=-2,-4,-6,\cdots;
\vspace{2mm}\\
%
\tilde{\bm{\Psi}}^{(7)} \quad\mbox{if} \quad\bm{m}=1,3,5,\cdots.
\end{array}
\right.
%
\end{array}
\label{even4K}
\end{align}
%
%
We find the same results if we take $\bm{l}=0$ or $\bm{l}=-1$. 
By taking $\bm{l}=-1$  we simply rearrange
 the previous expansions. We obtain, for instance,  
\begin{equation*}
%
%
\mathring{\bm{\Psi}}^{(1)}\Big|_{\bm{l}=-1}= \tilde{\bm{\Psi}}^{(5)}\Big|_{\bm{l}=0},\;
\tilde{\bm{\Psi}}^{(1)}\Big|_{\bm{l}=-1}=\mathring{\bm{\Psi}}^{(5)}
\Big|_{\bm{l}=0},\;
\mathring{\bm{\Psi}}^{(5)}\Big|_{\bm{l}=-1}=\tilde{\bm{\Psi}}^{(1)}\Big|_{\bm{l}=0},\;
\tilde{\bm{\Psi}}^{(5)} \Big|_{\bm{l}=-1}=\tilde{\bm{\Psi}}^{(1)} \Big|_{\bm{l}=0}.
 \end{equation*}
 Other examples, for fixed values of  $\bm{m}$ and $\bm{l}$, are presented in Appendix C.}

%
%

%
%
%

\subsection{Finite series of hypergeometric functions 
${\psi}^{(i)}(u)$ when $\bm{l}$ is an integer 
and $\bm{m}\neq -1/2$ is half 
 an odd integer }

 If $\bm{m}$ is half an odd positive integer, finite-series solutions ${\psi}^{(i)}(u)$
with $0\leq n\leq\bm{m}-1/2$ 
are obtained by inserting in (\ref{second-ganguly-3}) 
the hypergeometric-function expansions ${H}^{(i)}(x)$ with  $i= 1,4,6,7$. In these cases,  
the series are finite 
because the factor $(n+\alpha-\beta)=(n-\bm{m}-1/2)$ implies
$\gamma_n^{(i)}=0$ for $n=\bm{m}+1/2$
as demanded in (\ref{truncation}).  
Since $\delta=1/2$, {according to (\ref{adendos-1-conv}) 
the four hypergeometric functions
converge when $z=1$ or $\mathrm{sn}^2u=1$, where $z={(a-1)x}/{(a-x)}=(1-k^2)\;\mathrm{sd}^2u$.}

 If $\bm{m}\neq -1/2$ is half an odd negative integer, finite-series solutions ${\psi}^{(i)}$
with $0\leq n\leq-\bm{m}-3/2$ 
result from ${H}^{(i)}$ for $i= 2,3,5,8$. Now the series terminate  at  $ n=-\bm{m}-3/2$ 
because $(n-\alpha+\beta)=(n+\bm{m}+1/2)$ implies
$\gamma_n^{(i)}=0$ for $n=-\bm{m}-1/2$
as demanded in (\ref{truncation}).  
Since $\delta=1/2$,  by (\ref{adendos-1-conv}) {the four solutions 
converge when $z=1$ ($\mathrm{sn}^2u=1$).}

 {
Eigenfunctions for negative and positive values of $\bm{m}$ are connected to each other by means of the substitutions (\ref{subs}). 
In both cases (positive and negative $\bm{m}$) the solutions 
are degenerate 
in the sense that there are two independent solutions with the same 
energies ${\cal E}$ for a given set of values for the parameters
($\bm{m}$, $\bm{l}$, $k^2$). {In addition, for a given expansion the 
	the Arscott 
	condition (\ref{autovalores}) is satisfied only
	if we impose restrictions on the values of $\bm{l}$ and $\bm{m}$.}{ However, by considering all the expansions, 
	we  find at least one expansion satisfying the condition for any allowed 
	value of 
($\bm{m}$, $\bm{l}$}).

{Here and in Sec. 5.4, in order to check the 
parity and periodicity properties
we consider the one-term expansions because in these cases the
hypergeometric functions can be written in terms of elementary functions rather than in terms of infinite series. For this we use \cite{nist}
\begin{eqnarray}
\label{elementary}
\begin{array}{l}
F\left(\mathrm{a},-\mathrm{a};\frac{1}{2};-z^2\right)=\frac{1}{2}
\left[\left( \sqrt{1+z^2}+z\right)^{2\mathrm{a}}+
\left(\sqrt{1+z^2}-z\right)^{2\mathrm{a}}\right],
\label{hiper-lame-1}
\vspace{3mm}\\
F\left(\mathrm{a},1-\mathrm{a};\frac{1}{2};-z^2\right)=
\frac{1}{2\sqrt{1+z^2}}
\left[\left( \sqrt{1+z^2}+z\right)^{2\mathrm{a}-1}+
\left(\sqrt{1+z^2}-z\right)^{2\mathrm{a}-1}\right],
\label{hiper-lame-2}\vspace{3mm}\\
F\left(\mathrm{a},1-\mathrm{a};\frac{3}{2};-z^2\right)=
\frac{1}{(2-4\mathrm{a})z}
\left[\left( \sqrt{1+z^2}+z\right)^{1-2\mathrm{a}}-
\left(\sqrt{1+z^2}-z\right)^{1-2\mathrm{a}}\right].
\label{hiper-lame-3}
\end{array}
\end{eqnarray}
These relations can also be applied to some two-term series.
Further, we find that such parity and periodicity properties hold for any finite-series expansion.}

%
%
\subsubsection{Finite series when $\bm{l}$ is an integer 
and $\bm{m}$ is half 
 an odd positive integer}

 First we write the eigenfunctions ${\psi}^{(1)}(u)$ and  
 ${\psi}^{(6)}(u)$. {Then, we proof the degeneracy by using the
matrix form for the recurrence relations. After this, we discuss the Arscott condition and, finally, 
we consider the eigenfunctions ${\psi}^{(4)}(u)$ and  
${\psi}^{(7)}(u)$.}
%
 %
 \begin{eqnarray}\label{adendos-a}
 \begin{cases} 
 {\psi}^{(1)}(u)\stackrel{\text{(\ref{adendos-1},\ref{second-ganguly-3})}}{=}
 \left(\mathrm{dn}\,u\right)^{\bm{m}}\displaystyle\sum_{n=0}^{\bm{m}-1/2}
 \begin{array}{l}
 {b}_{n}^{(1)}\left[(1-k^2)\;\mathrm{sd}^2u\right]^n\times
 %
 \end{array}\\
 \begin{array}{l}
\tilde{F}\left[n+\frac{\bm{l}-\bm{m}+1}{2},
 \frac{\bm{m}-\bm{l}-1}{2};n+\frac{1}{2};
 (1-k^2)\;\mathrm{sd}^2u\right],\qquad
\bm{m}=\frac{1}{2},\frac{3}{2},\frac{5}{2},\cdots,
\quad
[\mbox{even,\; {4K}}],
 \end{array}
 \vspace{3mm}\\
 %
 {\alpha}_{n}^{(1)}=
\left(1-\frac{1}{k^2}\right)(n+1),\;
 %
{\beta}_{n}^{(1)}=\left(\frac{1}{k^2}-2\right)n^2+
 \left[\left(\frac{1}{k^2}-1\right)(\bm{l-m}+1)+\bm{m}\right]n 
 +
 \vspace{3mm}\\
 \frac{(\bm{l}+1-\bm{m}
 )^2-{\cal E}}{4k^2}+
\frac{2\bm{lm}-\bm{l}+\bm{m}-(\bm{m}-1)^{2}}{4},\quad{\gamma}_{n}^{(1)}=
 \left(n+\frac{\bm{l-m-1}}{2}\right)
 \left(n+\frac{\bm{l-m}}{2}\right) \left(n-\bm{m}-\frac{1}{2}\right);
 \end{cases}
 \end{eqnarray}
 %
%
%
%
%
%
\begin{eqnarray}\label{adendos-b}
\begin{cases}
{\psi}^{(6)}(u)
\stackrel{\text{(\ref{adendos-6},\ref{second-ganguly-3})}}{=}
\mathrm{sn}u(\mathrm{dn}u)^{m-1}\displaystyle\sum_{n=0}^{\bm{m}-{1}/{2}}
\begin{array}{l}
{b}_{n}^{(6)}\left[(1-k^2)\;\mathrm{sd}^2u\right]^n\times
\end{array}\vspace{2mm}\\
 %
 \tilde{F}\left[n-\frac{\bm{l}+\bm{m}-1}{2},
 \frac{\bm{m}+\bm{l}+1}{2};n+\frac{3}{2};
 (1-k^2)\;\mathrm{sd}^2u\right],\quad
 \bm{m}=\frac{1}{2},\frac{3}{2},\frac{5}{2},\cdots ,\;
[\mbox{odd,\;{ 4K}}],
 \vspace{3mm}\\
%
 %
 {\alpha}_{n}^{(6)}=
\left(1-\frac{1}{k^2}\right)(n+1),
 \quad
 %
  {\beta}_{n}^{(6)}=\left(\frac{1}{k^2}-2\right)n^2+
 \left[\left(1-\frac{1}{k^2}\right)(\bm{l+m})+\bm{m}-1\right]n +
 \vspace{3mm}\\
\frac{(\bm{l}+\bm{m})^2-{\cal E}}{4k^2}+
\frac{\bm{l}-2\bm{lm}+\bm{m}-(\bm{m}-1)^2}{4},\qquad {\gamma}_{n}^{(6)}=
 \left(n-\frac{\bm{l+m+1}}{2}\right)
 \left(n-\frac{\bm{l+m}}{2}\right) \left(n-\bm{m}-\frac{1}{2}\right).
\end{cases}
\end{eqnarray}

For $\bm{m}=1/2$  we obtain two independent eigenfunctions,
 \begin{eqnarray*}
 \begin{array}{l}
 {\psi}^{(1)}(u)\Big|_{\bm{m}=1/2}=
 {b}_{0}^{(1)}\sqrt{\mathrm{dn}\,u}\;
\tilde{F}\left[\frac{1}{2}\left(\bm{l}+\frac{1}{2}\right),
 -\frac{1}{2}\left(\bm{l}+\frac{1}{2}\right);\frac{1}{2};
 (1-k^2)\;\mathrm{sd}^2u\right],\vspace{3mm}\\
 %
%
%
{\psi}^{(6)}(u)\Big|_{\bm{m}=1/2}={b}_{0}^{(6)}
\frac{\mathrm{sn}u}{\sqrt{\mathrm{dn}u}}
\;
\tilde{ F}\left[-\frac{1}{2}\left(\bm{l}-\frac{1}{2}\right),
 \frac{1}{2}\left(\bm{l}+\frac{3}{2}\right);\frac{3}{2};
 (1-k^2)\;\mathrm{sd}^2u\right],
 \end{array}
\end{eqnarray*}
which are degenerate because both correspond to the same energy ${\cal E}$, that is,
\begin{eqnarray*}\begin{array}{l}
 {\beta}_{0}^{(1)}=0,\;\quad  {\beta}_{0}^{(6)}=0 \quad\Rightarrow\quad{\cal E}=\left(\bm{l}+\frac{1}{2}\right)^2+\frac{ k^2}{4}, \quad
\left[\bm{m}=\frac{1}{2}\right].
\end{array}
\end{eqnarray*}
%
To prove the degeneracy for $\bm{m}\geq 3/2$ , 
first we redefine ${b}_{n}^{(6)}$ as
\letra
\begin{eqnarray}\begin{array}{l}
{b}_{n}^{(6)}=\left(\frac{k^2}{1-k^2}\right)^n\Gamma\left(n-\frac{\bm{l+m}-1}{2}\right)
\Gamma\left(n-\frac{\bm{l+m}-2}{2}\right)\hat{b}_{n},
\end{array}\end{eqnarray}
 which implies that the $\hat{b}_{n}$ obey the relations
 $
 \hat{\alpha}_{n}\hat{b}_{n+1}+\hat{\beta}_{n}\hat{b}_{n}+
\hat{\gamma}_{n}\hat{b}_{n-1}=0$  ($\hat{b}_{-1}=0$) where
\begin{eqnarray}\label{degeneracy-1}
 &&\begin{array}{l}
 \hat{{\alpha}}_{n}=-\left(n-\frac{\bm{l+m-1}}{2}\right)
 \left(n-\frac{\bm{l+m}-2}{2}\right)(n+1),
 \qquad
  \hat{\beta}_{n}={\beta}_{n}^{(6)},\end{array}\nonumber
 \vspace{2mm}\\
  &&\begin{array}{l}
 \hat{{\gamma}}_{n}=-\left(1-\frac{1}{k^2}\right)
 \left(n-\bm{m}-\frac{1}{2}\right).\end{array}
\end{eqnarray} 
We show the degeneracy by writing 
the relations for $b_{n}^{(1)}$ and $\hat{b}_{n}$
in the forms $\mathbb{A}\vec{b}=0$ and $\mathbb{B}\vec{\hat{b}}=0$, where
\begin{eqnarray*}
\begin{array}{l}
\vec{b}=\left(b_{0}^{(1)},b_{1}^{(1)},\cdots,b_{\bm{m}-\frac{1}{2}}^{(1)}\right)^{t},\qquad
\vec{\hat{b}}=\left(\hat{b}_{0},\hat{b}_{1},\cdots,\hat{b}_{\bm{m}-\frac{1}{2}}\right)^{t},
\end{array}
\end{eqnarray*}
while $\mathbb{A}$ and $\mathbb{B}$ 
denote the matrices 
\begin{eqnarray*}
\mathbb{A}=
\left[
\begin{array}{ccccccccc}
{\beta}_{0}^{(1)} & {\alpha}_{0}^{(1)}\ &0   &  & \vspace{2mm}                 \vspace{2mm} \\
{\gamma}_{1}^{(1)}&{\beta}_{1}^{(1)}   & {\alpha}_{1}^{(1)} &              \vspace{2mm}  \\
   &  \ddots       &     \ddots        &   \ddots             &        \vspace{2mm}  \\
                     & & {\gamma}_{\bm{m}-\frac{3}{2}}^{(1)}&
                      {\beta}_{\bm{m}-\frac{3}{2}}^{(1)}&{\alpha}_{\bm{m}-\frac{3}{2}}^{(1)}\vspace{2mm}\\
                   &   &    0    &{\gamma}_{\bm{m}-\frac{1}{2}}^{(1)} & {\beta}_{\bm{m}-\frac{1}{2}}^{(1)}
\end{array}
\right],\;
{\mathbb{B}}=
\left[
\begin{array}{ccccccccc}
\hat{\beta}_{0} & \hat{\alpha}_{0}\ &0   &  & \vspace{2mm}                 \vspace{2mm} \\
\hat{\gamma}_{1}&\hat{\beta}_{1}   & \hat{\alpha}_{1} &              \vspace{2mm}  \\
   &  \ddots       &     \ddots        &   \ddots             &        \vspace{2mm}  \\
                     & & \hat{\gamma}_{\bm{m}-\frac{3}{2}}&
                      \hat{\beta}_{\bm{m}-\frac{3}{2}}&\hat{\alpha}_{\bm{m}-\frac{3}{2}}\vspace{2mm}\\
                   &   &    0    &\hat{\gamma}_{\bm{m}-\frac{1}{2}} & \hat{\beta}_{\bm{m}-\frac{1}{2}}
\end{array}
\right].
\end{eqnarray*}
Then (\ref{adendos-a}), (\ref{adendos-b}) and (\ref{degeneracy-1}) imply 
that both matrices have the same elements since
\antiletra
\begin{eqnarray}
{\mathbb{B}}=
\left[
\begin{array}{ccccccccc}
\beta_{\bm{m}-\frac{1}{2}}^{(1)} & \gamma_{\bm{m}-\frac{1}{2}}^{(1)}\ &0   &  & \vspace{2mm}                 \vspace{2mm} \\
\alpha_{\bm{m}-\frac{3}{2}}^{(1)}&\beta_{\bm{m}-\frac{3}{2}}^{(1)}   & \gamma_{\bm{m}-\frac{3}{2}}^{(1)} &              \vspace{2mm}  \\
   &  \ddots       &     \ddots        &   \ddots             &        \vspace{2mm}  \\
                     & & \alpha_{1}^{(1)}& \beta_{1}^{(1)}&
                     \gamma_{1}^{(1)}\vspace{2mm}\\
                   &   &    0    &\alpha_{0}^{(1)} & \beta_{0}^{(1)}
\end{array}
\right].
\end{eqnarray}
The degeneracy results from the fact that $\det{\mathbb{A}}=$
$\det{\mathbb{B}}$.  In effect, let $\mathbb{U}$ be a 
square matrix of order $\left(\bm{m}+\frac{1}{2}\right)$, 
 having $1$'s on
the antidiagonal as the only nonzero elements, that is,
\begin{eqnarray}\label{antidiagonal}
\mathbb{U}=\mathbb{U}^{-1}=\left[
\begin{array}{ccc}
  &   &   1    \\
            & . &  \\
1 & &
\end{array}
\right],\qquad \det{\mathbb{U}}=\pm 1,
\end{eqnarray}
where $\mathbb{U}^{-1}$ is the matrix inverse to $\mathbb{U}$.
Then, we find that 
\begin{eqnarray}
{\mathbb{B}}=\mathbb{U}\mathbb{A}\mathbb{U}\qquad \Rightarrow\qquad \det{\mathbb{A}}=\det{\mathbb{B}},
\end{eqnarray}
 where the last equality follows from the properties of the
determinants. The above approach has already been used
to get the degeneracy of wavefunctions for a problem which 
is ruled by a reduced confluent Heun equation \cite{volcano}.
%


Finally, in (\ref{adendos-a}) and (\ref{adendos-b}) 
$	{{\alpha}}_{n}^{(1)}=	{{\alpha}}_{n}^{(6)}<0$. Thence, 
the Arscott condition (\ref{autovalores}) for $\bm{m}\geq3/2$ requires that ${{\gamma}}_{n}^{(1)}<0$ and ${{\gamma}}_{n}^{(6)}<0$.
%
From the expressions for ${{\gamma}}_{n}^{(1)}$ and 
${{\gamma}}_{n}^{(6)}$ these conditions are equivalent to
\begin{eqnarray}
\begin{array}{l}
f(n)=\left(n+\frac{\bm{l-m}-1}{2}\right)
\left(n+\frac{\bm{l-
		m}}{2}\right)>0\quad \mbox{for}\quad{\psi}^{(1)},\end{array}
\nonumber\\
\begin{array}{l}
g(n)=\left(n-\frac{\bm{l+m}+1}{2}\right)
\left(n-\frac{\bm{l+m}}{2}\right)>0
\quad	\mbox{for}\quad{\psi}^{(6)}
\end{array}
\end{eqnarray}
for any admissible value of $n$, that is, $1\leq n\leq{\bm{m}}-\frac{1}{2}$. However,
these conditions are satisfied only
for special values of $\bm{l}$ and $\bm{m}$. 
%
%
As an illustration, for the Lam\'e equation ($\bm{l}=0$, $\bm{l}=-1$), we find that
\begin{eqnarray}\label{LAME-1}
\begin{array}{l}
\mbox{for}\; \bm{l}=0,\quad f(n)=g(n)>0
\quad \mbox{ only  if }\quad
\bm{m}=\frac{5}{2},\frac{9}{2},\frac{13}{2},
\cdots; \vspace{3mm}\\
\mbox{for}\; \bm{l}=-1\; (f\neq g), \quad
f(n)>0 \mbox{ and } g(n)>0\quad \mbox{ only if }\quad
\bm{m}=\frac{3}{2},\frac{7}{2},\frac{11}{2},
\cdots. 
\end{array}\end{eqnarray}
So, we get two degenerate finite-series  solutions for the Lam\'e equation when $\bm{m}$ is half an odd positive integer 
-- for $\bm{m}=1/2$ see solutions after Eq. (\ref{adendos-b}). These are not the finite series proposed by Ince \cite{ince3}. 

On the other side, 
\begin{eqnarray}\label{intervals}
\begin{array}{l}
f(n)\leq 0 \;\mbox{ if }\; \frac{\bm{m-l}}{2} \leq n\leq\frac{\bm{m-l}}{2}+
\frac{1}{2}, \quad\mbox{and }\;
g(n)\leq 0 \; \mbox{ if }\; \frac{\bm{m+l}}{2} \leq n\leq\frac{\bm{m+l}}{2}+
\frac{1}{2}.
\end{array}\end{eqnarray}
Then,
Arscott condition holds if the above 
intervals are empty (no admissible $n$). 
In particular we can satisfy
the condition by demanding that, for a given
value of $\bm{m}$, the upper bounds of the intervals (\ref{intervals}) are less than 1, that is, $\bm{m-l}<1$ and  $\bm{m+l}<1$. Since $\bm{m}=3/2,5/2.\cdots$, the Arscott condition
stands if
\begin{eqnarray}\begin{array}{l}
\bm{l>m-1}	\mbox{ for }{\psi}^{(1)}
\; (\bm{l} \mbox{ is positive  integer});
\; \bm{l<1-m}	\mbox{ for }{\psi}^{(6)}
\; (\bm{l} \mbox{ negative integer}).
\end{array}\end{eqnarray}
In these cases, the degeneracy proof fails because $\bm{l}$ takes different values in each expansion.

{On the other hand, the degenerate eigenfunctions ${\psi}^{(4)}(u)$ and  
${\psi}^{(7)}(u)$ are given by}
%
 %
 \begin{eqnarray}\label{adendos-aa}
 \begin{cases} 
 {\psi}^{(4)}(u)\stackrel{\text{(\ref{adendos-4},\ref{second-ganguly-3})}}{=}\operatorname{sn}{u}
 \left(\mathrm{dn}\,u\right)^{\bm{m-1}}\displaystyle\sum_{n=0}^{\bm{m}-1/2}
 \begin{array}{l}
 {b}_{n}^{(4)}\left[(1-k^2)\;\mathrm{sd}^2u\right]^n\times
 \end{array}\\
 \begin{array}{l}
 \tilde{F}\left[n+\frac{\bm{l}-\bm{m}+2}{2},
 \frac{\bm{m}-\bm{l}}{2};n+\frac{3}{2};
 (1-k^2)\;\mathrm{sd}^2u\right],\quad
\bm{m}=\frac{1}{2},\frac{3}{2},\frac{5}{2},\cdots,\quad
[\mbox{odd,\; {4K}}],
 \end{array}
 \vspace{3mm}\\
 %
 {\alpha}_{n}^{(4)}=
\left(1-\frac{1}{k^2}\right)(n+1),\;
 %
{\beta}_{n}^{(4)}=\left(\frac{1}{k^2}-2\right)n^2+
 \left[\left(\frac{1}{k^2}-1\right)(\bm{l-m}+1)+
 \bm{m}-1\right]n 
 +
 \vspace{3mm}\\
 \frac{(\bm{l}+1-\bm{m}
 )^2-{\cal E}}{4k^2}+
\frac{2\bm{l}\bm{m}-\bm{m}^2+5\bm{m}-2-\bm{l}}{4},\quad{\gamma}_{n}^{(4)}=
 \left(n+\frac{\bm{l-m+1}}{2}\right)
 \left(n+\frac{\bm{l-m}}{2}\right) \left(n-\bm{m}-\frac{1}{2}\right);
 \end{cases}
 \end{eqnarray}
 %
%
%
%
%
%
\begin{eqnarray}\label{adendos-bb}
\begin{cases}
{\psi}^{(7)}(u)
\stackrel{\text{(\ref{adendos-7},\ref{second-ganguly-3})}}{=}
(\operatorname{dn}u)^{m}\displaystyle\sum_{n=0}^{\bm{m}-{1}/{2}}
\begin{array}{l}
{b}_{n}^{(7)}\left[(1-k^2)\;\mathrm{sd}^2u\right]^n\times
\end{array}\vspace{2mm}\\
 %
 \tilde{F}\left[n-\frac{\bm{l}+\bm{m}}{2},
 \frac{\bm{m}+\bm{l}}{2};n+\frac{1}{2};
 (1-k^2)\;\mathrm{sd}^2u\right],\quad
 \bm{m}=\frac{1}{2},\frac{3}{2},\frac{5}{2},\cdots ,
 \;[\mbox{even,\;{ 4K}}],
 \vspace{3mm}\\
%
 %
 {\alpha}_{n}^{(7)}=
\left(1-\frac{1}{k^2}\right)(n+1),
 \quad
 %
  {\beta}_{n}^{(7)}=\left(\frac{1}{k^2}-2\right)n^2+
 \left[\left(1-\frac{1}{k^2}\right)(\bm{l+m})+
 \bm{m}\right]n +
 \vspace{3mm}\\
\frac{(\bm{l}+\bm{m})^2-{\cal E}}{4k^2}+
\frac{\bm{l}+\bm{m}-2\bm{lm}-\bm{m}^2}{4},\qquad
 {\gamma}_{n}^{(7)}=
 \left(n-\frac{\bm{l+m+2}}{2}\right)
 \left(n-\frac{\bm{l+m+1}}{2}\right) \left(n-\bm{m}-\frac{1}{2}\right).
\end{cases}
\end{eqnarray}
%

%
For one-term series  ($\bm{m}=1/2$}),  ${\psi}^{(4)}$ and ${\psi}^{(7)}$ are not independent respect to  ${\psi}^{(1)}$ and ${\psi}^{(6)}$, since
 \begin{eqnarray*}
 \begin{array}{l}
 {\psi}^{(4)}(u)={\psi}^{(6)}(u)=
 {{b}_{0}^{(4)}}
 \sqrt{\operatorname{dn}{u}}\left[\left( \frac{\operatorname{cn}{u}+\sqrt{k^2-1}
 \operatorname{sn}{u}}{\operatorname{dn}{u}}\right)^{\bm{l}+\frac{1}{2}}-
 \left( \frac{\operatorname{cn}{u}-\sqrt{k^2-1}
 \operatorname{sn}{u}}{\operatorname{dn}{u}}\right)^{\bm{l}+\frac{1}{2}}
\right],\qquad \bm{m}=\frac{1}{2},
\vspace{3mm}\\
 %
 %
 %
 %
{\psi}^{(7)}(u)={\psi}^{(1)}(u)=
 {{b}_{0}^{(7)}\sqrt{\operatorname{dn}{u}}}
 \left[\left( \frac{\operatorname{cn}{u}+\sqrt{k^2-1}
 \operatorname{sn}{u}}{\operatorname{dn}{u}}\right)^{\bm{l}+\frac{1}{2}}+
 \left( \frac{\operatorname{cn}{u}-\sqrt{k^2-1}
 \operatorname{sn}{u}}{\operatorname{dn}{u}}\right)^{\bm{l}+\frac{1}{2}
}\right],\qquad \bm{m}=\frac{1}{2},
\end{array}
\end{eqnarray*}
which are odd and even solutions, respectively, both having period $4K$.
The previous expressions follow from relations 
(\ref{elementary}).
To prove the degeneracy for $\bm{m}\geq 3/2$ , 
first we redefine ${b}_{n}^{(7)}$ as
\begin{eqnarray*}\begin{array}{l}
{b}_{n}^{(7)}=\left(\frac{k^2}{1-k^2}\right)^n\Gamma\left(n-\frac{\bm{l+m}-1}{2}\right)
\Gamma\left(n-\frac{\bm{l+m}}{2}\right)\tilde{b}_{n}^{(7)}
\end{array}\end{eqnarray*}
and obtain the relations
\begin{eqnarray*}\begin{array}{l}
-\left(n-\frac{\bm{l+m}-1}{2}\right)
\left(n-\frac{\bm{l+m}}{2}\right)(n+1)
\;
\tilde{b}_{n+1}^{(7)}+  {\beta}_{n}^{(7)}\;\tilde{b}_{n}^{(7)}+
\frac{1-k^2}{k^2}\left(n-\bm{m}-\frac{1}{2}\right)\tilde{b}_{n-1}^{(7)}=0.
\end{array}
\end{eqnarray*}
 Thence, if  $\mathbb{C}$ and $\mathbb{D}$ 
 represent the matrices corresponding to the recurrence relations for 
  ${b}_{n}^{(4)}$ and $\tilde{b}_{n}^{(7)}$,  we find that
  ${\mathbb{D}}=\mathbb{U}\mathbb{C}\mathbb{U}$, where
  $\mathbb{U}$ is given in (\ref{antidiagonal}). Thus,
  ${\psi}^{(4)}$ and ${\psi}^{(7)}$ are degeneate
  because they satisfy the same characteristic equation. 
  
  %
%


For  $\bm{m}=\frac{3}{2}$  the two-terms series
$\psi^{(1)}$, $\psi^{(4)}$,  $\psi^{(6)}$ and $\psi^{(7)}$  are all of them degenerate 
and each one corresponds to two real values,  ${\cal E}^ {(\pm)}$, for the energy 
${\cal E}$ (regardless of the Arscott condition). In fact, from the characteristic equations ${\alpha}_{0}^{(i)}{\gamma}_{1}^{(i)}-
 {\beta}_{0}^{(i)}{\beta}_{1}^{(i)}=0$, we find
\begin{eqnarray*}\begin{array}{l}
 %
 {\cal E}^{(\pm)}=\bm{l}^2+\bm{l}+\frac{5}{2}+
 \frac{5}{2} k^2\pm\sqrt{4(1-k^2)(\bm{l}^2+\bm{l})+1}, \qquad
\bm{m}=\frac{3}{2}.
\end{array}
\end{eqnarray*}
The even solutions, $\psi^{(1)}$ and $\psi^{(7)}$,  can be written in term of elementary functions as
 \begin{eqnarray*}
 \begin{array}{l}
\psi^{(1)}(u)=
 {b}_{0}^{(1)}(\operatorname{dn}{u})^{\frac{3}{2}}\bigg\{
 \frac{1}{\Gamma\left(1/2\right)}
 \left[\left( \frac{\operatorname{cn}{u}+\sqrt{k^2-1}
 \operatorname{sn}{u}}{\operatorname{dn}{u}}\right)^{\bm{l}-\frac{1}{2}}+
 \left( \frac{\operatorname{cn}{u}-\sqrt{k^2-1}
 \operatorname{sn}{u}}{\operatorname{dn}{u}}\right)^{\bm{l}-\frac{1}{2}
}\right]+\\
 \frac{\sqrt{k^2-1}\quad\beta_{0}^{(1)}}{\Gamma\left(3/2\right)(1+2\bm{l})\;\alpha_{0}^{(1)}}
 \frac{\operatorname{sn}{u}}{\operatorname{dn}{u}}
 \left[\left( \frac{\operatorname{cn}{u}+\sqrt{k^2-1}
 \operatorname{sn}{u}}{\operatorname{dn}{u}}\right)^{\bm{l}+\frac{1}{2}}-
 \left( \frac{\operatorname{cn}{u}-\sqrt{k^2-1}
 \operatorname{sn}{u}}{\operatorname{dn}{u}}\right)^{\bm{l}+\frac{1}{2}
}\right]\bigg\},\quad \bm{m}=\frac{3}{2},
\end{array}
\end{eqnarray*}
 \begin{eqnarray*}
 \begin{array}{l}
\psi^{(7)}(u)=
 {b}_{0}^{(7)}(\operatorname{dn}{u})^{\frac{3}{2}}\bigg\{
 \frac{1}{\Gamma\left(1/2\right)}
 \left[\left( \frac{\operatorname{cn}{u}+\sqrt{k^2-1}
 \operatorname{sn}{u}}{\operatorname{dn}{u}}\right)^{\bm{l}+\frac{3}{2}}+
 \left( \frac{\operatorname{cn}{u}-\sqrt{k^2-1}
 \operatorname{sn}{u}}{\operatorname{dn}{u}}\right)^{\bm{l}+\frac{3}{2}
}\right]+\\
 \frac{\sqrt{k^2-1}\quad\beta_{0}^{(7)}}{\Gamma\left(3/2\right)(1+2\bm{l})\;\alpha_{0}^{(7)}}
 \frac{\operatorname{sn}{u}}{\operatorname{dn}{u}}
 \left[\left( \frac{\operatorname{cn}{u}+\sqrt{k^2-1}
 \operatorname{sn}{u}}{\operatorname{dn}{u}}\right)^{\bm{l}+\frac{1}{2}}-
 \left( \frac{\operatorname{cn}{u}-\sqrt{k^2-1}
 \operatorname{sn}{u}}{\operatorname{dn}{u}}\right)^{\bm{l}+\frac{1}{2}
}\right]\bigg\},\qquad \bm{m}=\frac{3}{2},
\end{array}
\end{eqnarray*}
 where there are two expressions for  $\beta_{0}^{(1)}$ and $\beta_{0}^{(7)}$
 corresponding to ${\cal E}={\cal E}^ {(\pm)}$.
  
                   										 
 As $	{{\alpha}}_{n}^{(4)}=	{{\alpha}}_{n}^{(7)}<0$ in (\ref{adendos-aa}) and (\ref{adendos-bb}), 
the Arscott condition (\ref{autovalores}) for $\bm{m}\geq3/2$ requires that ${{\gamma}}_{n}^{(4)}<0$ and ${{\gamma}}_{n}^{(7)}<0$ or,  equivalently, that
\begin{eqnarray}
&&\begin{array}{l}
F(n)=\left(n+\frac{\bm{l-m}+1}{2}\right)
\left(n+\frac{\bm{l-
		m}}{2}\right)>0\quad \mbox{for}\quad{\psi}^{(4)},\end{array}
\nonumber\vspace{2mm}\\
&&\begin{array}{l}
G(n)=\left(n-\frac{\bm{l+m}+2}{2}\right)
\left(n-\frac{\bm{l+m+1}}{2}\right)>0
\quad	\mbox{for}\quad{\psi}^{(7)}.
\end{array}
\end{eqnarray}
These conditions are satisfied only
for special values of $\bm{l}$ and $\bm{m}$; for the Lam\'e equation ($\bm{l}=0$, $\bm{l}=-1$), we find that
\begin{eqnarray}\label{LAME-2}
\begin{array}{l}
\mbox{for}\; \bm{l}=0,\quad F(n)>0 \quad\mbox{and}\quad G(n)>0
\quad \mbox{ only  if }\quad
\bm{m}=\frac{3}{2},\frac{7}{2},\frac{13}{2},
\cdots; \vspace{3mm}\\
\mbox{for}\; \bm{l}=-1, \quad
F(n)= G(n)>0\quad \mbox{ only if }\quad
\bm{m}=\frac{5}{2},\frac{9}{2},\frac{13}{2},
\cdots. 
\end{array}\end{eqnarray}
By comparing (\ref{LAME-1}) and  (\ref{LAME-2}),  we see that 
for a given choice of ($\bm{l},\bm{m}$):  (i) if the solutions ${\psi}^{(1)}$ and 
${\psi}^{(6)}$  satisfy the Arscott condition, the solutions ${\psi}^{(4)}$ and 
${\psi}^{(7)}$  do not satisfy;  (i) if the solutions ${\psi}^{(1)}$ and 
${\psi}^{(6)}$  do not satisfy the Arscott condition, the solutions ${\psi}^{(4)}$ and 
${\psi}^{(7)}$ satisfy the condition. Therefore, for the Lam\'e equation
there are solutions satisfying the Arscoott condition for all admissible values of 
($\bm{l},\bm{m}$).

The above conclusion is not restricted to Lam\'e equation.  Since
\begin{eqnarray}\label{intervals-2}
\begin{array}{l}
F(n)\leq 0 \;\mbox{ if }\; \frac{\bm{m-l-1}}{2} \leq n\leq\frac{\bm{m-\bm{l}}}{2}, \quad\mbox{and }\;
G(n)\leq 0 \; \mbox{ if }\; \frac{\bm{m+l+1}}{2} \leq n\leq\frac{\bm{m+l+2}}{2}, 
\end{array}\end{eqnarray}
Arscott condition for ${\psi}^{(4)}$ and 
${\psi}^{(7)}$ holds if the above 
intervals are empty similarly to the previous cases. As $n$ is an integer, in the  
intervals (\ref{intervals}) and  (\ref{intervals-2}) we can take the sign $<$
instead of $\leq$. Thus, for  ${\psi}^{(4)}$ and 
${\psi}^{(1)}$ we have, respectively,
\begin{eqnarray*}
\begin{array}{l}{\psi}^{(4)},
{\psi}^{(1)}:
F(n)\leq 0 \;\mbox{ if }\; \frac{\bm{m-l-1}}{2} < n<\frac{\bm{m-\bm{l}}}{2}, \quad\mbox{and }\;
f(n)\leq 0 \; \mbox{ if }\; \frac{\bm{m-l}}{2} < n<\frac{\bm{m-l+1}}{2}.
\end{array}\end{eqnarray*}
Hence, at least one of the intervals is empty for $n=2,3,4,\cdots$. Analogously,  for  ${\psi}^{(6)}$ and 
${\psi}^{(7)}$ we have
\begin{eqnarray*}
\begin{array}{l}{\psi}^{(6)},
{\psi}^{(7)}:
g(n)\leq 0 \;\mbox{ if }\; \frac{\bm{m+l}}{2} < n<\frac{\bm{m+\bm{l}}+1}{2}, \quad\mbox{and }\;
G(n)\leq 0 \; \mbox{ if }\; \frac{\bm{m+l+1}}{2} < n<\frac{\bm{m+l+2}}{2}.
\end{array}\end{eqnarray*}
%

%
\subsubsection{Finite series when $\bm{l}$ is an integer 
and $\bm{m}\neq-1/2$ is half 
 an odd negative integer}
 %
 

 If $\bm{m}\neq -1/2$ is half an odd negative integer, finite-series solutions ${\psi}^{(i)}$
with $0\leq n\leq-\bm{m}-3/2$ 
result from ${H}^{(i)}$ for $i= 2,3,5,8$. Now the series terminate  at  $ n=-\bm{m}-3/2$ 
because $(n-\alpha+\beta)=(n+\bm{m}+1/2)$ implies
$\gamma_n^{(i)}=0$ for $n=-\bm{m}-1/2$
as demanded in (\ref{truncation}).  
Since $\delta=1/2$,  by (\ref{adendos-1-conv}){ the four solutions 
converge when $z=1$ ($\mathrm{sn}^2u=1$).}


The eigenfunctions ${\psi}^{(2)}(u)$ and  ${\psi}^{(5)}(u)$ are
%
%
%
 %
 \begin{eqnarray}
 \begin{cases} 
 {\psi}^{(2)}(u)\stackrel{\text{(\ref{adendos-2},\ref{second-ganguly-3})}}{=}\displaystyle
 \frac{\mathrm{sn}\,u}{\mathrm{dn}^{\bm{m}+2}u}\sum_{n=0}^{-\bm{m}-3/2}
 \begin{array}{l}
 {b}_{n}^{(2)}\left[(1-k^2)\;\mathrm{sd}^2u\right]^n
\times\end{array}\\
 %
%
\tilde{F}\left[n+\frac{\bm{l}+\bm{m}+3}{2},
 -\frac{\bm{m}+\bm{l}+1}{2};n+\frac{3}{2};
 (1-k^2)\;\mathrm{sd}^2u\right],\quad
\bm{m}=-\frac{3}{2},-\frac{5}{2},\cdots, \;
[\mbox{odd,\; 4K}],
 \vspace{3mm}\\
 %
 {\alpha}_{n}^{(2)}=
\left(1-\frac{1}{k^2}\right)(n+1),\;\;
 %
{\beta}_{n}^{(2)}=\left(\frac{1}{k^2}-2\right)n^2+
 \left[\bm{l}-\frac{\bm{l+m}+2}{k^2}\right]n +
   \frac{(\bm{l}+\bm{m}+2
 )^2-{\cal E}}{4k^2}
 \vspace{3mm}\\
 %
-
\frac{2\bm{lm}+3\bm{l}+\bm{m}^{2}+7\bm{m}+8}{4},\;
 \;\;{\gamma}_{n}^{(2)}=
 \left(n+\frac{\bm{l+m+1}}{2}\right)
 \left(n+\frac{\bm{l+m}+2}{2}\right) \left(n+\bm{m}+\frac{1}{2}\right);
 \end{cases}
 \end{eqnarray}
 %
%
%
\begin{eqnarray}
\begin{cases}
{\psi}^{(5)}(u)
\stackrel{\text{(\ref{adendos-5},\ref{second-ganguly-3})}}{=}
\displaystyle
\frac{1}{\mathrm{dn}^{m+1}u}\displaystyle\sum_{n=0}^{-\bm{m}-{3}/{2}}
\begin{array}{l}
{b}_{n}^{(5)}\left[(1-k^2)\;\mathrm{sd}^2u\right]^n\times
\end{array}\vspace{2mm}\\
 %
 \tilde{F}\left[n+\frac{\bm{m}-\bm{l}+1}{2},\frac{\bm{l}-\bm{m}-1}{2};n+\frac{1}{2};
 (1-k^2)\;\mathrm{sd}^2u\right],\quad
 \bm{m}=-\frac{3}{2},-\frac{5}{2},\cdots, \;
[\mbox{even,\;4K}],
 \vspace{3mm}\\
%
 %
 {\alpha}_{n}^{(5)}=
\left(1-\frac{1}{k^2}\right)(n+1),
 \quad
 %
  {\beta}_{n}^{(5)}=\left(\frac{1}{k^2}-2\right)n^2-
 \left[\bm{l}+\frac{\bm{m}-\bm{l}+1}{k^2}\right]n +
\frac{(\bm{l}-\bm{m}-1)^2-{\cal E}}{4k^2}
 \vspace{3mm}\\
-
\frac{(\bm{l}-\bm{m}-1)^2+\bm{l}+\bm{m}+2-(\bm{l}+1)^2}{4},\qquad {\gamma}_{n}^{(5)}=
 \left(n+\frac{\bm{m-l}}{2}\right)
 \left(n+\frac{\bm{m-l}-1}{2}\right) \left(n+\bm{m}+\frac{1}{2}\right).
\end{cases}
\end{eqnarray}
We can verify that ${\psi}^{(2)}$ and  ${\psi}^{(5)}$ result, respectively, from ${\psi}^{(6)}$ and  ${\psi}^{(1)}$ by means of the substitutions (\ref{subs}). By this reason we do not discuss 
the degeneracy and the Arscott condition.
%

On the other side, the eigenfunctions ${\psi}^{(3)}(u)$ and  ${\psi}^{(8)}(u)$ are
%
%
%
%
\begin{eqnarray}
\begin{cases}
{\psi}^{(3)}(u)
\stackrel{\text{(\ref{adendos-3},\ref{second-ganguly-3})}}{=}
(\operatorname{dn}u)^{-m-1}\displaystyle\sum_{n=0}^{-\bm{m}-{3}/{2}}
\begin{array}{l}
{b}_{n}^{(3)}\left[(1-k^2)\;\mathrm{sd}^2u\right]^n\times
\end{array}\vspace{2mm}\\
 %
 \tilde{F}\left[n+\frac{\bm{l}+\bm{m}+2}{2},
 -\frac{\bm{m}+\bm{l}+2}{2};n+\frac{1}{2};
 (1-k^2)\;\mathrm{sd}^2u\right],\quad
 \bm{m}=\frac{1}{2},\frac{3}{2},\frac{5}{2},\quad
 \cdots \;[\mbox{even,\;{ 4K}}],
 \vspace{3mm}\\
%
 %
 {\alpha}_{n}^{(3)}=
\left(1-\frac{1}{k^2}\right)(n+1),
 \qquad
 %
  {\beta}_{n}^{(3)}=\left(\frac{1}{k^2}-2\right)n^2+
 \left[\left(\frac{1}{k^2}-1\right)(\bm{l+m}+2)
 -\bm{m}-1\right]n 
 \vspace{3mm}\\
+\frac{(\bm{l}+\bm{m}+2)^2-{\cal E}}{4k^2}-
\frac{1}{4}\left[\bm{l}+\bm{m}+2+2\bm{(l+1)(m+1)}+(\bm{m+1})^2\right],\vspace{3mm}\\
 {\gamma}_{n}^{(3)}=
 \left(n+\frac{\bm{l+m}}{2}\right)
 \left(n+\frac{\bm{l+m+1}}{2}\right) \left(n+\bm{m}+\frac{1}{2}\right);
\end{cases}
\end{eqnarray}
%
 %
 \begin{eqnarray}
 \begin{cases} 
 {\psi}^{(8)}(u)\stackrel{\text{(\ref{adendos-8},\ref{second-ganguly-3})}}{=}\operatorname{sn}{u}
 \left(\mathrm{dn}\,u\right)^{\bm{-m-2}}\displaystyle\sum_{n=0}^{-\bm{m}-3/2}
 \begin{array}{l}
 {b}_{n}^{(8)}\left[(1-k^2)\;\mathrm{sd}^2u\right]^n\times
 \end{array}\\
 \begin{array}{l}
\tilde{F}\left[n-\frac{\bm{l}-\bm{m}-2}{2},
 \frac{\bm{l}-\bm{m}}{2};n+\frac{3}{2};
 (1-k^2)\;\mathrm{sd}^2u\right],\quad
\bm{m}=-\frac{3}{2},-\frac{5}{2},-\frac{7}{2},\cdots,\quad
[\mbox{odd,\; {4K}}],
 \end{array}
 \vspace{3mm}\\
 %
 {\alpha}_{n}^{(8)}=
\left(1-\frac{1}{k^2}\right)(n+1),\quad
 %
{\beta}_{n}^{(8)}=\left(\frac{1}{k^2}-2\right)n^2+
 \left[\left(1-\frac{1}{k^2}\right)(\bm{l-m}-1)
 -\bm{m}-2\right]n 
 +
 \vspace{3mm}\\
 \frac{(\bm{l}-1-\bm{m}
 )^2-{\cal E}}{4k^2}+
\frac{2(\bm{l}+1)(\bm{m}+1)-(\bm{m}+1)^2-5\bm{m}
	-6+\bm{l}}{4},\vspace{3mm}\\
{\gamma}_{n}^{(8)}=
 \left(n-\frac{\bm{l-m-1}}{2}\right)
 \left(n-\frac{\bm{l-m}}{2}\right) \left(n+\bm{m}+\frac{1}{2}\right).
 \end{cases}
 \end{eqnarray}
 %
%
%
%
%
%
%
 ${\psi}^{(3)}$ and  ${\psi}^{(8)}$ result, respectively, from ${\psi}^{(7)}$ and  ${\psi}^{(4)}$ by means of the substitutions (\ref{subs}).


%
%

\subsection{Finite series $\bm{\Psi}^{(i)}(u)$ of hypergeometric functions when $\bm{m}$  is an integer and $\bm{l}\neq-1/2$ is half an odd integer}

{
	These solutions are obtained from the expansions 
	$\bm{H}^{(i)}(x)$ given in Section 4.3; {alternatively, they can be generated from 
		the above solutions (Sec. 5.3) through  the substitutions (\ref{associado-gan-2}). }
{If $\bm{l}\neq-1/2$ is half an odd integer, then the factor
 $(n-\alpha-\beta+\gamma+\delta)=(n-\bm{l}-1/2)$ 
 implies that four of the coefficients $\bm{\gamma}_n^{(i)}$ vanish for $n=\bm{l}+1/2$. These give finite series with
 $0\leq n\leq \bm{l}-\frac{1}{2}$ if $\bm{l}$ is positive; {according to (\ref{adendos-2-conv})
 the expansions converge for all values of $u$.
 Similarly,  the factor
 $(n+\alpha+\beta-\gamma-\delta)=(n+\bm{l}+1/2)$ 
 implies that four expansions terminate at $n=-\bm{l}-{3}/{2}$ if  $\bm{l}\leq-{3}/{2}$,
 and by 
 (\ref{adendos-2-conv}) the  eigenfunctions are bounded for all permissible values of $u$.}}

{Due to the aforementioned connection with solutions of Section 5.3,
the  eigenfunctions are  degenerate and a given expansion 
satisfies the Arscott 
condition (\ref{autovalores}) only for special  values of $\bm{m}$
and $\bm{l}$.  However, once more, by considering all the expansions, 
	we  find at least one expansion satisfying the condition for any allowed 
	value of 
($\bm{m}$, $\bm{l}$)}.
 

%
%
\subsubsection{Finite series when $\bm{m}$ is an integer 
and $\bm{l}$ is half 
 an odd positive integer}

Now we write the degenerate eigenfunctions $\bm{\Psi}^{(5)}(u)$ and  
$\bm{\Psi}^{(7)}(u)$. By using the Eqs. (\ref{sd-cd-5})) 
we find they can be obtained from ${\Psi}^{(1)}(u)$ and 
${\Psi}^{(6)}(u)$. 
%
%
\begin{eqnarray}\label{adendos-5-associada}
\begin{cases}
\bm{\Psi}^{(5)}(u)
\stackrel{\text{(\ref{h-bold-5},\ref{second-ganguly-3})}}{=}
\displaystyle\frac{1}{
	\mathrm{dn}^{\bm{l}}u}\;\sum_{n=0}^{\bm{l}-1/2}\begin{array}{l} 
\bm{b}_{n}^{(5)}\;\mathrm{cn}^{2n}{u}\;
\tilde{F}\left(n-\frac{\bm{l}-\bm{m}-1}{2},
\frac{\bm{l}-\bm{m}-1}{2};n+\frac{1}{2};\mathrm{cn}^2{u}\right)
\end{array}\\
\hspace{1.6cm}\begin{array}{l}\stackrel{\text{(\ref{sd-cd-5})}}{=}{\Psi}^{(1)}(u+K)\Big|_{\bm{l}\leftrightarrow\bm{m}},
\hspace{1.0cm}\mathrm{l}=\frac{1}{2},\frac{3}{2},\frac{5}{2},\cdots,
\quad\left[\mbox{even, period } 4K\right]
\end{array}\vspace{3mm}\\
\begin{array}{l}
\bm{\alpha}_{n}^{(5)}={\alpha}_{n}^{(1)}\Big|_{\bm{l}\leftrightarrow\bm{m}}={\alpha}_{n}^{(1)},\qquad 
\bm{\beta}_{n}^{(5)}={\beta}_{n}^{(1)}\Big|_{\bm{l}\leftrightarrow\bm{m}}
\qquad\bm{\gamma}_{n}^{(5)}={\gamma}_{n}^{(1)}\Big|_{\bm{l}\leftrightarrow\bm{m}};\end{array}
\end{cases}
\end{eqnarray}
%
%
%
%
%
\begin{eqnarray}\label{adendos-7-associada}
\begin{cases}
\bm{\Psi}^{(7)}(u)\stackrel{\text{(\ref{h-bold-7},\ref{second-ganguly-3})}}{=}\displaystyle
\frac{\mathrm{cn}\,u}{\mathrm{dn}^{\bm{l}}u}\sum_{n=0}^{\bm{l}-1/2}\begin{array}{l}
\bm{b}_{n}^{(7)}\;\mathrm{cn}^{2n}{u}\;
\tilde{F}\left(n-\frac{\bm{l}+\bm{m}-1}{2},
\frac{\bm{l}+\bm{m}+1}{2};n+\frac{3}{2};\mathrm{cn}^2{u}\right)
\end{array}
\\
\hspace{1.5cm}\begin{array}{l}\stackrel{\text{(\ref{sd-cd-5})}}{=}{\Psi}^{(6)}(u+K)\Big|_{\bm{l}\leftrightarrow\bm{m}},
\hspace{1.0cm}\mathrm{l}=\frac{1}{2},\frac{3}{2},\frac{5}{2},\cdots,
\quad\left[\mbox{odd
	, period } 4K\right]
\end{array}\vspace{3mm}\\
\begin{array}{l}
\bm{\alpha}_{n}^{(7)}= {\alpha}_{n}^{(6)}\Big|_{\bm{l}\leftrightarrow\bm{m}}={\alpha}_{n}^{(6)}, 
\qquad \bm{\beta}_{n}^{(7)}={\alpha}_{n}^{(6)}\Big|_{\bm{l}\leftrightarrow\bm{m}},
\qquad\bm{\gamma}_{n}^{(7)}={\gamma}_{n}^{(6)}\Big|_{\bm{l}\leftrightarrow\bm{m}},
\end{array}.
\end{cases}
\end{eqnarray}

On the other side,  degenerate eigenfunctions 
$\bm{\Psi}^{(6)}(u)$ and  
$\bm{\Psi}^{(8)}(u)$ result from $\bm{H}^{(6)}(u)$ and  
$\bm{H}^{(8)}(u)$. By  Eqs. (\ref{sd-cd-5})) 
we see that they can also be obtained from ${\Psi}^{(7)}(u)$ and 
${\Psi}^{(4)}(u)$, respectively. 
%
%
\begin{eqnarray}\label{adendos-5-associada}
\begin{cases}
\bm{\Psi}^{(6)}(u)
\stackrel{\text{(\ref{h-bold-6},\ref{second-ganguly-3})}}{=}
\displaystyle\frac{1}{
	\mathrm{dn}^{\bm{l}}u}\;\sum_{n=0}^{\bm{l}-1/2}\begin{array}{l} 
\bm{b}_{n}^{(6)}\;\mathrm{cn}^{2n}{u}\;
\tilde{F}\left(n-\frac{\bm{l}+\bm{m}}{2},
\frac{\bm{l}+\bm{m}}{2};n+\frac{1}{2};\mathrm{cn}^2{u}\right)
\end{array}\\
\hspace{1.5cm}\begin{array}{l}\stackrel{\text{(\ref{sd-cd-5})}}{=}{\Psi}^{(7)}(u+K)\Big|_{\bm{l}\leftrightarrow\bm{m}},
\hspace{1.0cm}\mathrm{l}=\frac{1}{2},\frac{3}{2},\frac{5}{2},\cdots,\quad\left[\mbox{even, period } 4K\right]
\end{array}\vspace{3mm}\\
\begin{array}{l}
\bm{\alpha}_{n}^{(6)}={\alpha}_{n}^{(7)}\Big|_{\bm{l}\leftrightarrow\bm{m}}={\alpha}_{n}^{(7)},\qquad 
\bm{\beta}_{n}^{(6)}={\beta}_{n}^{(7)}\Big|_{\bm{l}\leftrightarrow\bm{m}}
\qquad\bm{\gamma}_{n}^{(6)}={\gamma}_{n}^{(7)}\Big|_{\bm{l}\leftrightarrow\bm{m}};\end{array}
\end{cases}
\end{eqnarray}
%

%

%
%
%
\begin{eqnarray}\label{adendos-7-associada}
\begin{cases}
\bm{\Psi}^{(8)}(u)\stackrel{\text{(\ref{h-bold-8},\ref{second-ganguly-3})}}{=}\displaystyle
\frac{\operatorname{cn}{u}}{\mathrm{dn}^{\bm{l}}u}\sum_{n=0}^{\bm{l}-1/2}\begin{array}{l}
\bm{b}_{n}^{(8)}\;\mathrm{cn}^{2n}{u}\;
\tilde{F}\left(n-\frac{\bm{l}-\bm{m}-2}{2},
\frac{\bm{l}-\bm{m}}{2};n+\frac{3}{2};\mathrm{cn}^2{u}\right)
\end{array}
\\
\hspace{1.5cm}\begin{array}{l}\stackrel{\text{(\ref{sd-cd-5})}}{=}{\Psi}^{(4)}(u+K)\Big|_{\bm{l}\leftrightarrow\bm{m}},
\hspace{1.0cm}\mathrm{l}=\frac{1}{2},\frac{3}{2},\frac{5}{2},\cdots,
\quad\left[\mbox{odd, period } 4K\right]
\end{array}\vspace{3mm}\\
\begin{array}{l}
\bm{\alpha}_{n}^{(8)}= {\alpha}_{n}^{(4)}\Big|_{\bm{l}\leftrightarrow\bm{m}}={\alpha}_{n}^{(4)},
\qquad \bm{\beta}_{n}^{(8)}={\alpha}_{n}^{(4)}\Big|_{\bm{l}\leftrightarrow\bm{m}},
\qquad\bm{\gamma}_{n}^{(8)}={\gamma}_{n}^{(4}\Big|_{\bm{l}\leftrightarrow\bm{m}},
\end{array}.
\end{cases}
\end{eqnarray}

For one-term series ($\bm{l}=1/2$),  only two expansions are independent since
\begin{eqnarray*}
	\begin{array}{l}
		\bm{\psi}^{(5)}(u)=\bm{\Psi}^{(6)}(u)=
		\frac{\bm{b}_{0}^{(5)}}{
			\sqrt{\operatorname{dn}{u}}}\left[\left( {\operatorname{cn}{u}-i
			\operatorname{sn}{u}}\right)^{\bm{m}+\frac{1}{2}}+
		\left( {\operatorname{cn}{u}+i
			\operatorname{sn}{u}}\right)^{\bm{m}+\frac{1}{2}}
		\right],\qquad \bm{l}=\frac{1}{2},\vspace{2mm}\\
		\bm{\psi}^{(7)}(u)=\bm{\Psi}^{(8)}(u)=
	\frac{\bm{b}_{0}^{(7)}}{
		\sqrt{\operatorname{dn}{u}}}\left[\left( {\operatorname{cn}{u}-i
		\operatorname{sn}{u}}\right)^{\bm{m}+\frac{1}{2}}-
\left( {\operatorname{cn}{u}+i
	\operatorname{sn}{u}}\right)^{\bm{m}+\frac{1}{2}}
\right],\qquad \bm{l}=\frac{1}{2}.	\end{array}
\end{eqnarray*}
These are even and odd solutions, respectively, both with 
period $4K$ and energy ${\cal E}=\left(\bm{m}+\frac{1}{2}\right)^2+\frac{ k^2}{4}$. We can also verify that, for $\bm{l}=3/2$,
the four two-term series are degenerate with 
energies $ {\cal E}^{(\pm)}=\bm{m}^2+\bm{m}+\frac{5}{2}+
\frac{5}{2} k^2\pm\sqrt{4(1-k^2)(\bm{m}^2+\bm{m})+1}$.

%
%
\subsubsection{Finite series when $\bm{m}$ is an integer 
and $\bm{l}\neq -1/2$ is half 
 an odd negative integer}

 For the eigenfunctions ${\bm\Psi}^{(1)}(u)$ and  ${\bm\Psi}^{(3)}(u)$ we find 	
 %
 %
 \begin{eqnarray}\label{adendos-1-associada}
 \begin{cases}
 \bm{\Psi}^{(1)}(u)
 \stackrel{\text{(\ref{h-bold-1},\ref{second-ganguly-3})}}{=}
 \displaystyle{
 	\mathrm{dn}^{\bm{l}+1}u}\;\sum_{n=0}^{-\bm{l}-3/2}\begin{array}{l} 
 \bm{b}_{n}^{(1)}\;\mathrm{cn}^{2n}{u}\;
 \tilde{F}\left(n+\frac{\bm{l}-\bm{m}+1}{2},
 \frac{\bm{m}-\bm{l}-1}{2};n+\frac{1}{2};\mathrm{cn}^2{u}\right)
 \end{array}\vspace{3mm}\\
 \begin{array}{l}
  \hspace{1.5cm}\stackrel{\text{(\ref{sd-cd-5})}}{=}{\psi}^{(5)}(u+K)\Big|_{\bm{l}\leftrightarrow\bm{m}},
 \hspace{1.0cm}\mathrm{l}=-\frac{3}{2},-\frac{5}{2},-\frac{7}{2},
 \cdots,\quad\left[\mbox{even, period } 4K\right]
 \end{array}\vspace{3mm}\\
 \begin{array}{l}
 \bm{\alpha}_{n}^{(1)}={\alpha}_{n}^{(5)}\Big|_{\bm{l}\leftrightarrow\bm{m}}={\alpha}_{n}^{(5)},\qquad 
 \bm{\beta}_{n}^{(1)}={\beta}_{n}^{(5)}\Big|_{\bm{l}\leftrightarrow\bm{m}}
 \qquad\bm{\gamma}_{n}^{(1)}={\gamma}_{n}^{(5)}\Big|_{\bm{l}\leftrightarrow\bm{m}}
 ;
 \end{array}
 \end{cases}
 \end{eqnarray}
 %
 %
 %
 %
 %
 \begin{eqnarray}
 \begin{cases}
 \bm{\Psi}^{(3)}(u)\stackrel{\text{(\ref{h-bold-3},\ref{second-ganguly-3})}}{=}\displaystyle
 {\mathrm{cn}\,u}\;{\mathrm{dn}^{\bm{l+1}}u}\sum_{n=0}^{-\bm{l}-3/2}\begin{array}{l}
 \bm{b}_{n}^{(3)}\;\mathrm{cn}^{2n}{u}\;
 \tilde{F}\left(n+\frac{\bm{l}+\bm{m}+3}{2},
 -\frac{\bm{l}+\bm{m}+1}{2};n+\frac{3}{2};\mathrm{cn}^2{u}\right)
 \end{array}
 \\
 \begin{array}{l}
 \hspace{1.5cm} \stackrel{\text{(\ref{sd-cd-5})}}{=}{\psi}^{(2)}(u+K)\Big|_{\bm{l}\leftrightarrow\bm{m}},
 \hspace{1.0cm}\mathrm{l}=-\frac{3}{2},-\frac{5}{2},-\frac{7}{2},\cdots,
 \quad\left[\mbox{odd, period } 4K\right]
 \end{array}\vspace{3mm}\\
 \begin{array}{l}
 \bm{\alpha}_{n}^{(3)}={\alpha}_{n}^{(2)}\Big|_{\bm{l}\leftrightarrow\bm{m}}={\alpha}_{n}^{(2)},
 \qquad \bm{\beta}_{n}^{(3)}={\beta}_{n}^{(2)}\Big|_{\bm{l}\leftrightarrow\bm{m}},
 \qquad\bm{\gamma}_{n}^{(3)}={\gamma}_{n}^{(2)}\Big|_{\bm{l}\leftrightarrow\bm{m}}
 .
 \end{array}
 \end{cases}
 \end{eqnarray}
${\bm\Psi}^{(1)}$ and  ${\bm\Psi}^{(3)}$ 
 follow also from ${\bm\Psi}^{(5)}$ and  ${\bm\Psi}^{(7)}$  by the substitutions (\ref{subs}). 

The  degenerate eigenfunctions 
($\bm{\Psi}^{(2)},\bm{\Psi}^{(4)}$), coming from  ($\bm{H}^{(2)},\bm{H}^{(4)}$), are given by 
%
 %
 \begin{eqnarray}\label{adendos-1-associada}
 \begin{cases}
 \bm{\Psi}^{(2)}(u)
 \stackrel{\text{(\ref{h-bold-2},\ref{second-ganguly-3})}}{=}
 \displaystyle{
 	\mathrm{dn}^{\bm{l}+1}u}\;\sum_{n=0}^{-\bm{l}-\frac{3}{2}}\begin{array}{l} 
 \bm{b}_{n}^{(2)}\;\mathrm{cn}^{2n}{u}\;
 \tilde{F}\left(n+\frac{\bm{l}+\bm{m}+2}{2},
- \frac{\bm{m}+\bm{l}+2}{2};n+\frac{1}{2};\mathrm{cn}^2{u}\right)
 \end{array}\vspace{3mm}\\
 \hspace{1.5cm}\begin{array}{l}
  \stackrel{\text{(\ref{sd-cd-5})}}{=}{\psi}^{(3)}(u+K)\Big|_{\bm{l}\leftrightarrow\bm{m}},
 \hspace{1.0cm}\bm{l}=-\frac{3}{2},-\frac{5}{2},-\frac{7}{2},
 \cdots,
\quad\left[\mbox{even, period } 4K\right]
 \end{array}\vspace{3mm}\\
 \begin{array}{l}
 \bm{\alpha}_{n}^{(2)}={\alpha}_{n}^{(3)}\Big|_{\bm{l}\leftrightarrow\bm{m}}= {\alpha}_{n}^{(3)},\qquad 
 \bm{\beta}_{n}^{(2)}={\beta}_{n}^{(3)}\Big|_{\bm{l}\leftrightarrow\bm{m}}
 \qquad\bm{\gamma}_{n}^{(2)}={\gamma}_{n}^{(3)}\Big|_{\bm{l}\leftrightarrow\bm{m}}
 ;
 \end{array}
 \end{cases}
 \end{eqnarray}
 %
 %
 %
 %
 %
 \begin{eqnarray}
 \begin{cases}
 \bm{\Psi}^{(4)}(u)\stackrel{\text{(\ref{h-bold-4},\ref{second-ganguly-3})}}{=}\displaystyle 
 {\mathrm{cn}\,u}\;{\mathrm{dn}^{\bm{l+1}}u}\sum_{n=0}^{-\bm{l}-\frac{3}{2}}\begin{array}{l}
 \bm{b}_{n}^{(4)}\;\operatorname{cn}^{2n}{u}\;
 \tilde{F}\left(n+\frac{\bm{l}-\bm{m}+2}{2},
 \frac{\bm{m}-\bm{l}}{2};n+\frac{3}{2};\mathrm{cn}^2{u}\right)
 \end{array}
 \\
 \hspace{1.5cm}\begin{array}{l}
  \stackrel{\text{(\ref{sd-cd-5})}}{=}{\psi}^{(8)}(u+K)\Big|_{\bm{l}\leftrightarrow\bm{m}},
 \hspace{8mm}\bm{l}=-\frac{3}{2},-\frac{5}{2},-\frac{7}{2},\cdots,
 \quad\left[\mbox{odd, period } 4K\right]
 \end{array}\vspace{3mm}\\
 \begin{array}{l}
 \bm{\alpha}_{n}^{(4)}={\alpha}_{n}^{(8)}\Big|_{\bm{l}\leftrightarrow\bm{m}}={\alpha}_{n}^{(8)},
 \qquad \bm{\beta}_{n}^{(4)}={\beta}_{n}^{(8)}\Big|_{\bm{l}\leftrightarrow\bm{m}},
 \qquad\bm{\gamma}_{n}^{(4)}={\gamma}_{n}^{(8)}\Big|_{\bm{l}\leftrightarrow\bm{m}}
 .
 \end{array}
 \end{cases}
 \end{eqnarray}
 Notice that ${\bm\Psi}^{(2)}$ and  ${\bm\Psi}^{(4)}$ 
 follow from ${\bm\Psi}^{(6)}$ and  ${\bm\Psi}^{(8)}$  by the substitutions (\ref{subs}). Beside this, f{\tiny }or 
 one-term series ($\bm{l}=-3/2$):
 $ \bm{\Psi}^{(1)}= \bm{\Psi}^{(2)}$ and
  $ \bm{\Psi}^{(3)}= \bm{\Psi}^{(4)}$.
 

%
\subsection{Infinite-series solutions}
							
To obtain infinite-series solutions we require that 
$\mathring{\gamma}_{n}^{(i)}\neq 0$ in the power series expansions
$\mathring{H}^{(i)}(x)$ written in section 3.1. 
This imposes 
restrictions on the parameters $\bm{l}$ and $\bm{m}$: one condition
on $\bm{l-m}$  and other on $\bm{l+m}$. 
Since $a=1/k^2>1$, the relations (\ref{convergence-primeiro}) imply that 
the eigenfunctions are convergent and bounded for any value of the independent variable, that is, for $0\leq x=\mathrm{sn}^2u\leq 1$. They are denoted by $\mathring{\Phi}^{(i)}$
while the series coefficients are denoted by $\mathring{b}_{n}^{(i)}$, 
\begin{eqnarray}\label{recurrenc-infinita}
\mathring{\alpha}_{n}^{(i)}\;\mathring{b}_{n+1}^{(i)}+
\mathring{\beta}_{n}^{(i)}\;
\mathring{b}_{n}^{(i)}+
\mathring{\gamma}_{n}^{(i)} \;\mathring{b}_{n-1}^{(i)}=0, \qquad 
\mathring{b}_{-1}^{(i)}=0,
\qquad  n\geq 0, 
\end{eqnarray}
$\mathring{\alpha}_{n}^{(i)}$,  $\mathring{\beta}_{n}^{(i)}$ and $\mathring{\gamma}_{n}^{(i)}$  being defined
in Section 5.1. 
Below

\begin{itemize}
\itemsep-3pt
 %
 \item   we write down the conditions to have infinite series;
 \item   we show that each  finite series  corresponds to four infinite-series expansions
 (same values for $\bm{l}$ and $\bm{m}$).
\end{itemize}
Therefore, for this family of quasi-exactly solvable potentials, for solutions of the 
Schr\"{o}dinger equation we have to regard both finite-series and infinite-series wavefunctions since the 
infinite series are convergent and bounded for all values of the independent variable, in opposition to the case mentioned in \cite{kalnins}.


%

%
%
 %
%

	%
	%
	%

 By inserting  
	 the expansions $\mathring{H}^{(i)}(x)$  into (\ref{second-ganguly-3}), we get four types of even and odd infinite-series 
	 solutions (periods $2K$ and $4K$), namely,
	\begin{eqnarray}\label{1}
	\begin{array}{r}
	\text{even,}\\ 
		2K:
		\end{array}
	\begin{cases}
	 \begin{array}{l}
	\mathring{\Phi}^{(1)}(u)
	\stackrel{\text{(\ref{S1},\ref{second-ganguly-3}})}{=}
	\mathrm{dn}^{\bm{l}+1}{u}\displaystyle 
	\sum_{n=0}^{\infty}\mathring{b}_{n}^{(1)}\;
	\mathrm{sn}^ {2n}{u},\quad \mbox{if }
	\begin{array}{l}\frac{\bm{l}-\bm{m}-1}{2}
	\mbox{ and } \frac{\bm{l}+\bm{m}}{2} \end{array} \mbox{ are not}\\
	\hspace{8.5cm}\mbox{ negative  integers}, \end{array}\vspace{2mm}\\
	\mathring{\Phi}^{(5)}(u)\stackrel{\text{(\ref{S5},\ref{second-ganguly-3}})}{=}
\displaystyle\frac{1}{
\mathrm{dn}^{\bm{l}}u}\; 
\sum_{n=0}^{\infty }\mathring{b}_{n}^{(5)}
\; \mathrm{sn}^{2n}u,\quad \mbox{if }
	\begin{array}{l}\frac{\bm{l}-\bm{m}+1}{2}
	\mbox{ and } \frac{\bm{l}+\bm{m}+2}{2} \end{array} \mbox{ are not}\\
	\hspace{8.5cm}\mbox{ positive  integers};
	%
	%
	\end{cases}
	\end{eqnarray}
	%
	%
	%
	\begin{eqnarray}\label{2}
		\begin{array}{r}
	\text{odd,}\\
	4K:\end{array}
	\begin{cases}
	\mathring{\Phi}^{(2)}(u)
	\stackrel{\text{(\ref{S2},\ref{second-ganguly-3}})}{=}
	\mathrm{sn}{u}\;\mathrm{dn}^{\bm{l}+1}{u}
	\displaystyle \sum_{n=0}^{\infty}\mathring{b}_{n}^{(2)}\;
	\mathrm{sn}^{2n}u,\quad \mbox{if }
	\begin{array}{l}
	\frac{\bm{l}-\bm{m}}{2} \mbox{ and } \frac{\bm{l}+\bm{m}+1}{2} \mbox{ are 
	not}\end{array}\\
	\hspace{8.5cm}\mbox{ negative  integers},\vspace{2mm}
	\\
	\mathring{\Phi}^{(6)}(u)\stackrel{\text{(\ref{S6},\ref{second-ganguly-3}})}{=}
\displaystyle  \frac{\mathrm{sn}{u}}{{\rm dn}^{\bm{l}}u}\; 
\sum_{n=0}^{\infty}\mathring{b}_{n}^{(6)}
\; {\rm sn}^{2n}u,\quad \mbox{if }
	\begin{array}{l}\frac{\bm{l}-\bm{m}}{2}
	\mbox{ and } \frac{\bm{l}+\bm{m}+1}{2} \end{array} \mbox{ are not}\\
	\hspace{8.5cm}\mbox{  positive  integers};
\\
	%
	%
	\end{cases}
	\end{eqnarray}
	%
	%
	%
	\begin{eqnarray}\label{3}
		\begin{array}{r}
	\text{even, }\\
	4K:\end{array}
	\begin{cases}
	\mathring{\Phi}^{(3)}(u)
	\stackrel{\text{(\ref{S3},\ref{second-ganguly-3}})}{=}
		\mathrm{cn}{u}\;\mathrm{dn}^{\bm{l}+1}u\displaystyle \sum_{n=0}^{\infty}\mathring{b}_{n}^{(3)}\;
	\mathrm{sn}^{2n}u,\quad \mbox{if }
	\begin{array}{l}
	\frac{\bm{l}-\bm{m}}{2} \mbox{ and } \frac{\bm{l}+\bm{m}+1}{2} \mbox{ are not}\end{array}\\
	 \hspace{8,5cm}\mbox{ negative integers},
	\vspace{2mm}
	\\
	\mathring{\Phi}^{(7)}(u)\stackrel{\text{(\ref{S7},\ref{second-ganguly-3}})}{=}
\displaystyle\frac{\mathrm{cn}{u}}{ {\rm dn}^{\bm{l}}u}\; 
\sum_{n=0}^{\infty}\mathring{b}_{n}^{(7)}\;
{\rm sn}^{2n}u,\quad \mbox{if }
	\begin{array}{l}\frac{\bm{l}-\bm{m}}{2}
	\mbox{ and } \frac{\bm{l}+\bm{m}+1}{2} \end{array} \mbox{ are not}\\
	\hspace{8,5cm}\mbox{  positive integers};
	%
	\end{cases}
	\end{eqnarray}
	%
	%
	%
	%
	\begin{eqnarray}\label{4}
	\begin{array}{r}\mbox{odd,}\\
	 2K:\end{array}
	\begin{cases}
	\mathring{\Phi}^{(4)}(u)
	\stackrel{\text{(\ref{S4},\ref{second-ganguly-3}})}{=}
	\mathrm{sn}{u}\;\mathrm{cn}{u}\; \mathrm{dn}^{\bm{l}+1}u
	\displaystyle \sum_{n=0}^{\infty}\mathring{b}_{n}^{(4)}\;
	\mathrm{sn}^{2n}u,\mbox{ if }
	\begin{array}{l}\frac{\bm{l}-\bm{m}+1}{2}
	\mbox{ and } \frac{\bm{l}+\bm{m}+2}{2}\mbox{ are}\end{array} \\
	\hspace{8.5cm}\mbox{not
	 negative integers},\vspace{2mm}\\
	\mathring{\Phi}^{(8)}(u)\stackrel{\text{(\ref{S8},\ref{second-ganguly-3}})}{=}\displaystyle
\frac{\mathrm{sn}{u}\;\mathrm{cn}{u}}{
{\rm dn}^{\bm{l}}u}\; 
\sum_{n=0}^{\infty}\mathring{b}_{n}^{(8)}
\;{\rm sn}^{2n}u, \mbox{ if }
	\begin{array}{l}\frac{\bm{l}-\bm{m}-1}{2}
	\mbox{ and } \frac{\bm{l}+\bm{m}}{2} \end{array} 
	\mbox{ are}\\
	\hspace{8.5cm}\mbox{ not positive  integers},
	%
	\end{cases}
	\end{eqnarray}
	The expansions in each pair are related to each other  by $\bm{l}\mapsto -\bm{l}-1$
	and  $\bm{m}\mapsto -\bm{m}-1$.	From the above conditions we {will} see that there are four types of infinite-series solutions corresponding to 
	a given finite-series solution.

		%
	%
	%
%
First, consider the cases in which either $\bm{l}$
or $\bm{m}$ is an integer and the other is half an odd integer, Sections 5.3 and 5.4.  
Then, the eight expansions  $\mathring{\Phi}^{(i)}(u)$  are given by infinite series which,  
however, reduce to four expansions because Eq. (\ref{ince}) implies that 
$\mathring{\Phi}^{(i+4)}(u)=$ $\mathring{\Phi}^{(i)}(u)$  for
$i=1,2,3,4$.

	%
%
	%
	%
	%
	%
		As examples for finite series  $	\left(\mathring{\psi}^{(i)}, \tilde{\psi}^{(i)}\right)$ in terms of $\mathrm{sn}^2u$,  we consider the same cases given in Section 5.1. Thus, the finite-series
		 $\mathring{\psi}^{(1)}$, given in (\ref{ex-01}),
		 is associated with the four infinite-series expansions: $\mathring{\phi}^{(2)}=\mathring{\phi}^{(6)}$, 
		 $\mathring{\phi}^{(3)}=\mathring{\phi}^{(7)}$, 
		 $\mathring{\phi}^{(5)}$ and 
		 $\mathring{\phi}^{(8)}$.
	%
For the second example, $\bm{m}=-1/2$. Then $\mathring{\psi}^{(j)}=\tilde{\psi}^{(j)}$
and, by using the values for $\bm{l}$ written in (\ref{m1/2-1}-\ref {m1/2-3}), 
we find four infinite-series expansions corresponding to each of the finite series according to 
\begin{eqnarray}\label{V.A,m-1/2}
\mathring{\psi}^{(i)}=\tilde{\psi}^{(i)}\quad \mapsto\quad
\mathring{\Phi}^{(i+4)},\qquad 
\mathring{\psi}^{(i+4)}=\tilde{\psi}^{(i+4)}\quad \mapsto\quad
\mathring{\Phi}^{(i)},\qquad 
%
%
\end{eqnarray}
where $i$ assumes the four values $i=1,2,3,4$ in each case. In the third example, we take $\bm{m}=0$ for Eq. (\ref{first-case}); thence, from the
values for $\bm{l}$ written in (\ref{m0-1}-\ref{m0-3}),  we obtain the correspondence
\begin{eqnarray}\label{V.A,m=0}
\begin{matrix}
\tilde{\psi}^{(i)} \quad \mapsto \quad\mathring{\Phi}^{(i+4)}, 
\qquad \tilde{\psi}^{(i+4)} \quad \mapsto\quad \mathring{\Phi}^{(i)}, 
\end{matrix}
\end{eqnarray}
where $i$ assumes the four values $i=1,2,3,4$ once  more.



{
	As an example of finite-series $	\left(\mathring{\bm\Psi}^{(i)}, \tilde{\bm\Psi}^{(i)}\right)$ in terms of $\mathrm{cn}^2u$, given in Section 5.2, we take the Lam\'e equation with $\bm{l}=0$. Then, by using the parameters written in (\ref{even2K}-\ref{even4K} ),
 we get the following four types of infinite-series solutions 
corresponding  each of the 16 finite series:
\begin{eqnarray}\label{V.lame}
&&\begin{array}{l}
\left( \mathring{\bm\Psi}^{(1)},\tilde{\bm\Psi}^{(1)}    \right),\;
\left( \mathring{\bm\Psi}^{(4)},\tilde{\bm\Psi}^{(4)}    \right),\;
\left( \mathring{\bm\Psi}^{(6)},\tilde{\bm\Psi}^{(6)}    \right),\;
\left( \mathring{\bm\Psi}^{(7)},\tilde{\bm\Psi}^{(7)}    \right)\;\mapsto
\;\mathring{\Phi}^{(2,3,5,8)};\qquad\end{array}\vspace{3mm}\\
&&\begin{array}{l}
\left( \mathring{\bm\Psi}^{(2)},\tilde{\bm\Psi}^{(2)}    \right),\;
\left( \mathring{\bm\Psi}^{(3)},\tilde{\bm\Psi}^{(3)}    \right),\;
\left( \mathring{\bm\Psi}^{(5)},\tilde{\bm\Psi}^{(5)}    \right),\;
\left( \mathring{\bm\Psi}^{(8)},\tilde{\bm\Psi}^{(8)}    \right)\;\mapsto
\;\mathring{\Phi}^{(1,4,6,7)}.\end{array}
\end{eqnarray}
%
	%
%
%
%
%
%
%
%
  %
%
%
%
%
%
More exemples are given in Appendix C.
}
	\section{Conclusions}

By treating the elliptic Darboux equation as 
an instance of Heun's general equation, we have applied  known 
transformations of
variables to generate new solutions for the latter equation. In Section 3 we have established two groups of power series 
expansions and, in Section 4, two groups of expansions in series of Gauss 
hypergeometric functions. Each group is constituted by eight expansions connect to each other by homotopic transformations of the dependent variable. After this, the solutions of the Heun equation can be particularized for Darboux equation and its special cases.
 
{The relations among the eight expansions of each group require further study. In Section 3 we have seen that the solutions 
 $\mathring{U}^{(i) }$ reduce to four expansions under certain circumstances. 
 On the other side, Ince \cite{ince2,ince3} has discussed these relations for  some solutions for the particular case of the Lam\'e equation. }

{The solutions we have considered} may be used to solve the stationary 
Schr\"odinger equation for 
quasi-exactly solvable (QES) potentials. For this purpose, firstly, in 
section 2.1, we have seen that the 
Schr\"odinger equation for certain elliptic potentials reduces
to Darboux equations.  
For particular values of the parameters, some finite-series solutions to each of such potentials 
are already known in the literature \cite{ganguly-1,khare2}, but finite series  turn out to be systematically generated from solutions for the Heun equation without the need of specifying
the parameters.  This procedure provides as well suitable infinite-series  eigenfunctions. 

As an example, in Section 5 we have considered the Schr\"odinger equation
for the associated Lam\'e potential  (\ref{ganguly-3}) which depends on the parameters $\bm{l}$ and $\bm{m}$. When either $\bm{m}-\bm{l}$
or $\bm{l}+\bm{m} $ is an integer, the solutions 
$\mathring{{H}}^{(i)}(x)$ and $\mathring{\bm{H}}^{(i)}(x)$ of the Heun equation lead
to finite-series eingenfunctions given by expansions in series
of $\mathrm{sn}^2u$ and $\mathrm{cn}^2u$, respectively.  To 
assure real and distinct energy eingenvalues, we use the
Arscott condition (\ref{autovalores}) which, in turn, imposes
further restrictions on $\bm{l}$ and $\bm{m}$ 
(such restrictions are written on the right-hand 
side of the expansions
given in Sections 5.1 and 5.2). Finite series in terms of $\mathrm{sn}^2u$ 
and $\mathrm{cn}^2u$ 
cannot satisfy simultaneously the Arscott condition 
because this condition requires that $\mathring{{\gamma}}_n^{(i)}>0$ and 
$\mathring{{\gamma}}_n^{(i)}<0$, respectively. This fact gives
a criterion to select the appropriate solutions for the 
associated Lam\'e equation, {as in the examples of
of Sections 5.1, 5.2 and  
Appendix C, where the use of each of the preceding  finite-series expansions has been illustrated.}  {From the solutions of Sections 5.1 and 5.2 we can generate new solutions by means of the substitutions (\ref{associado-gan-2}).}

On the other hand, when either $\bm{l}$
or $\bm{m}$ is an integer and the other is half an odd integer,  we have found {eigenfunctions  by using the hypergeometric-function expansions ${H}^{(i)}(x)$ and 
${\bm{H}}^{(i)}(x)$ given in Section 4. If $\bm{l}$
or $\bm{m}$ is a positive or a negative integer all the expansions afford finite-series solutions bounded for all values of the independent variable $u$.
 Thus, if} $\bm{l}$ is an integer and $\bm{m}\neq -1/2$ is half 
 an odd integer we have the degenerate finite-series solutions given in Section 5.3; for $\bm{l}=-1$ or $\bm{l}=0$, we get new solutions
 for the Lam\'e equation. If $\bm{m}$ is an integer and $\bm{l}\neq -1/2$ is half an odd integer,
 we have the degenerate finite-series solutions given in Section 5.4.
  {In fact,  the solutions of Sections 5.3 and 5.4 are connected  
 by means of the substitutions (\ref{associado-gan-2}).}
 The eigenfunctions are degenerate in the sense that there are two independent eigenfunctions corresponding to the same energy.
 The degeneracy has been shown by writing the recurrence relations in the matrix form. 
 
 However, 
 Khare and Sukhatme \cite{khare,khare2} have pointed out another type of degeneracy which 
 is not obtained by using expansions in series of hypergeometric functions. In fact, this type of 
 degenerate solutions comes from expansions in 
 series of  $\mathrm{cn}^2u$ or $\mathrm{sn}^2u$
 for particular values for $\bm{l}$ and $\bm{m}$,
 as explained in Appendix C.

The previous examples show how to use the solutions of the Heun equation to find finite-series
solutions for the elliptic Darboux equation {or, more precisely, for the associated Lam\'e equation. Furthermore, 
to each finite-series solution, in Section 5.5 we have seen that this procedure   
also 
gives four periodic 
infinite-series solutions which are bounded and convergent for any value of the independent variable. Hitherto, relations among these  have not been determined
even for the known case of the Lam\'e equation \cite{ince3}. 
%
%
{We note that the existence of such periodic  
solutions for the associated Lam\'e equation can also be studied by using expansions in series of trigonometric functions like the ones considered in  
References \cite{ince3,magnus, volkmer}. In Appendix D
we explain how these solutions could be obtained as particular instances of solutions of the Heun equation (\ref{heun})}. 


{ 
Anyway, 
 in the present case, infinite series as solutions for quasi-exactly solvable problems cannot be discarded 
 by asserting that the wavefunctions are unbounded, as 
 sometimes occurs
 \cite{kalnins}. Besides this, it is known that the infinite series are important  in band theory of solids.  Indeed, these are the only type of eigenstates admitted by a model proposed by Slater in which the Schr\"odinger equation reduces to a Mathieu equation \cite{slater} (this does not admit finite-series solutions). }

Although we have considered only the associated Lam\'e potential, the solutions of the Heun equation can also  be applied for the other potentials presented in Section 2.1. 
{To finalize, we mention some supplementary issues.} {In the first place, for the Heun equation, there  are other expansions in series of hypergeometric functions \cite{svartholm,erdelyi2,erdelyi3,armenio} which, however, are not appropriate for
the problems of Section 2.1. The same is true 
 respecting a  recent study on Darboux equation \cite{et-al}. 
 
In the second place, one of the Erd\'elyi expansions in {series of}  hypergeometric functions seems to constitute a suitable start-point for generating new solutions for the associated Lam\'e equation. To this end, from the inception we have to modify the recurrence relations for series coefficients as in Appendix D and, after that, use the homotopic transformations as well as the fractional substitutions (\ref{17}-\ref{65}).} {Similarly, by using modified recurrence relations we see that the Svartholm solution  \cite{svartholm} leads to
 expansions in series of trigonometric functions for the associated Lam\'e equation. However, in order to compare  solutions
obtained from Svartholm's and Erdelyi's expansions with the ones presented in this article, it would be necessary to find out if the finite-series expansions fulfill the Arscott condition (\ref{autovalores}).}

 \section*{Acknowledgement}
 
%
This work was funded by Minist\'erio da Ci\^encia, Tecnologia e Inova\c{c}\~ oes of Brazil. The author thanks 
L\'ea Jaccoud El-Jaick for computations of section 
4.1 and for reading the manuscript. This work does not have any conflicts of interest.

%

	%

%
\appendix
\section{Homotopic transformations}
\protect\label{A}
\setcounter{equation}{0}
\renewcommand{\theequation}{A.\arabic{equation}}

The 8 homotopic transformations (\ref{arscott-maier}) are:
\begin{align}
&\begin{array}{l}
T_1H(x)=H(x)=H(a,q;\alpha,\beta,\gamma,\delta;x),\quad (\text{identity})
\end{array}\label{t1}
\vspace{2mm}\\
&\begin{array}{l}
T_{2}H(x)=
x^{1-\gamma}H\big[a,q-(\gamma-1)(\delta a+\epsilon);{\beta-\gamma+1},
\alpha-\gamma+1,2-\gamma,\delta;x \big],
\end{array}\label{t2}
\vspace{2mm}\\
%
&\begin{array}{l}
T_{3}H(x)= (1-x)^{1-\delta}H\big[a,q-(\delta-1)\gamma a;{\beta-\delta+1},
\alpha-\delta+1,\gamma,2-\delta;x \big],\end{array}\label{t3}
\vspace{2mm}\\
%
%
&\begin{array}{l}
T_{4}H(x)= x^{1-\gamma}(1-x)^{1-\delta}\times
\vspace{2mm}\\
%
H\left[a,q-(\gamma+\delta-2)a-(\gamma-1)\epsilon;
{\alpha-\gamma-\delta+2},
\beta-\gamma-\delta+2,2-\gamma,2-\delta;x \right],\end{array}
\label{t4}
\end{align}
%
%
\begin{align}
&\begin{array}{l}
T_{5}H(x)=
\left[1-\frac{x}{a} \right]^{1-\epsilon}
H\big[a,q-\gamma(\alpha+\beta-\gamma-\delta);
{-\alpha+\gamma+\delta},
-\beta+\gamma+\delta,
\gamma,\delta;x \big],\end{array}
\label{t5}
\vspace{2mm}\\
&\begin{array}{l}
T_{6}H(x)= \left[1-\frac{x}{a} \right]^{1-\epsilon} x^{1-\gamma}
\times
\vspace{2mm}\\
\hspace{.8cm}
H\big[a,q-\delta(\gamma-1)a-\alpha-\beta+\delta+1;
{-\beta+\delta+1},-\alpha+\delta+1,2-\gamma,\delta;x \big],
\end{array}
\label{t6}\vspace{2mm}\\
 &\begin{array}{l}
 T_{7}H(x)=  \left[1-\frac{x}{a} \right]^{1-\epsilon}(1-x)^{1-\delta}
\times
\vspace{2mm}\\
\hspace{.4cm}
H\Big[a,q-\gamma[(\delta-1)a+\alpha+\beta-\gamma-\delta];
{-\beta+\gamma+1},-\alpha+\gamma+1,\gamma,2-\delta;x \Big],
\end{array}\label{t7}
\vspace{2mm}\\
&\begin{array}{l}
T_{8}H(x)=\left[1-\frac{x}{a} \right]^{1-\epsilon}
\displaystyle x^{1-\gamma}
(1-x)^{1-\delta}
\times
\vspace{2mm}\\
\hspace{1cm}
H\big[a,q-(\gamma+\delta-2)a
-\alpha-\beta+\delta+1;{2-\alpha},2-\beta,2-\gamma,2-\delta;x \big].
\end{array}\label{t8}
\end{align}

\section{Cases solved by hypergeometric functions}
\protect\label{B}
\setcounter{equation}{0}
\renewcommand{\theequation}{B.\arabic{equation}}

In the four groups of solutions there are cases where the 
recurrence relations for the series coefficients present
only two terms.  For such cases, by combining results given in Ref. \cite{vidunas}
with transformations of Maier's table \cite{maier}, we find that the Heun equation
(\ref{heun}) is reduced to the hypergeometric equation (\ref{hypergeometric}).

In (\ref{S1-S8}) we have seen that the powers expansions (\ref{S1}-\ref{S8})
present two-term recurrence relations in the following three cases:
\begin{eqnarray*}
\begin{array}{l}
a=0;\quad
a=-1,\quad \epsilon=\delta,\quad  q=0\quad \mbox{and}\quad
a=-1,\quad  \alpha+\beta=1+\gamma, \quad
q=\gamma(1-\delta). 
\end{array}
\end{eqnarray*}
For $a=0$ the Heun equation equation reads
\begin{eqnarray}
x(x-1)\frac{d^{2}H}{dx^{2}}+\left[(\alpha+\beta+1)x-(\gamma+\epsilon)\right]
\frac{dH}{dx}+\left[ \alpha \beta -\frac{q}{x}\right]H=0, \quad (a=0)
\end{eqnarray}
and the substitution
\begin{eqnarray*}
H(x)=x^{k_{1}}F(x), \qquad k_{1}=\frac{1}{2}\left[1-\gamma-\epsilon
-\sqrt{(1-\gamma-\epsilon)^2-4q}\right]
\end{eqnarray*}
gives the hypergeometric equation
\begin{eqnarray*}
x(1-x)\frac{d^{2}F}{dx^{2}}+\big[(2 k_{1}+\gamma+\epsilon)
-(2k_{1}+\alpha+\beta+1)x\big]\frac{dF}{dx}
- (k_{1}+\alpha)(k_{1}+ \beta) F=0.
\end{eqnarray*}
For the second case ($ a=-1,\; \epsilon=\delta,\;q=0$), the Heun equation (\ref{heun}) reads 
\begin{eqnarray*}
\frac{d^{2}H(x)}{dx^{2}}+\left[\frac{\gamma}{x}+
\frac{\alpha+\beta+1-\gamma}{2(x-1)}+\frac{\alpha+\beta+1-\gamma}{2(x+1)}\right]\frac{dH(x)}{dx}+
 \frac{\alpha \beta }{(x-1)(x+1)}H(x)=0,
\end{eqnarray*}
since $\epsilon=\delta=(\alpha +\beta+1-\delta)/2$. It is already known that
the substitutions \cite{vidunas}
\begin{eqnarray}
z=x^2,\qquad u(z)=H(x),\quad [a=-1,\; \epsilon=\delta,\;  q=0]
\end{eqnarray}
reduce the above equation to
\begin{eqnarray}
z(1-z)\frac{d^{2}u(z)}{dz^{2}}+\left[\frac{1}{2}(\gamma+1)-\frac{1}{2}(\alpha+\beta+2)z
\right]\frac{du(z)}{dz}-\frac{\alpha\beta}{4}\;u(z)=0,
\end{eqnarray}
which is the hypergeometric equation (\ref{hypergeometric}) with 
 $\mathrm{a}=\alpha/2$, $\mathrm{b}=\beta/2$ and $\mathrm{c}=(1+\gamma)/2$.
 For the third case, first we transform the Heun equation (\ref{heun}) by using the homotopic 
 substitution $T_3$, that is, we take
\begin{eqnarray*}
H(x)=(1-x)^{1-\delta}h(x),\qquad [a=-1,\;\gamma=\alpha+\beta-1,\; q=\gamma(1-\delta)].
\end{eqnarray*} 
Then, the Heun equation leads to
\begin{eqnarray*}
\frac{d^{2}h(x)}{dx^{2}}+\left[\frac{\alpha+\beta-1}{x}+
\frac{2-\delta}{x-1}+\frac{2-\delta}{x+1}\right]\frac{dh(x)}{dx}+
 \frac{(\alpha-\delta+1)( \beta-\delta+1) }{(x-1)(x+1)}h(x)=0.
\end{eqnarray*}
Now, by the substitutions \cite{vidunas}
\begin{eqnarray}
\begin{array}{r}
z=x^2,\qquad u(z)=h(x)\quad\Rightarrow\quad
H(x)=(1-x)^{1-\delta}u(z),\end{array}
\vspace{2mm}\\
 \left[a=-1,\;\gamma=\alpha+\beta-1,\; q=\gamma-\gamma\delta\right]]
 \nonumber
\end{eqnarray}
the above equation gives the hypergeometric equation 
\begin{eqnarray}
\begin{array}{l}
z(1-z)\frac{d^{2}u(z)}{dz^{2}}+\left[\frac{1}{2}(\alpha+\beta)-\frac{1}{2}(\alpha+\beta+4-2\delta)z
\right]\frac{du(z)}{dz}-\frac{(\alpha-\delta+1)(\beta-\delta+1)}{4}\;u(z)=0,
\end{array}
\end{eqnarray}
with parameters
 $\mathrm{a}=(\alpha-\delta+1)/2$, $\mathrm{b}=(\beta-\delta+1)/2$ and 
 $\mathrm{c}=(\alpha+\beta)/2$.

According to (\ref{SS1-SS8}) and (\ref{HH1-HH8}) the expansions 
$\mathring{\bm{H}}^{(i)}(x)$ and $H^{(i)}(x)$  
have two-term recurrence relations if 
\begin{eqnarray*}
\begin{array}{l}
a=1;\quad
a=2,\quad \alpha+\beta+1=2\gamma+\delta,\quad  q=\alpha\beta\quad \mbox{and}\quad\vspace{2mm}\\
a=2,\quad \alpha+\beta=1+\delta,\quad 
q=\alpha\beta+(\gamma-1)\delta .
\end{array}
\end{eqnarray*}
%
%
%
For $a=1$ we have
\begin{eqnarray}
x(x-1)\frac{d^{2}H}{dx^{2}}+\big[(\alpha+\beta+1)x-\gamma\big]
\frac{dH}{dx}+\frac{\alpha \beta x-q}{x-1}H=0,\qquad (a=1)
\end{eqnarray}
and the change of variable
\begin{eqnarray}
H(x)=(x-1)^{k_{2}}G(x), \qquad k_{2}=\frac{1}{2}\left[\gamma-\alpha-\beta
-\sqrt{(\gamma-\alpha-\beta)^2-4\alpha\beta+4q}\right]
\end{eqnarray}
yields an hypergeometric equation, namely,
\begin{eqnarray}
x(1-x)\frac{d^{2}G}{dx^{2}}+\big[\gamma
-(2 k_{2}+\alpha+\beta+1)x\big]\frac{dG}{dx}
- (k_{2}+\alpha)(k_{2}+ \beta) G=0.
\end{eqnarray}
For the second case the Heun equation (\ref{heun}) becomes
\begin{eqnarray*}
\frac{d^{2}H(x)}{dx^{2}}+\left[\frac{\gamma}{x}+
\frac{\alpha+\beta+1-2\gamma}{x-1}+\frac{\gamma}{x-2}\right]\frac{dH(x)}{dx}+
 \frac{\alpha \beta }{x(x-2)}H(x)=0,\qquad\vspace{3mm}\\ 
 \left[a=2,\;\alpha+\beta=2\gamma+\delta,\;q=\alpha\beta\right].
\end{eqnarray*}
It is known that the substitutions \cite{vidunas}
\begin{eqnarray}\begin{array}{r}
z=x(2-x),\qquad u(z)=H(x),\quad\left[a=2,\;\alpha+\beta=2\gamma+\delta,\;q=\alpha\beta\right]
\end{array}\end{eqnarray}
transform the above equation into the hypergeometric equation (\ref{hypergeometric})
\begin{eqnarray}
\begin{array}{l}
z(1-z)\frac{d^{2}u(z)}{dz^{2}}+\left[\gamma-\frac{1}{2}(\alpha+\beta+2)z
\right]\frac{du(z)}{dz}-\frac{\alpha\beta}{4}\;u(z)=0,
\end{array}
\end{eqnarray}
with 
 $\mathrm{a}=\alpha/2$, $\mathrm{b}=\beta/2$ and $\mathrm{c}=\gamma$.
 For the third case, first we transform the Heun equation (\ref{heun}) by using the homotopic 
 substitution $T_2$, that is, we take
\begin{eqnarray*}
\begin{array}{l}
H(x)=x^{1-\gamma}h(x),\qquad [a=2,\;\alpha+\beta=1+\delta,\; q=\alpha\beta+(\gamma-1)\delta].
\end{array}
\end{eqnarray*}
Then, the Heun equation leads to
\begin{eqnarray*}
\begin{array}{l}
\frac{d^{2}h(x)}{dx^{2}}+\left[\frac{2-\gamma}{x}+
\frac{\delta}{x-1}+\frac{2-\gamma}{x-2}\right]\frac{dh(x)}{dx}+
 \frac{(\alpha-\gamma+1)( \beta-\gamma+1) }{x(x-2)}\;h(x)=0.
 \end{array}
\end{eqnarray*}
By the substitutions \cite{vidunas}
\begin{eqnarray}
\begin{array}{r}
z=x(2-x),\qquad u(z)=h(x)\quad \Rightarrow\quad
H(x)=x^{1-\gamma}u(z),\end{array}\vspace{2mm} \\
\left[a=2,\;\alpha+\beta=1+\delta,\; q=\alpha\beta+(\gamma-1)\delta\right]
\nonumber
\end{eqnarray}
we get the hypergeometric equation
\begin{eqnarray}
\begin{array}{l}
z(1-z)\frac{d^{2}u(z)}{dz^{2}}+\left[2-\gamma-\frac{1}{2}(\alpha+\beta+2-2\gamma)z
\right]\frac{du(z)}{dz}-\frac{(\alpha-\gamma+1)(\beta-\gamma+1)}{4}\;u(z)=0,
¨\end{array}
\end{eqnarray}
  with 
 $\mathrm{a}=(\alpha-\gamma+1)/2$, $\mathrm{b}=(\beta-\gamma+1)/2$ and 
 $\mathrm{c}=2-\gamma$.

%
Finally, by (\ref{bbH}) the expansions $\bm{H}^{(i)}(x)$
present two-term recurrence relations if  
\begin{eqnarray*}
\begin{array}{l}
a=2,\quad  \beta=\alpha+1-\delta,\quad q=\alpha\gamma\quad \mbox{and}\quad
a=2,\quad  \alpha=\beta+1-\delta,\quad
q=\beta\gamma.
\end{array}
\end{eqnarray*}
For the first case, 
in Eq. (\ref{heun}) we use the fractional substitution
\begin{eqnarray*}
\begin{array}{l}
H(x)=\left(1-\frac{x}{a}\right)^{-\alpha}f(y), \qquad y=\frac{(1-a)x}{x-a},\qquad [a=2,\; \beta=\alpha+1-\delta,\; q=\alpha\gamma]
\end{array}
\end{eqnarray*}
which is Maier the transformation $M_{17}$. Thence
\begin{eqnarray*}
\frac{d^{2}f(y)}{dy^{2}}+\left[\frac{\gamma}{y}+
\frac{\delta}{y-1}+\frac{\delta}{y+1}\right]\frac{df(y)}{dy}-
 \frac{\alpha (1+\alpha-\gamma-2\delta) }{(y-1)(y+1)}f(y)=0.
\end{eqnarray*}
The changes
\begin{eqnarray}
\begin{array}{r}
z=y^2=\left(\frac{x}{2-x}\right)^2, \quad u(z)=f(y)\quad
\Rightarrow\quad H(x)=\left(1-\frac{x}{2}\right)^{-\alpha}u(z),\end{array}
\vspace{2mm}\\
\left[a=2,\; \beta=\alpha+1-\delta,\; q=\alpha\gamma\right]\nonumber
\end{eqnarray}
give the hypergeometric equation
\begin{eqnarray}
\begin{array}{l}
z(1-z)\frac{d^{2}u(z)}{dz^{2}}+\left[\frac{1+\gamma}{2}-\frac{1+\gamma+2\delta}{2}z
\right]\frac{du(z)}{dz}+\frac{\alpha(1+\alpha-\gamma-2\delta)}{4}\;u(z)=0,
¨\end{array}
\end{eqnarray}
whose parameters are $\mathrm{a}=\alpha/2$, $\mathrm{b}=(\gamma+2\delta-\alpha-1)/2$
and $\mathrm{c}=(1+\gamma)/2$. The second case differs from the first
one by the interchange $\alpha\longleftrightarrow\beta$. Thus, 
the changes
\begin{eqnarray}
\begin{array}{r}
z=y^2=\left(\frac{x}{2-x}\right)^2, \; u(z)=f(y)\;
\Rightarrow\; H(x)=\left(1-\frac{x}{2}\right)^{-\beta}u(z),\end{array}
\vspace{2mm}\\
\left[a=2,\; \alpha=\beta+1-\delta,\; q=\beta\gamma\right]\nonumber
\end{eqnarray}
give the hypergeometric equation
\begin{eqnarray}
\begin{array}{l}
z(1-z)\frac{d^{2}u(z)}{dz^{2}}+\left[\frac{1+\gamma}{2}-\frac{1+\gamma+2\delta}{2}z
\right]\frac{du(z)}{dz}+\frac{\beta(1+\beta-\gamma-2\delta)}{4}\;u(z)=0,
¨\end{array}
\end{eqnarray}
whose parameters are $\mathrm{a}=\beta/2$, $\mathrm{b}=(\gamma+2\delta-\beta-1)/2$
and $\mathrm{c}=(1+\gamma)/2$.

%
%
%

\section{Some degenerate power-series solutions}

\protect\label{C}
\setcounter{equation}{0}
\renewcommand{\theequation}{C.\arabic{equation}}
We consider three cases of the Schr\"odinger equation ({\ref{associado-gan}) 
whose solutions are obtained as particular cases of the power-series solutions
given in Section 5.  For each case, we will find
 \begin{itemize}
 \itemsep-1pt
\item  
 degenerate finite-series eigenfunctions which satisfy the Arscott criterion;
 \item   
 four infinite-series eigenfunctions. 
 \end{itemize}
The cases are: (1)
 $\bm{l}=1/2$ and $\bm{m}=3/2$, (2) $\bm{l}=1/2$ and $\bm{m}=5/2$, 
 (3) $\bm{l}=1/2$ and $\bm{m}=7/2$. 
 For these cases  the finite series have at most three terms whose coefficients are computed from the recurrence relations
 (\ref{recurrence1}) by means of 
  \begin{eqnarray}
b_1=-b_0\;\frac{\beta_0}{\alpha_0}; 
\qquad b_2=b_0\;\frac{1}{\alpha_1}
\left(\frac{\beta_0\beta_1}{\alpha_0}-\gamma_1\right),
 \end{eqnarray}
where $b_0$ is the coefficient of the first term. The characteristic equations (\ref{matriz}) reduces to
 \begin{eqnarray}
\beta_0=0,\qquad  \beta_0\;\beta_1-\alpha_0\;\gamma_1=0, \qquad  
\beta_0\;\beta_1\;\beta_2-\alpha_1\; \beta_0\;\gamma_2-\alpha_0 \;\beta_2\;\gamma_1 =0,
 \end{eqnarray}
for series with one, two and three terms respectively.

 Khare and Sukhatme \cite{khare,khare2} have found solutions for these
 cases by integrating the  Schr\"odinger equation directly.
Here, the solutions are obtained from the expansions in series of $\mathrm{cn}^2u$ ($\mathring{\bm\Psi}^{(i)}$ and $\tilde{\bm\Psi}^{(i)}$ written in
{Section 5.2)}; these satisfy the Arscott condition for series with  two or more terms,  in opposition to $\mathring\psi^{(i)}$ and  $\tilde\psi^{(i)}$.

	%
	%

%
\subsection*{First case: $\bm{l}=1/2$ and $\bm{m}=3/2$ }
For this case, $\bm{m}-\bm{l}=1$ and $\bm{m}+\bm{l}=2$. Thence, (\ref{ps1-C}-\ref{ps8-D}}) imply the 
%
\begin{eqnarray}\mbox{finite-series expansions}:\;
\mathring{\bm\Psi}^{(1)}\;(1 \mbox{ term}),\quad 
\tilde{\bm\Psi}^{(5)}\;(2 \mbox{ terms}),\quad
\tilde{\bm\Psi}^{(8)}\;(1 \mbox{ term}),
\end{eqnarray}
while (\ref{1}-\ref{4}) give four 
\begin{eqnarray}\mbox{infinite-series expansions}:\;
\mathring{\Phi}^{(2)} \stackrel{\text{(\ref{ince})}}{=}
	\mathring{\Phi}^{(6)}, \quad
	\mathring{\Phi}^{(3)}\stackrel{\text{(\ref{ince})}}{=}\mathring{\Phi}^{(7)},\quad
	\mathring{\Phi}^{(4)},\quad
	\mathring{\Phi}^{(5)}.
\end{eqnarray}
The one-term expansions and their respective energies read
\begin{eqnarray}
&&\begin{array}{l}
\mathring{\bm\Psi}^{(1)}(u)\stackrel{\text{(\ref{ps1-C}})}{=}
\bm{b}_{0}^{(1)}\;\mathrm{dn}^{3/2}u,\hspace{1cm} {\cal E}=\frac{9}{4} k^2;\end{array}\vspace{2mm}\\
&&\begin{array}{l}
 \tilde{\bm\Psi}^{(8)}(u)\stackrel{\text{(\ref{ps2-D}})}{=}
\bm{b}_{0}^{(8)}\;\frac{\mathrm{sn}u\;\mathrm{cn}u}{\mathrm{dn}^{1/2}u},
\;\;\qquad{\cal E}=4+\frac{1}{4} k^2.
\end{array}
\end{eqnarray}
The two-term expansion $\mathring{\bm\Psi}^{(5)}(u)$ represents two solutions, but one is equivalent to the
previous $\mathring{\bm\Psi}^{(1)}(u)$, that is,
\begin{eqnarray}
&&\begin{array}{l}
\tilde{\bm\Psi}^{(5)}_{\;1}(u)\stackrel{\text{(\ref{ps5-D}})}{=}
\frac{C_1}{\mathrm{dn}^{1/2}u}\left(1+\frac{k^2}{1-k^2}\mathrm{cn}^2u\right)=
\frac{C_1}{(1-k^2)\mathrm{dn}^{1/2}u}\left(1-k^2\mathrm{sn}^2u\right)
\stackrel{\text{(\ref{imply}})}{=}
\mathring{\bm\Psi}^{(1)}(u)
,
\end{array}\nonumber
\vspace{2mm}\\
&&\begin{array}{l}
\tilde{\bm\Psi}^{(5)}_{\;2}(u)\stackrel{\text{(\ref{ps5-D}})}{=}
\frac{C_2}{\mathrm{dn}^{1/2}u}\left(1-2\mathrm{cn}^2u\right)
,\hspace{1.5cm} {\cal E}=4+\frac{1}{4} k^2.
\end{array}
\end{eqnarray}
Therefore, there are three independent eigensolutions. 
$\tilde{\bm\Psi}^{(8)}(u)$ and $\tilde{\bm\Psi}^{(5)}_{\;2}(u)$
are degenerate because have the same energy.	
%
%

\subsection*{Second case: $\bm{l}=1/2$ and $\bm{m}=5/2$ }

Now $\bm{m}-\bm{l}=2$ and $\bm{l}+\bm{l}=3$. Then,
proceeding as in the above case,  we obtain the 
\begin{eqnarray}
&&\begin{array}{l}
\mbox{finite series}:\;
\mathring{\bm\Psi}^{(2)}\mbox{ and }
\mathring{\bm\Psi}^{(3)}\;(1 \mbox{ term}),\;
\tilde{\bm\Psi}^{(6)} \mbox{ and } 
\tilde{\bm\Psi}^{(7)}\;(2 \mbox{ terms});\quad\end{array}\vspace{3mm}\\
&&\begin{array}{l}
\mbox{infinite series }:\;
\mathring{\Phi}^{(1)},\quad
	\mathring{\Phi}^{(4)},\quad
	\mathring{\Phi}^{(5)},\quad
	\mathring{\Phi}^{(8)}.\end{array}
\end{eqnarray}
	The one-term finite series give two eigenfunctions, 
\begin{eqnarray}
&&\displaystyle\mathring{\bm\Psi}^{(2)}(u)\stackrel{\text{(\ref{ps2-C}})}{=}
\bm{b}_{0}^{(2)}\;\mathrm{sn}u\;\mathrm{dn}^{3/2}u,\hspace{8mm} {\cal E}=1+\frac{25}{4} k^2;\qquad\vspace{2mm}\\
&&\displaystyle \mathring{\bm\Psi}^{(3)}(u)\stackrel{\text{(\ref{ps3-C}})}{=}
\bm{b}_{0}^{(3)}\; \mathrm{cn}u\;{\mathrm{dn}^{3/2}u},
\qquad{\cal E}=1+\frac{9}{4} k^2.
\end{eqnarray}
The two-term series $\tilde{\bm\Psi}^{(6)}(u)$ represents two eigenfunctions but one of them repeats
the above solution $\mathring{\bm\Psi}^{(2)}(u)$, 
\begin{eqnarray}
&&
\tilde{\bm\Psi}^{(6)}_{\;1}(u)\stackrel{\text{(\ref{ps6-D}})}{=}
C_1\;\frac{ \mathrm{sn}u}{\mathrm{dn}^{1/2}u}\left(1+\frac{k^2}{1-k^2}\mathrm{cn}^2u\right)=\mathring{\bm\Psi}^{(2)}(u),
\nonumber \\
&& \tilde{\bm\Psi}^{(6)}_{\;2}(u)\stackrel{\text{(\ref{ps6-D}})}{=}
C_2\;\frac{ \mathrm{sn}u}{\mathrm{dn}^{1/2}u}\left(1-4\mathrm{cn}^2u\right),\qquad
{\cal E}=9+\frac{1}{4} k^2.
\end{eqnarray}
Similarly, the two-term series $\tilde{\bm\Psi}^{(7)}(u)$ represent two eigenfunctions but one of them repeats
the above solution $\mathring{\bm\Psi}^{(3)}(u)$, 	
\begin{eqnarray}
&&
\begin{array}{l}
\displaystyle
\tilde{\bm\Psi}^{(7)}_{\;1}(u)\stackrel{\text{(\ref{ps7-D}})}{=}
D_1\; \frac{ \mathrm{cn}u}{\mathrm{dn}^{1/2}u}
\left(1+\frac{k^2}{1-k^2}\mathrm{cn}^2u\right)=
\displaystyle\mathring{\bm\Psi}^{(3)}(u)
,
\end{array}
\nonumber
\vspace{	3mm}\\  \nonumber\\
&&\displaystyle
\begin{array}{l}
 \tilde{\bm\Psi}^{(7)}_{\;2}(u)\stackrel{\text{(\ref{ps7-D}})}{=}
 D_2\;\frac{ \mathrm{cn}u}{\mathrm{dn}^{1/2}u}\left(1-\frac{4}{3}\mathrm{cn}^2u\right),\qquad
{\cal E}=9+\frac{1}{4} k^2 \mbox{ (degenerate ver acima)};.
\end{array}
\end{eqnarray}
Therefore, in this case	there are only four independent solutions, being
$\tilde{\bm\Psi}^{(6)}_{\;2}(u)$ and $ \tilde{\bm\Psi}^{(7)}_{\;2}(u)$ degenerate solutions.

%
%

\subsection*{Third case: $\bm{l}=1/2$ and $\bm{m}=7/2$ }

In this case $\bm{m}-\bm{l}=3$ and $\bm{m}+\bm{l}=4$ . Thence we get the
the following finite and infinite series:
\begin{eqnarray}
&&\mbox{finite}:\;
\mathring{\bm\Psi}^{(1)}\;(2 \mbox{ terms}),\;
\mathring{\bm\Psi}^{(4)}\;(1 \mbox{ term}),\;
\tilde{\bm\Psi}^{(5)}\;(3 \mbox{ terms}),\;
\tilde{\bm\Psi}^{(8)}\;(2 \mbox{ terms});\qquad
\vspace{3mm}\\
&&\mbox{infinite}:\;
\mathring{\Phi}^{(2)},\quad
	\mathring{\Phi}^{(3)},\quad
	\mathring{\Phi}^{(6)},\quad
	\mathring{\Phi}^{(7)}.
\end{eqnarray}
	The one-term series reads
\begin{eqnarray}
\begin{array}{l}\displaystyle
\mathring{\bm\Psi}^{(4)}(u)\stackrel{\text{(\ref{ps4-C}})}{=}
\bm{b}_{0}^{(4)}\;\mathrm{sn}u\;\mathrm{cn}u\;\mathrm{dn}^{3/2}u,\hspace{1cm} {\cal E}=4+\frac{25}{4} k^2.
\end{array}
\end{eqnarray}
The two-term expansions $\mathring{\bm\Psi}^{(1)}(u)$ represent two
eigenfunctions $\mathring{\bm\Psi}^{(1)}_{\;\pm}(u)$ corresponding to the energies ${\cal E}_{\pm}$,
\begin{eqnarray}
&&\begin{array}{l}
\mathring{\bm\Psi}^{(1)}_{\;\pm}(u)
\stackrel{\text{(\ref{ps1-C}})}{=}
C_{\pm}\;{\mathrm{dn}^{3/2}u}
\left[1+\frac{7k^2-2\pm\sqrt{4+25k^4-4k^2}}{2(1-k^2)}\;\mathrm{cn}^2u\right],\end{array} \quad\vspace{3mm}\\
&& \begin{array}{l}
\hspace{3cm}{\cal E}_{\pm}=2+\frac{29}{4}k^2\mp\sqrt{4+25k^4-4k^2}.\nonumber
\end{array}
\end{eqnarray}
%
%
%
%
The expansion $\mathring{\bm\Psi}^{(8)}(u)$ also represents two
eigenfunctions but one of these is identical to $\mathring{\bm\Psi}^{(4)}(u)$. We  find	
\begin{eqnarray}
&&
\begin{array}{l}
\displaystyle
\tilde{\bm\Psi}^{(8)}_{\;1}(u)\stackrel{\text{(\ref{ps8-D}})}{=}
D_1\; \frac{\mathrm{sn}u\; \mathrm{cn}u}{\mathrm{dn}^{1/2}u}
\left[1+\frac{k^2}{1-k^2}\mathrm{cn}^2u\right]=
\displaystyle\mathring{\bm\Psi}^{(4)}(u)
,
\end{array}
\nonumber
\vspace{	3mm}\\  \nonumber\\
&&
\begin{array}{l}
\displaystyle
 \tilde{\bm\Psi}^{(8)}_{\;2}(u)\stackrel{\text{(\ref{ps8-D}})}{=}
 D_2\;\frac{\mathrm{sn}u\; \mathrm{cn}u}{\mathrm{dn}^{1/2}u}\left[1-2\mathrm{cn}^2u\right],\qquad
{\cal E}=16+\frac{1}{4} k^2 .
\end{array}
\end{eqnarray}
On the other hand, as $\mathring{\bm\Psi}^{(4)}(u)$ is a three-term expansion, it represents three eigensolutions
having the form
\begin{eqnarray}
&&\begin{array}{l}\displaystyle
\tilde{\bm\Psi}^{(5)}(u)\stackrel{\text{(\ref{ps5-D}})}{=}
\frac{\tilde{\bm{b}}_{\;0}^{(5)}}{\mathrm{dn}^{1/2}u}
\left\{1-\frac{2k^2}{k^2-1}\;\tilde{\bm{\beta}}_{0}^{(5)}\;\mathrm{cn}^2u+
\right.\end{array}\nonumber\\
&&\hspace{2.2cm}\begin{array}{l}
\left.
\frac{k^2}{3(k^2-1)}\left[4+\frac{2k^2}{(k^2-1)}\;\tilde{\bm{\beta}}_{0}^{(5)}\left(\tilde{\bm{\beta}}_{0}^{(5)}
+\frac{1}{k^2}-\frac{3}{2}\right)\right]
 \mathrm{cn}^4u\right\}, 
\end{array}
\end{eqnarray}
where $\tilde{\bm{\beta}}_{0}^{(5)}$ is determined from the characteristic equation			
\begin{eqnarray}\label{cha}
\begin{array}{l}
\left[\tilde{\bm{\beta}}_{0}^{(5)}\right]^3+ \left(\frac{5}{k^2}-\frac{17}{2}\right)\left[\tilde{\bm{\beta}}_{0}^{(5)}\right]^2
+ \left(\frac{43k^4-48k^2+8}{2k^4}\right)\tilde{\bm{\beta}}_{0}^{(5)}
+2\left(1-\frac{1}{k^2}\right)
\left(\frac{4}{k^2}-7\right)=0,\end{array}\\
\begin{array}{l}
\tilde{\bm{\beta}}_{0}^{(5)}= \frac{65}{16}-\frac{{\cal E}}{4k^2}.\nonumber
\end{array}
\end{eqnarray}
The Arscott condition implies three real and distinct values for
$\tilde{\bm{\beta}}_{0}^{(5)}$. We find that,	
%
$\mbox{if }{\cal E}=16+k^2/4\mbox{ then }
\tilde{\bm{\beta}}_{0}^{(5)}= 4(k^2-1)/{k^2}$	
%
and Eq. (\ref{cha})	is fulfilled. The eigenfunction takes the form
\begin{eqnarray}
\begin{array}{l}\displaystyle
\tilde{\bm\Psi}^{(5)}_{\;1}(u)=\frac{\bar{D}_1}{\mathrm{dn}^{1/2}u}
\left(1-8\;\mathrm{sn}^2u\; \mathrm{cn}^2u\right), \qquad
{\cal E}=16+\frac{k^2}{4}
\end{array}
\end{eqnarray}
and in conjunction with $ \tilde{\bm\Psi}^{(8)}_{\;2}(u)$  constitute a pair of degenerate eigenfunctions as stated in
\cite{khare}. In fact, the solution mentioned in \cite{khare} 
comes from expansions in series of $\mathrm{sn}^2u$, namely,
\begin{eqnarray*}
\begin{array}{l}\displaystyle
\tilde{\psi}^{(5)}(u)
\stackrel{\text{(\ref{ps5-B}})}{=}
\frac{\tilde{{b}}_{\;0}^{(5)}}{\mathrm{dn}^{1/2}u}
\left\{1-\frac{1}{2}\left({\cal E}-\frac{k^2}{4}\right)
\mathrm{sn}^2u+ \right.\vspace{3mm} \\
\hspace{2.2cm}\left.
\frac{k^2}{3}\left[4+\frac{1}{8k^2}
\left({\cal E}-\frac{k^2}{4}\right)\left({\cal E}-4
-\frac{9}{4}k^2\right)\right]
 \mathrm{sn}^4u\right\}, 
\end{array}
\end{eqnarray*}
which satisfy the characteristic equation			
\begin{eqnarray*}
\begin{array}{l}
\lambda^3-4 \left(\frac{7}{2}k^2+5\right)\lambda^2
+ 16\left(\frac{3}{2}k^4+16k^2+4\right)\lambda
-128k^2 
\left(3{k^2}+4\right)=0,\qquad 
\lambda= {\cal E}-\frac{k^2}{4}.
\end{array}
\end{eqnarray*}
These solutions do not obey the Arscott condition, but we can see that $\tilde{\psi}^{(5)}(u)=\tilde{\bm\Psi}^{(5)}(u)$  if 
${\cal E}=16+k^2/4$.


%
%
%
%
	%
%
%
%

\section{Svartholm's and Erdelyi's solutions}
\protect\label{D}
\setcounter{equation}{0}
\renewcommand{\theequation}{D.\arabic{equation}}


Svartholm \cite{svartholm}
and Erd\'elyi \cite{erdelyi2,erdelyi3} have supposed that the coefficients of their expansions in series of hypergeometric functions
satisfy recurrence relations having
the form given in (\ref{recurrence1}).
Such conjecture has been repeated in subsequent literature \cite{nist,kristensson,arscott-ronveaux,{erdelyi-3}}, including Erd\'elyi himself \cite{erdelyi-3}. However, there are three forms for the recurrence relations, namely,
%
%
\begin{align}
&\begin{array}{l}\label{r1}
\alpha_{0}b_{1}+\beta_{0}b_{0}=0,\qquad
\alpha_{n}b_{n+1}+\beta_{n}b_{n}+\gamma_{n}b_{n-1}=0,
\quad (n\geq 1),
\end{array}\vspace{3mm}\\
&\begin{array}{r}\label{r2}
\begin{cases}
\alpha_{0}b_{1}+\beta_{0}b_{0}=0,
\quad
\alpha_{1}b_{2}+\beta_{1}b_{1}+
\left[\gamma_{1}+\alpha_{-1}\right]b_{0}=0,\quad\vspace{2mm}\\
\alpha_{n}b_{n+1}+\beta_{n}b_{n}+\gamma_{n}b_{n-1}=0,\quad
(n\geq2),
\end{cases}
\end{array}\vspace{3mm}\\
&\begin{array}{l}\label{r3}
\alpha_{0}b_{1}+\left[\beta_{0}+\alpha_{-1}\right]b_{0}=0,
\qquad
\alpha_{n}b_{n+1}+\beta_{n}b_{n}+\gamma_{n}b_{n-1}=0,\qquad(n\geq 1)
\end{array}
\end{align}
%
which reduce to  (\ref{recurrence1}) only if 
$\alpha_{-1}=0$.

The above 
recurrences also occur for some expansions in series of hypergeometric functions and in series of Coulomb wave functions
for solutions of confluent and double-confluent 
Heun equations \cite{eu-2002}. In fact,
the procedure used in  Appendix A of reference  \cite{eu-2002} can be easily adapted
for the expansions of Svartholm 
and Erd\'elyi. 


As an example we consider the  
Erd\'elyi  solutions \cite{erdelyi3}
%
%
%
\begin{eqnarray}\label{erdelyi-00}
\begin{array}{l}
H(x)=\displaystyle\sum_{n=0}^{\infty}b_{n}P_n(x)=\displaystyle\sum_{n=0}^{\infty}b_{n}
F(n+\lambda,-n+\mu;\gamma; x),   \end{array}\vspace{2mm}\\
\left[\gamma\neq 0,-1,-2,\cdots,\quad \mu-\lambda\neq 1,2,3,\cdots\right]\nonumber
\end{eqnarray}
where $\lambda$ and $\mu$ depend on the parameters of the Heun equation and  are connected  by
$\lambda+\mu=\gamma+\delta-1$. The condition $\mu-\lambda\neq1,2,3,\cdots$ avoids two identical hypergeometric functions for different values of $n$. 
For $H(x)$ given in  (\ref{erdelyi-00}), Eq. (\ref{heun}) leads to
%
\begin{eqnarray}\label{erdelyi-2}
\displaystyle\sum_{n=0}^{\infty}{b}_{n}
\left[ \alpha_{n-1}\ P_{n-1}(x)
+\beta_{n}\ P_{n}(x)+
\gamma_{n+1}\ P_{n+1}(x)\right]
=0, 
\end{eqnarray}
which can be written in  Erd\'elyi's  notation \cite{erdelyi3} by taking  $x=z$, $b_n=c_n$, $\alpha_n=	M_n$,  $\beta_n=L_n$ and 
$\gamma_n=K_n$: 
for example \cite{nist}
\begin{eqnarray*}\begin{array}{l}
\alpha_{n}=M_n=-\frac{(n-\alpha+\lambda+1)(n-\beta+\lambda+1)
	(n-\gamma+\lambda+1)(n-\mu+1)}
{\left(2n+\lambda-\mu+1\right)
	\left(2n+\lambda-\mu+2\right)}.\end{array}
\end{eqnarray*}
In order to get the three forms for the recurrence relations, we  rewrite Eq. (\ref{erdelyi-2}) as
\begin{eqnarray}\label{expandido}
&&\alpha_{-1}\ b_{0}\ P_{-1}(x)+
\big[\alpha_{0}\ b_{1}+\beta_{0}\ b_{0}\big]P_{0}(x)
+\big[\alpha_{1}\ b_{2}+\beta_{1}\ b_{1}+
\gamma_{1}\ b_{0}\big]P_{1}(x)\nonumber\\
&&+\displaystyle\sum_{n=2}^{\infty}
\big[\alpha_{n}\ b_{n+1}+\beta_{n}\ b_{n}+
\gamma_{n}\ b_{n-1}\big]
P_{n}(x)=0.
\end{eqnarray}
After this, we specify $\lambda$ and $\mu$ and equate to zero 
the coefficients of each independent function $P_n(x)$. Next we
regard only two choices for ($\lambda,\mu$).



In the first place, for the expansion mentioned at the end of Section 6  we 
take $\lambda=\alpha$ and   $\mu=\gamma+\delta-1-\alpha$. In this manner
we obtain the expansion $H^{(1)}(x)$,
\begin{eqnarray}\label{erdelyi-1}
\begin{array}{l}
H^{(1)}(x)=\displaystyle\sum_{n=0}^{\infty}b_{n}^{(1)}
F(n+\alpha,
-n-\alpha-1+\gamma+\delta;\gamma;x),\end{array}\quad\\
\begin{array}{l}
\left[\gamma \mbox{ and } 2+2\alpha-\gamma-\delta\neq 0,-1,-2,\cdots\right]\end{array}, \nonumber
\end{eqnarray}
%
for which the recurrence relations
for $b_n^{(1)}$ are given by 
%
%
\begin{eqnarray}\label{erdelyi-rec}
&&\mbox{Eqs.}\; (\ref{r1})\; \mbox{ if }\; 2\alpha-\gamma-\delta\neq0,\ -1; \qquad
\mbox{Eqs.}\; (\ref{r2}) \mbox{ if }  2\alpha-\gamma-\delta=-1;\nonumber
\vspace{2mm}\\
&&\mbox{Eqs.}\; (\ref{r3})\; \mbox{ if } \; 2\alpha-\gamma-\delta=0 ,
\end{eqnarray}
%
with $\alpha_{n}$, $\beta_{n}$ and $\gamma_{n}$
defined as \cite{erdelyi3}
\begin{eqnarray}\label{erdelyi-3}
\begin{cases}
%
%
\begin{array}{l}
\alpha_{n}^{(1)}=-\frac{(n+1)(n+1+\alpha-\beta)
	(n+1+\alpha-\gamma)(n+2+\alpha-\gamma-\delta)}
{\left(2n+2+2\alpha-\gamma-\delta\right)
	\left(2n+3+2\alpha-\gamma-\delta\right)},
	\hspace{3.5cm} [n\geq -1]
\vspace{3mm}\\
\beta_{n}^{(1)}=\left(\frac{1}{2}-a\right)\Big[n(n+2\alpha-\gamma-\delta+1)+
\alpha(\alpha+1-\gamma-\delta)\Big]-q+\frac{1}{2}\alpha\beta
+\vspace{2mm}\\
\hspace{1cm}
\frac{1}{8}
(\gamma-\delta)\left[2\alpha+2\beta-\gamma-\delta+
 \frac{(\gamma+\delta-2)(2\alpha-\gamma-\delta)(2\beta-\gamma-\delta)}
{(2n+2\alpha-\gamma-\delta)(2n+2+2\alpha-\gamma-\delta)}\right],\quad [n\geq 0]\vspace{3mm}\\
\gamma_{n}^{(1)}=-
\frac{(n+2\alpha-\gamma-\delta)
	(n+\alpha-\delta)
	(n+\alpha-1)(n+\alpha+\beta-\gamma-\delta)}
{(2n+2\alpha-\gamma-\delta)(2n+2\alpha-\gamma-\delta-1)},\hspace{3.4cm}[n\geq 1]
\end{array}
\end{cases}
\end{eqnarray}
%
%
As to the recurrence relations which are missing  in 
Erd\'ely, they result because there are terms linearly dependent in  
Eq. (\ref{expandido}), namely,
\begin{eqnarray*}
&&\begin{array}{l}
\mbox{if }     2\alpha-\gamma-\delta=-1 , \quad P_{-1}(x)=P_1(x)=F(\alpha-1,\alpha+1;\gamma;x)  \Rightarrow  \mbox{Eqs.}\; (\ref{r2}) ;\end{array}  \vspace{2mm}\\
&&\begin{array}{l}
\mbox{if } 2\alpha-\gamma-\delta=0,\qquad P_{-1}(x)=P_0(x)=F(\alpha-1;\alpha;\gamma;x)    \Rightarrow  \mbox{Eqs.}\; (\ref{r3}).  \end{array}
\end{eqnarray*}
 Relations (\ref{erdelyi-rec}) 
take the place
of (5.2), (5.3) and (5.4) given in  Erd\'elyi's paper
\cite{erdelyi3}. 
%
%
%

In the second place,  the Svartholm  solution arises when we put
$\lambda=\gamma+\delta-1$ and $\mu=0$  in (\ref{erdelyi-00}). Then the Erd\'elyi  
expansion  (\ref{erdelyi-00}) becomes
\begin{eqnarray}\label{svartholm}
&&\begin{array}{l}
\tilde{H}^{(1)}(x)=
\displaystyle\sum_{n=0}^{\infty}\tilde{b}_{n}^{(1)}\
F\left( n+\gamma+\delta-1,-n;
\gamma;x\right),\end{array}
\vspace{2mm}\\
&&\hspace{2cm}
\begin{array}{l}
\left[\gamma\neq 0,-1,-2,\cdots; \quad \gamma+\delta\neq 0,-1,-2, \cdots \right]\nonumber
\end{array}
\end{eqnarray}
where the coefficients $\tilde{b}_{n}^{(1)}$ satisfy the recursions
\begin{eqnarray}\label{svartholm-rec}
&&\mbox{Eqs.}\; (\ref{r1})\; \mbox{ if }\; \gamma+\delta\neq 1,2; \qquad
\mbox{Eqs.}\; (\ref{r2}) \mbox{ if }  \gamma+\delta=1;\nonumber
\vspace{2mm}\\
&&\mbox{Eqs.}\; (\ref{r3})\; \mbox{ if } \; \gamma+\delta=2 ,
\end{eqnarray}
with
\begin{eqnarray}\label{svartholm-2}
\begin{cases}
\begin{array}{l}
\tilde\alpha_{n}^{(1)}=
-\frac{\left( n+1\right) \left( n+\gamma+\delta-\alpha\right)
\left( n+\gamma+\delta-\beta\right)
\left( n+\delta\right) }
{\left(2n+\gamma+\delta \right)\left(2n+\gamma+\delta+1\right)},
\hspace{4.5cm}
[n\geq -1]
\vspace{2mm}\\
\tilde\beta_{n}^{(1)}=
\left[\frac{1}{2} -a\right]
 n\left[ n+\gamma+\delta-1\right]-q+\frac{1}{2}\alpha\beta+
\vspace{2mm}\\
\hspace{1.1cm}
\frac{1}{8}(\gamma-\delta)
\Big[
2\alpha+2\beta-\gamma-\delta 
+
\frac{(\gamma+\delta-2)(2\alpha-\gamma-\delta)
(2\beta-\gamma-\delta)}
{(2n+\gamma+\delta-2)(2n+\gamma+\delta)}\Big],
\quad [n\geq 0]\vspace{2mm}\\
\tilde\gamma_{n}^{(1)}=-\frac{\left( n+\alpha-1\right)
\left( n+\beta-
1\right) \left( n+\gamma+\delta-2\right)
\left( n+\gamma-1\right) }
{\left(2n+\gamma+\delta-3\right) \left(2n+\gamma+\delta-2\right) },
\hspace{4.2cm} [n\geq 1]
\end{array}
\end{cases}
\end{eqnarray}
%

%

By the homotopic transformations we can generate eight 
Svartholm-type solutions,  $\tilde{H}^{(i)}(x)$. These give series
of trigonometric functions for the associated Lam\'e equation. 
For this, we consider the definitions of the  basic elliptic  functions \cite{erdelyi-2}: 
\begin{eqnarray} \label{defs}
\begin{array}{l}
\operatorname {sn} u = \sin{v},\;\;
\operatorname {cn} u = \cos{v},\;\;
\operatorname {dn} u = \sqrt {1-k^2\sin^2 v}
\;
\mbox{for}\;
{u}
=\int_0^v
\frac{ d\theta} 
{\sqrt {1-k^2 \sin^2 \theta}}.
\end{array}
\end{eqnarray}
In addition, for the associated Lam\'e equation ($x=\operatorname {sn}^2 u=\sin^2v$), the relations \cite{nist}
\begin{eqnarray}\label{fourier}
&\begin{array}{l}
F\left(-a,a;\frac{1}{2}; \sin^2v\right)=\cos(2av),\qquad
F\left(a,1-a;\frac{3}{2}; \sin^2v\right)=
\frac{\cos[(2a-1)v]}
{\cos{v}},\end{array}
\nonumber\vspace{2mm}\\
&
\begin{array}{l}
F\left(1-a,a;\frac{3}{2}; \sin^2v\right)=\frac{\sin[(2a-1)v]}{(2a-1)\sin{v}},\quad
F\left(a,2-a;\frac{3}{2}; \sin^2v\right)=
\frac{\sin[(2a-2)v]}{(a-1)\sin(2v)},
\end{array}
\end{eqnarray}
allow to get the trigonometric series. 
For example, for $\gamma=\delta
 ={1}/{2}$ we find 
\begin{eqnarray}\label{trigonometric}
\begin{array}{l}
\tilde{H}^{(1)}(x)=
\displaystyle\sum_{n=0}^{\infty}\begin{array}{l}
\tilde{b}_{n}^{(1)}
F\left(-n,n;\frac{1}{2};x\right)=
\displaystyle\sum_{n=0}^{\infty}\tilde{b}_{n}^{(1)}\cos{(2nv)},\quad x=\operatorname {sn}^2 u = \sin^2{v},\end{array} 
%
\end{array}
\end{eqnarray}
where the coefficients $\tilde{b}_{n}^{(1)}$  satisfy  the relations (\ref{r2}) with
\begin{eqnarray}
\begin{cases}
\begin{array}{l}
\tilde\alpha_{n}^{(1)}=-\frac{1}{4}
\left( n+1-\alpha\right)
\left( n+1-\beta\right),\quad
[n\geq -1], \qquad
\vspace{2mm}\\
\tilde\beta_{n}^{(1)}=
\left(\frac{1}{2} -a\right)
 n^2-q+\frac{1}{2}\alpha\beta,
\qquad [n\geq 0]\vspace{2mm}\\
\tilde\gamma_{n}^{(1)}=-\frac{1}{4}{\left( n+\alpha-1\right)
\left( n+\beta-
1\right)  },
\qquad [n\geq 1]
\end{array}
\end{cases}
\end{eqnarray}
%
%
%
%
%
%
%
where $a=1/k^2$. For the Lam\'e equation ($\gamma=\delta=\epsilon
 ={1}/{2}$), the expansions in trigonometric functions are equivalent to  the solutions found by Ince in 1940 \cite{ince3}.

%
%
%
Finally we find that, by performing the limits (\ref{formal})  
	for the confluent Heun equation (\ref{CHE}), the solutions (\ref{erdelyi-1})-(\ref{erdelyi-rec})
	 and (\ref{svartholm})-(\ref{svartholm-rec}) remain formally unaltered while the coefficients  (\ref{erdelyi-2}) and   (\ref{svartholm-2}) give, respectively, 
%
%
%
%
%
%
%
%
\begin{eqnarray}\label{che-11}
\begin{cases}
\begin{array}{l}
\alpha_{n}^{(1)}=-\rho\;\frac{(n+1)
	(n+1+\alpha-\gamma)(n+2+\alpha-\gamma-\delta)}
{\left(2n+2+2\alpha-\gamma-\delta\right)
	\left(2n+3+2\alpha-\gamma-\delta\right)},
	\hspace{3.5cm} [n\geq -1],
\vspace{3mm}\\
\beta_{n}^{(1)}=-n(n+2\alpha-\gamma-\delta+1)-
\alpha(\alpha+1-\gamma-\delta)+\sigma-\frac{1}{2}\alpha\rho
+\vspace{2mm}\\
\hspace{1.3cm}\frac{1}{4}\rho
(\gamma-\delta)\left[
 \frac{(\gamma+\delta-2)(\gamma+\delta-2\alpha)}
{(2n+2\alpha-\gamma-\delta)(2n+2+2\alpha-\gamma-\delta)}-1\right],\hspace{1.7cm} [n\geq 0],\vspace{3mm}\\
\gamma_{n}^{(1)}=\rho\; \frac{(n+2\alpha-\gamma-\delta)
	(n+\alpha-\delta)
	(n+\alpha-1)}
{(2n+2\alpha-\gamma-\delta)(2n+2\alpha-\gamma-\delta-1)},\hspace{4.5cm} [n\geq 1].
\end{array}
\end{cases}
\end{eqnarray}
\begin{eqnarray}\label{dche-2}
\begin{cases}
\begin{array}{l}
\tilde\alpha_{n}^{(1)}=
-\rho\frac{\left( n+1\right) \left( n+\gamma+\delta-\alpha\right)
\left( n+\delta\right) }
{\left(2n+\gamma+\delta \right)\left(2n+\gamma+\delta+1\right)},
\hspace{4.5cm}
[n\geq -1]
\vspace{2mm}\\
\tilde\beta_{n}^{(1)}=-
 n\left[ n+\gamma+\delta-1\right]+\sigma-\frac{1}{2}\alpha\rho-
\vspace{2mm}\\
\hspace{1.1cm}
\frac{1}{4}\rho(\gamma-\delta)
\Big[1
+
\frac{(\gamma+\delta-2)(2\alpha-\gamma-\delta)}
{(2n+\gamma+\delta-2)(2n+\gamma+\delta)}\Big],
\quad [n\geq 0]\vspace{2mm}\\
\tilde\gamma_{n}^{(1)}=\rho\frac{\left( n+\alpha-1\right)
 \left( n+\gamma+\delta-2\right)
\left( n+\gamma-1\right) }
{\left(2n+\gamma+\delta-3\right) \left(2n+\gamma+\delta-2\right) },
\hspace{4.2cm} [n\geq 1]
\end{array}
\end{cases}
\end{eqnarray}
Note that the three forms for the recurrence relations for the CHE have already been 
obtained \cite{eu-2002} independently of the solutions for the Heun equation, as aforementioned.

%
%


\begin{thebibliography}{plain}
	%
	%
	\bibitem{darboux}Darboux, G.: Sur une \'equation lin\'eaire.  
	{ C. R. Acad. Sci.} { 94}, 1645-1648 (1882).
	%
	\bibitem{humbert}Humbert, P.: { Fonctions de Lam\'e  et fonctions de
		Mathieu.}  M\'emorial des Sciences Math\'ematique, Vol. X, Gauthier-Villards (1926).
	%
	%
		\bibitem{nist} 
	Olver, F.W.J., Lozier, D.V.,
 Boisvert,  R.F., Clark  C.W. (editors): { NIST Handbook of Mathematical Functions}. Cambridge University Press (2010).
	%
	%
	\bibitem{heun}Heun, K.: Zur Theorie der Riemannschen
	Functionen zweiter Ordnung mit vier Verzweigungspunkten. 
	{Math. Ann.}{ 33}, 161-179 (1899).
	%
	\bibitem{maier}Maier, R.S.: {The 192 solutions of the Heun equation}. { Mathematics of Computation} { 76}, 811-843 (2007).
	%
	%
	\bibitem{erdelyi1}Erd\'elyi, A., Magnus, W., Oberhettinger, F.,  Tricomi, F.G.: 
	{Higher Transcendental Functions}  
	vol. 1. McGraw-Hill, New York (1953).
	%
		%
\bibitem{maier-2}Maier, R.S.: On reducing the Heun equation to the hypergeometric equation. {J. Diff. Equations} { 213}, 171-203 (2005).
%
\bibitem{vidunas}Vidunas, R., Filipuk, G.: Parametric transformations between the Heun and Gauss hypergeometric functions. {  Funkcialaj Ekvacioj} { 56}, 271-321 (2013). 
%
	%
	\bibitem{kristensson}
Kristensson, G.:{ Second Order Differential
	Equations: 
	Special Functions and Their Classification}. Springer (2010).
	%
	\bibitem{turbiner1}
Turbiner, A.V.: Quasi-exactly-solvable problems and $sl(2)$
algebra. { Commun. Math. Phys.} { 118}, 467-474 (1988).
		%
\bibitem{ushveridze1}Ushveridge, A.G.: Quasi-exactly solvable models in
quantum mechanics. { Sov. J. Part. Nucl.}  
{ 20}, 504-528 (1989). 
%
			%
\bibitem{kalnins}Kalnins, E.G, Miller, M.,  Pogosyan, G.S.: Exact and quasiexact solvability of second-order superintegrable
quantum systems: I. Euclidian space preliminaries.   
{J. Math. Phys.} {47}, 033502 (2006). 
%
\bibitem{svartholm} Svartholm, N.: Die L\"{o}sung der Funchs'chen
Differentialgleisung zweiter Ordnung durch Hypergeometrische Polynome.  
{Math. Ann.}  {116}, 413-421 (1939).
%
%
\bibitem{erdelyi2}Erd\'elyi, A.:  
The Fuchsian equation
of second order with four singularities. { Duke
Math. J.}  {9}, 48-58 (1942). 
%
%
\bibitem{erdelyi3}Erd\'elyi, A.: Certain
expansions of solutions of the
Heun equation. { Q. J. Math. (Oxford)} { 15}, 62-69 (1944).
%
%
		\bibitem{erdelyi-2}Erd\'ely, A.,  Magnus, W., Oberhettinger, F.,  Tricomi, F.G.:{ Higher Transcendental Functions} vol. 2.  
	McGraw-Hill, New York (1953).
%
\bibitem{ganguly-1}
Ganguly, A.: Associated Lam\'e equation, periodic potentials and sl(2,R).  { Mod. Phys. 
Lett.  A} { 15}, 1923-1930 (2002). 
%
\bibitem{ganguly-2}
Ganguly, A.: Associated Lam\'e  and various other new classes of elliptic potentials from sl(2,R) and
related orthogonal polynomials. { J. Math. Phys.} { 43}, 1980-1999 (2002). 
	%
	%
	\bibitem{khare} Khare, A., Sukhatme, U.: New solvable and quasiexactly solvable periodic potentials.  
	{J. Math. Phys.} { 40}, 5473-5494 (1999).
%
\bibitem{khare2}Khare, A., Sukhatme, U.: Some exact results for mid-band and zero band-gap states
	of associated Lam\'e potentials. {J. Math. Phys.} { 42}, 5652-5664 (2001).
%
	%
	\bibitem{arscott-ronveaux}Ronveaux, A. (editor): {Heun's
		differential equations}. Oxford University Press, New York (1995).
%
	%
\bibitem{lea}El-Jaick, L.J., Figueiredo, B.D.B.: Transformations of Heun's equation and its integral relations.   
 {J. Phys. A: Math. Theor.} { 44}, 075204 (2011). 
%
%
	\bibitem{erdelyi-3}Erd\'elyi, A., Magnus, W., Oberhettinger, F., Tricomi, F.G.:  
	{Higher Transcendental Functions}  
vol. 3. McGraw-Hill, New York (1955).
	%
	\bibitem{gautschi}Gautschi, W.: Computational aspects
	of three-term recurrence relations.  {SIAM Rev.} 
	{9}, 24-82 (1967).
	%
	\bibitem{arscott}Arscott, F.M.: {Periodic Differential Equations}. Pergamon Press (1964).
	%
	\bibitem{arscott-2}Arscott, F.M.:  
	Latent roots of tri-diagonal matrices.    
	{ Edinburgh Mathematical Notes} {44}, 5-7 (1961).
	%
\bibitem{watson}Whittaker, E.T., Watson, G.N.: {A Course of Modern Analysis}.   Cambridge University Press (1954).
	%
	\bibitem{knopp}Knopp, K.:{ Infinite Sequences and Series}. Dover (1956).
	%
\bibitem{liu}Liu, J.W.: Analytical solutions to the 
generalized spheroidal wave
equation and the Green's function of one-electron diatomic molecules.       
{J. Math. Phys.} {33}, 4026-4036 (1992).
%
	%
\bibitem{hodge}Hodge, D.B.: Eigenvalues and eigenfunctions of the spheroidal wave equation.  
{ J. Math. Phys.} {11}, 2308-2312 (1970). 
%
\bibitem{falloon}Falloon, P.E., Abbott, P.C., Wang, J.B.:    
Theory and computation of spheroidal wavefunctions.    
{J. Phys. A: Math. Gen.} {36}, 5477–5495 (2003).
	%
\bibitem{Gradshteyn}Gradshteyn, I.S., Ryzhik, I. M.: {Table of Integrals, Series, and Products}. Elsevier (2014).
%
	\bibitem{ince2}Ince, E.L.: The periodic Lam\'e functions. {Proc. Roy. Soc.  Edinburgh} { 60}, 47-63 (1940).
	%
%
%
	%
	\bibitem{decarreau1}Decarreau, A., Dumont-Lepage, M.G., 
	Maroni, P., Robert, A., Ronveaux,
	A.: Formes canoniques
	des \'equations confluentes de
	l'\'equation de Heun. {Ann. Soc. Sci. Brux.} 
	{T92(I-II)}, 53-78 (1978). 
	%
	%
	\bibitem{baber}Baber, W. G., Hass\'e, H.R.: The two centre
	problem in wave mechanics. {Proc. Cambr. Philos. Soc.} {25},  
	564-581 (1935).
	%
	%
	%
%
%
\bibitem{fisher}Fisher, E.: Some differential
equations involving three-term recursion formulas. {Phil. Mag.} {24}, 245-256 (1937).
%
	%
%
%
	%
\bibitem{slater}Slater, J.C.:   
A soluble problem in energy bands. {Phys. Rev.} {87}, 807-835 (1951).
%
	\bibitem{matveev}Matveev, V.B.,  
	 Smirnov,A.O.: On the link between the Sparre equation and
	Darboux-Treibich-Verdier equation. {Lett. Math. Phys.} {76}, 283-295
	(2006).  
%
\bibitem{volcano}El-Jaick, L.J., 
 Figueiredo, B.D.B.: A limit of the confluent Heun equation and the Schr\"{o}dinger equation for an inverted potential and for an electric dipole.{ J. Math.  Phys.} {50}, 123511 (2009).
%
	\bibitem{ince3}Ince, E.L.: Further investigations into the periodic
	Lam\'e functions. {Proc. Roy. Soc.  Edinburgh} {60}, 83-99 (1940).
	%
	%
	%
%
\bibitem{magnus} Magnus, W., Winkler, S.:{ Hill's Equation}. Wiley, New York (1966).
%
%
\bibitem{volkmer}
Volkmer, H.: Coexistence of periodic solutions of Ince's equation. 
{Analysis} {23}, 97-105 (2003).
%
%
%
%
\bibitem{armenio}Ishkhanyan, T.A.,  Shahverdyan, T.A., 
 Ishkhanyan, A. M.: Expansions of the solutions
of the general Heun equation governed by two-term recurrence relations for
coefficients. {Adv. High Energy Phys. }  426367 (2018).
%
\bibitem{et-al} 
Chiang, Y.M.,  A.Ching, A., Tsan, C.Y.: 
Symmetries of the Darboux Equation. 
 { Kumamoto J.Math.} { 31}, 
 15-48 (2018).
\bibitem{hulten} 
Karayer, H., Demirhan, D.,  B\"{u}y\"{u}kkili\c{c}, F.:  A particular solution of Heun equation for Hulthen and Woods-Saxon potentials. {Ann. Phys.}  
{ 526}, 527-532 (2014).
%
\bibitem{eu-2002}Figueiredo, B.D.B.:  
On some solutions to generalized spheroidal wave equations and applications. {J. Phys. A: Math. Gen.}  {35}, 2877 (2002).
%
\end{thebibliography}
\end{document}